\documentclass[iop,numberedappendix,appendixfloats,preprint2,tighten]{emulateapj}

\usepackage{subfigure}
\usepackage{transparent}
\usepackage{graphicx} 
\usepackage{chngcntr} 
\usepackage{multirow}
\usepackage{subscript} 
\usepackage[usenames]{color}
\usepackage[T1]{fontenc}

\def\Plus{\texttt{+}}

\renewcommand{\thefootnote}{\alph{footnote}}

\received{}
\revised{}
\accepted{}
\shortauthors{Goldman et al.}
\shorttitle{``DUSTiNGS V: $P$--$L$ relation''}

\begin{document}

\title{An Infrared Census of DUST in Nearby Galaxies with \textit{S\MakeLowercase{pitzer}} (DUST\MakeLowercase{i}NGS): \\ V. The Period-Luminosity relation for dusty metal-poor AGB stars} 

\email{$\dagger$ Email: sgoldman@stsci.edu} 

\author{S.~R.\ Goldman\altaffilmark{1}$^{\dagger}$} 
\author{M.~L.\ Boyer\altaffilmark{1}} 
\author{K.~B.~W.\ McQuinn\altaffilmark{2,3}} 
\author{P.~A. \ Whitelock\altaffilmark{4,5}} 
\author{I.~McDonald\altaffilmark{6}}
\author{J.~Th.~van~Loon\altaffilmark{7}} 
\author{E.~D.\ Skillman \altaffilmark{8}} 
\author{R.~D.\ Gehrz\altaffilmark{8}} 
\author{A.~ Javadi\altaffilmark{9}} 
\author{G.~C.\ Sloan\altaffilmark{1,10}} 
\author{O.~C. \ Jones\altaffilmark{11}} 
\author{M.~A.~T.\,Groenewegen\altaffilmark{12}}
\author{J.~W.\ Menzies\altaffilmark{4}} 

\altaffiltext{1}{Space Telescope Science Institute, 3700 San Martin Drive, Baltimore, MD 21218, USA}

\altaffiltext{2}{University of Texas at Austin, McDonald Observatory, 2515 Speedway, Stop C1402, Austin, Texas 78712 USA}

\altaffiltext{3}{Rutgers University, Department of Physics and Astronomy, 136 Frelinghuysen Road, Piscataway, NJ 08854, USA}

\altaffiltext{4}{South African Astronomical Observatory, PO Box 9, 7935 Observatory, South Africa}

\altaffiltext{5}{Department of Astronomy, University of Cape Town, 7701 Rondebosch, South Africa} 

\altaffiltext{6}{Jodrell Bank Centre for Astrophysics, Alan Turing Building, University of Manchester, M13 9PL, UK}

\altaffiltext{7}{Lennard-Jones Laboratories, Keele University, ST5 5BG, UK}

\altaffiltext{8}{Minnesota Institute for Astrophysics, School of Physics and Astronomy, 116 Church Street S. E., University of Minnesota, Minneapolis, MN 55455, USA}

\altaffiltext{9}{School of Astronomy, Institute for Research in Fundamental Sciences (IPM), PO Box 19395-5531, Tehran, Iran}

\altaffiltext{10}{University of North Carolina Chapel Hill, Chapel Hill, NC 27599-3255}

\altaffiltext{11}{UK Astronomy Technology Centre, Royal Observatory, Blackford Hill, Edinburgh EH9 3HJ, UK}

\altaffiltext{12}{Koninklijke Sterrenwacht van Belgi\"e, Ringlaan 3, B-1180 Brussels, Belgium}

\begin{abstract}
The survey for DUST In Nearby Galaxies with \textit{Spitzer} (DUSTiNGS) has identified hundreds of candidate dust-producing Asymptotic Giant Branch (AGB) stars in several nearby metal-poor galaxies. We have obtained multi-epoch follow-up observations for these candidates with the \textit{Spitzer Space Telescope} and measured their infrared (IR) lightcurves. This has allowed us to confirm their AGB nature and investigate pulsation behavior at very low metallicity. We have obtained high-confidence pulsation periods for 88 sources in seven galaxies. We have confirmed DUSTiNGS variable star candidates with a 20\% success rate, and determined the pulsation properties of 19 sources already identified as Thermally-Pulsing AGB (TP-AGB) stars. We find that the AGB pulsation properties are similar in all galaxies surveyed here, with no discernible difference between the DUSTiNGS galaxies (down to 1.4\% solar metallicity; [Fe/H]\,=\,$-1.85$) and the far more metal-rich Magellanic Clouds (up to 50\% solar metallicity; [Fe/H]\,=\,$-0.38$). These results strengthen the link between dust production and pulsation in AGB stars and establish the IR Period-Luminosity ($P$\,--\,$L$) relation as a reliable tool ($\pm$\,4\%) for determining distances to galaxies, regardless of metallicity.
\end{abstract}

 
\section{INTRODUCTION}
\label{sec:intro} 

\setcounter{footnote}{0}

Variable Stars on the asymptotic giant branch (AGB) are known to show a linear correlation between the logarithm of the period and luminosity \citep{Gerasimovic1928}. This relationship reveals details about the stellar physics that drives AGB evolution and, like the period-luminosity ($P$--$L$) relationship for Cepheids and RR Lyrae, is a useful distance indicator. As such, the AGB $P$--$L$ relationship has been studied extensively over the years \citep{Feast1989,Hughes1990,Ita2004,Glass2009,Soszynski2009}. Many of these studies focus only on Galactic and Magellanic AGB variables, since AGB stars are difficult to identify and resolve in more distant galaxies, mainly due to  extinction by circumstellar dust that almost always accompanies large amplitude pulsation. As a result, the properties of the AGB $P$--$L$ relation are not well constrained at low metallicity, and some uncertainty remains regarding its usefulness as a distance indicator in metal-poor galaxies and/or metal-poor galaxy halos \citep{Feast2002}.

Strong AGB pulsations and dust production are known to be tightly linked \citep{Lagadec2008,Sloan2016,McDonald2018,McDonald2019}. As a star evolves through the AGB phase, the strength of pulsations grows, which simultaneously levitates more material to large circumstellar radii where it condenses into dust. AGB stars can either be oxygen-rich (M-type) or carbon-rich (carbon stars), producing silicate-rich dust and carbonaceous dust, respectively. The processes dictating the envelope chemistry are third dredge-up events (TDUs) and hot-bottom burning (HBB), which depend primarily on a star's initial mass \citep[see the reviews by ][]{Herwig2005,Karakas2014a}, and metallicity. Carbon stars produce enough carbon internally that they can produce significant amounts of dust regardless of their initial metallicity \citep{Sloan2012}, while oxygen-rich stars require heavier elements (e.g. Fe, Mg, Al, Si) which must be produced by a previous generation of stars or the byproducts of TDU and HBB \citep[e.g.][]{Sloan2010,Bladh2015}. 

How AGB mass loss, dust production, and evolution are affected by metallicity is still unclear. Variability studies in the Large (LMC) and Small (SMC) Magellanic Clouds have produced large samples of long-period variables (LPVs), but these samples span a narrow range in metallicity. Additional works have  discovered large samples of LPVs in globular clusters, and smaller samples in nearby dwarf galaxies (see \S\ref{section:metal poor LPV samples}). Here we present the first large-scale IR survey of LPVs in nearby galaxies, reaching lower metallicity than ever before. 

\begin{deluxetable*}{lccccccc}
\tablewidth{\linewidth}
\tabletypesize{\normalsize}
\tablecolumns{8}
\tablecaption{Surveys of LPVs in nearby galaxies, including this work \label{table:galaxies}}

\tablehead{
&
\multirow{2}{*}{$d$}&
&
\multirow{2}{*}{12\,+\,log(O/H)}&
\multirow{2}{*}{$M$\textsubscript{V}}&
\colhead{Number of sources} &
\multirow{2}{*}{Number from} &
\multirow{2}{*}{Previous work} \\
\colhead{Galaxy} &
\multirow{2}{*}{(Mpc)} &
\colhead{[Fe/H]}& 
\multirow{2}{*}{(mag)} &
\multirow{2}{*}{(mag)} &
\colhead{with high-confidence}&
\multirow{2}{*}{previous work} &
\multirow{2}{*}{Reference} \\
&&&&& \colhead{fit periods} &
}
\startdata    
And\,IX    & 0.77 & $-2.20 \pm 0.20$ &  $\ldots$ & $-8.1 \pm 1.\rlap{1}$         & 0 & $\ldots$  \\
DDO\,216   & 0.92 & $-1.40 \pm 0.02$ & $7.93 \pm 0.13$ & $-12.2 \pm 0.2$        & 5  & $\ldots$ \\
Fornax\rlap{\tablenotemark{a}}     & 0.15 & $-0.99 \pm 0.01$ & $\ldots$ & $-13.4 \pm 0.3$        & $\ldots$ & 7 & \citet{Whitelock2009} \\ 
IC\,10     & 0.79 & $-1.28$          & $8.19 \pm 0.15$ & $-15.0 \pm 0.2$        & \llap{1}6 & $\ldots$ \\   
IC\,1613   & 0.76 & $-1.6 \pm 0.2\rlap{0}$  & $7.62 \pm 0.05$ & $-15.2 \pm 0.2$        & \llap{1}5 & 9 & \nocite{Menzies2015}Menzies et al. (2015)\\
Leo I\rlap{\tablenotemark{a}}      & 0.25 & $-1.43 \pm 0.01$ &    $\ldots$  & $-12.0 \pm$ 0.3 & $\ldots$ & \llap{2}6 & \citet{Menzies2010} \\
NGC\,147   & 0.68 & $-1.1 \pm 0.1\rlap{0}$  &    $\ldots$  & $-14.6 \pm 0.1$        & 8  & \llap{16}8 & \citet{Lorenz2011} \\ 
NGC\,185   & 0.62 & $-1.3 \pm 0.1\rlap{0}$  &   $\ldots$   & $-14.8 \pm 0.1$        & \llap{2}9 & \llap{41}9 & \citet{Lorenz2011} \\
NGC 6822\rlap{\tablenotemark{a}}   & 0.46 & $-1.0 \pm 0.5$   &    $\ldots$   & $-15.2 \pm 0.2$        &  $\ldots$  & \llap{50}\Plus & \citet{Whitelock2012} \\
Phoenix\rlap{\tablenotemark{a}}    & 0.42 & $\,\llap{$-$}1.37 \pm 0.2$  &   $\ldots$ & $-9.9 \pm 0.\rlap{4}$  &  $\ldots$  & 1 & \citet{Menzies2008} \\
Sag\,DIG   & 1.07 & $-2.1 \pm 0.2\rlap{0}$  & $7.26-7.50$     & $-11.5 \pm 0.3$        & 0  & 3 & \citet{Whitelock2018} \\ 
Sculptor\rlap{\tablenotemark{a}}   & 0.09 & $-1.68 \pm 0.01$ &    $\ldots$  & $-11.1 \pm 0.5$        &  $\ldots$  & 2 & \citet{Menzies2011} \\
Sextans\,A & 1.43 & $-1.85$          & $7.54 \pm 0.06$ & $-14.3 \pm 0.1$        & 6 & $\ldots$ \\   
Sextans\,B & 1.43 & $-1.6$           & $7.53 \pm 0.05$ & $-14.5 \pm 0.2$        & 0 & $\ldots$  \\          
WLM        & 0.93 & $-1.27 \pm 0.04$ & $7.83 \pm 0.06$ & $-14.2 \pm 0.1$        & 9 & $\ldots$
\enddata 
\tablenotetext{}{\small{\textbf{Note.} Distances, [Fe/H], and $M$\textsubscript{V} are from \citet{McConnachie2012} and references therein; the metallicity for Sextans B is from \citet{Bellazzini2014}. \citet{Kirby2017} also derived a higher metallicity for Sag DIG of [Fe/H]=$-1.88^{+0.13}_{-0.09}$, based on RGB star spectroscopy. ISM gas-phase oxygen abundances (12\,$+$\,log(O/H)) are from \citet{Mateo1998}, \citet{Lee2006}, and \citet{Saviane2002}. Alternative name for DDO 216: Pegasus dwarf Irregular.\\ 
\tablenotemark{a}Galaxies not analyzed in this work.\\ }}
\end{deluxetable*}

\subsection{Long-Period Variable stars}

The driving force behind AGB pulsations is poorly understood. While sources within the instability strip of the Hertzsprung Russell (HR) diagram (e.g., Cepheid variables or RR Lyrae stars) pulsate as a result of a gravity-opacity instability known as the $\kappa$-mechanism, the large convection cells within an AGB star would likely disrupt both spherical symmetry and this mechanism \citep{Liljegren2018}.

Red giants and supergiants (RSGs) and AGB stars can follow several sequences on the $P$--$L$ diagram often labeled A through E \citep{Wood1999, Ita2004}\footnote{These sequences have also been labeled as 1--4, D, and E \citep[e.g.,][]{Riebel2010}.}. Many of the sources on these sequences pulsate in multiple modes, with secondary periods falling on the other sequences \citep[see][]{Trabucchi2018}. The B and C$^{\prime}$ sequences are composed of Red Giant Branch (RGB) and AGB first overtone radial pulsators. The dusty and evolved AGB stars, or Mira variables, primarily lie along the fundamental mode \citep[sequence 1;][]{Riebel2010}, also known as sequence C \citep{Wood1999}, however some ($\sim$\,30\%) lie along sequence D, with pulsation periods between $500-2000$ days. These have been referred to as Long Secondary Periods (LSPs), yet are clearly the dominant mode in some evolved stars \citep{Nicholls2009,Trabucchi2017}; the sequence will be discussed further in Section \ref{sec: lsp}. 

% . This sequence is known to be related to mass loss \citep{Wood2009} but how remains unclear

\subsection{Dust at low metallicity}

AGB stars can have considerably different lifetimes as a result of their different masses. The main-sequence lifetime of AGB progenitors (low- and intermediate-mass stars; 0.8$-$8 M$_\odot$) is between 0.1 and 12 Gyr, after which they typically spend 20\% of that time as red giants, $\sim$\,$1\%$ of that as early-AGB stars, and $\sim$\,$0.1\%$  as Thermally-Pulsing AGB (TP-AGB) stars \citep{Marigo2008,Marigo2017,Javadi2011b,Javadi2017}. On the TP-AGB, these stars will produce the most dust and contribute the most mass back to the ISM \citep[see review by ][]{Hoefner2018}. Recent works exploring the metallicity dependence of dust production in carbon stars have produced mixed results. A strong dependence was originally suggested by \citet{Loon2000} and corroborated  by \citet{Loon2006} and \citet{vanLoon2005}, while \citet{McDonald2011,Sloan2012,Sloan2016} found little to no dependence. \citet{Nanni2013}, \citet{Nanni2014}, and \citet{Ferrarotti2006} have given estimates on the metallicity dependence of the dust production using theoretical models. Work within the Galaxy and the Magellanic Clouds has also allowed us to study the effect of metallicity on the mass loss of oxygen-rich AGB stars \citep{Loon2000,Loon2006,Goldman2017}, with results showing little-to-no effect on the measured mass loss rates.

The DUST in Nearby Galaxies with \textit{Spitzer} survey \citep[DUSTiNGS;][hereafter Paper I]{Boyer2015a} searched for dust-producing AGB stars in 50 nearby (<\,1.5\,Mpc) metal-poor ($-2.7$\,<\,[Fe/H]\,<\,$-$1.0) dwarf galaxies using \textit{Spitzer Space Telescope} \citep{Werner2004,Gehrz2007} InfraRed Array Camera \citep[IRAC;][]{Fazio2004} channels 1 and 2. The survey discovered hundreds of candidate dust-producing AGB stars at metallicities as low as 0.6\% solar and provided no evidence for a strong metallicity dependence in overall dust production \citep[][hereafter Paper II]{Boyer2015}. Observations at wavelengths longer than $\lambda$\,=\,5\,$\mu$m, where thermal emission from circumstellar dust dominates the IR spectral energy distribution (SED), will be required to confirm this. \citet[][hereafter Paper IV]{Boyer2017} identified 146 carbon- and oxygen-rich type stars by exploiting the strength of the water features in M-type stars and the CN+C$_2$ features found in carbon stars. Though most (120) of these sources were classified as carbon rich, 26 were identified as M-type. These observations showed that dust is produced both by carbon- and oxygen-rich AGB stars over the full metallicity range spanned by DUSTiNGS. This suggests that metal-poor high-mass AGB stars can produce dust as early as 30\,Myr after forming (for a 10 M$_{\odot}$ star), while lower-mass carbon stars form dust after roughly 0.3--3.6 Gyr \citep{vanLoon2005}. AGB stars are therefore likely important contributors of dust in the early Universe. This work also led to the discovery of a potential dust-producing super-AGB star in IC 10 with an assumed mass $\sim$\,$8-12$\,M$_{\odot}$ and strong water absorption indicative of an AGB star. Super-AGB stars are more massive (6\,M$_{\odot} \lesssim M \lesssim$ 9\,M$_{\odot}$) AGB stars that are capable of fusing carbon and developing a degenerate oxygen-neon core. There is evidence that they can be dusty \citep{Javadi2013}, and produce the ONeMg white dwarfs that are responsible for neon nova explosions \citep{Evans2012}. These stars may also be capable of ending in an electron-capture supernova without developing an iron core like the observationally similar RSGs \citep{Doherty2015}. We did not detect variability in this source, due to a lack of temporal coverage (2-epochs).

\subsection{Metal-poor LPV samples} 
\label{section:metal poor LPV samples}

Much of the information known about metal-poor LPVs is from the Optical Gravitational Lensing Experiment \citep[OGLE;][]{Udalski1997}, Massive Compact Halo Object \citep[MACHO;][]{Alcock1997}, and \textit{Spitzer} Surveying the Agents of a Galaxy's Evolution surveys \citep[SAGE;][]{Meixner2006,Gordon2011,Riebel2010,Riebel2015}, with a handful of surveys in other galaxies (shown in Table \ref{table:galaxies}). While a large number of metal-rich samples exists \citep{Huang2018,Yuan2018}, the majority of the more metal-poor LPVs have been found in Leo I, NGC 185, NGC 147, and NGC 6822, galaxies all only slightly more metal-poor than the SMC. The most metal-poor sources to date were found by \citet{McDonald2010} who discovered two sequence D variables in the globular cluster M15 at [Fe/H]\,=\,$-$2.37 dex \citep{Harris1996}, and by \citet{Whitelock2018} who discovered three LPVs in the Sagittarius Dwarf Irregular Galaxy (Sag DIG). The metallicity of Sag DIG has been measured in both stars, using red giants \citep[{[Fe/H]\,=\,$-1.88^{+0.13}_{-0.09}$};][]{Kirby2017} and isochrones \citep[{[Fe/H]\,=\,$-$2.1};][]{Momany2002}, and also the gas \citep[{12\,$+$\,log(O/H) = 7.26 $-$ 7.50;}][]{Skillman1989,Saviane2002}. One of these LPVs in Sag DIG has a pulsation period of 950 days, indicating a very late stage of evolution, and was found to be oxygen-rich (Paper IV). Variables have also been detected in globular clusters \citep{Clement2001,Feast2002,Lebzelter2005}. However, their low mass limits these sources to the lower regions of the $P$--$L$ sequences.

It is difficult to study the $P$--$L$ relation at lower metallicities because so few LPVs have been discovered in this regime. The DUSTiNGS survey initially identified several LPV candidates using two-epoch photometry. Here, we follow up with additional epochs and provide a larger sample to populate the $P$--$L$ diagram over a large metallicity range ($-$1.27\,>\,[Fe/H]\,>\,$-$1.85). DUSTiNGS is the first large-scale IR survey to identify the dustiest evolved stars in these galaxies. These stars can be obscured in the near-IR and optical. Observing in the IR ensures that all of the prominent dust producers are detected.

%%%%%%%%%%%%%%%%%%%%%%%%%%%%%%%%%%%%%%%%%%
\begin{deluxetable*}{rcccccc}
\tablewidth{0.8\linewidth}
\tabletypesize{\normalsize}
\tablecolumns{8}
\tablecaption{The archival \textit{Spitzer} observations \label{table:archival_spitzer_obs}}
\tablehead{
\multirow{2}{*}{Galaxy}&
\multirow{2}{*}{PID}&
\multirow{2}{*}{AOR}&
\colhead{R.A.}&
\colhead{Decl.}&
\colhead{Start Date}&
\colhead{$t_{\rm exp}$}\\
\colhead{}&
\colhead{}&
\colhead{}&
\colhead{(J2000)}&
\colhead{(J2000)}&
\colhead{(UT)}&
\colhead{(h)}
}
\startdata
IC 10   & 69    & 442496\rlap{0} & 00h20m24.50s & +59d17m30.0s & 2004-Jul-23 & 0.12 \\
IC 10   & 61001 & \llap{3}320422\rlap{4} & 00h20m24.00s & +59d18m14.0s & 2010-Jan-29 & 0.92 \\
IC 10   & 61001 & \llap{3}320396\rlap{8} & 00h20m24.00s & +59d18m14.0s & 2010-Feb-19 & 0.87 \\
IC 10   & 61001 & \llap{3}320345\rlap{6} & 00h20m24.00s & +59d18m14.0s & 2010-Mar-10 & 0.95 \\
IC 10   & 61001 & \llap{3}320294\rlap{4} & 00h20m24.00s & +59d18m14.0s & 2010-Sep-09 & 0.87 \\
IC 10   & 61001 & \llap{3}320243\rlap{2} & 00h20m24.00s & +59d18m14.0s & 2010-Oct-04 & 0.83 \\
IC 10   & 61001 & \llap{3}320192\rlap{0} & 00h20m24.00s & +59d18m14.0s & 2010-Oct-14 & 0.83 \\
IC 1613 & 61001 & \llap{3}318400\rlap{0} & 01h04m58.20s & +02d09m44.0s & 2010-Jan-26 & 0.97  
\enddata
\tablenotetext{}{\small{\textbf{Note.} The full catalog ($n$\,=\,139) is available for download in the electronic version and on VizieR. Information on the Program IDs (PIDs) and individual observations (AORs) can be found through the \textit{Spitzer Heritage Archive}.\\}}
\end{deluxetable*}

\section{Data and Observations}

We construct lightcurves using 3.6 and 4.5\,$\mu$m imaging data from the IRAC on board \textit{Spitzer}, with a mix of programs from both the cryogenic and post-cryogenic phase. Data include new and archival observations from 2003 to 2017. 

\subsection{DUSTiNGS}
\begin{figure}
\includegraphics[width=\linewidth]{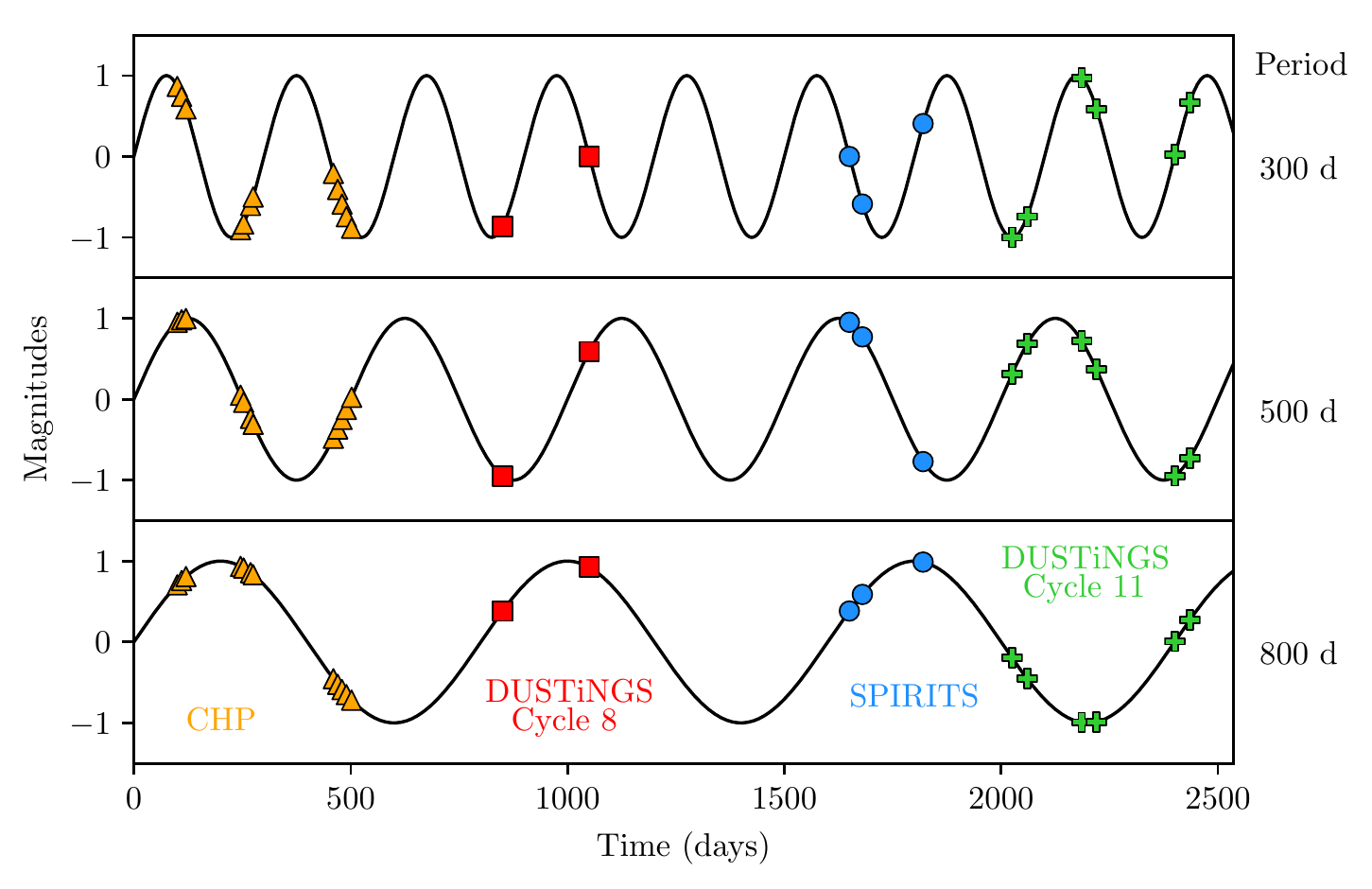}
\caption{An artificial lightcurve showing the two initial DUSTiNGS observations (red) and the six follow-up observations (green) fit with pulsation periods of 300, 500, and 800 days. Also shown is the cadence of SPIRITS (blue) and CHP (orange) surveys; the cadence of the remaining archival programs is shown in Table \ref{table:archival_spitzer_obs}. \\}
\label{fig:dustings_example}
\end{figure}

The DUSTiNGS data include both the original Cycle 8 data obtained in 2011--2012 (PID: 80063) and data obtained during the Cycle 11 follow-up program (PID: 11041) in 2015--2016. Lightcurves are sparsely sampled owing to the spacing of the \textit{Spitzer} visibility windows for the DUSTiNGS galaxies, which are roughly 4--6 weeks long. There are typically two windows each year, separated by approximately 6 months. The Cycle 8 program, described in Paper I, obtained two epochs, one in each visibility window. The Cycle 11 program obtained six additional epochs, with a pair of observations at the beginning and end of each of the three consecutive visibility windows. The cadence is illustrated in Figure \ref{fig:dustings_example}.

\subsection{Archival Data} 

By taking advantage of rich archival history of the \textit{Spitzer Space Telescope}, we have been able to use data from 11 observing programs (Table \ref{table:archival_spitzer_obs}). Most of the archival data that we use here are from two programs: 

\paragraph{SPIRITS}The \textit{SPitzer} InfraRed Intensive Transients Survey \citep[SPIRITS; ][]{Kasliwal2017} program (Cycles $10-12$, PID: 10136, 11063) was aimed at discovering explosive transients, eruptive variables, and new IR events lacking optical counterparts. The SPIRITS targets include a few of the DUSTiNGs galaxies: IC 1613, NGC 147, NGC 185, Sextans A, and Sextans B. These observations were taken between the original DUSTiNGS epochs and the follow-up DUSTiNGS observations, filling a gap in our temporal coverage (Figure \ref{fig:dustings_example}). Additional SPIRITS epochs cover the same epochs covered by the DUSTiNGS Cycle 11 observations and are included here.

\paragraph{The Carnegie Hubble Program}The Carnegie Hubble Program \citep[CHP; Cycle 6, PID: 61001;][]{Freedman2011} was aimed at determining distances to nearby galaxies using Cepheid variables. The CHP observations, taken between July 2009 and March 2010, preceded both the DUSTiNGS and SPIRITS observations. The time between each epoch is $\sim$\,10\,d given the focus on short-period variability (Figure \ref{fig:dustings_example}). 

We also use data from eight additional programs (individual observations are listed in Table \ref{table:archival_spitzer_obs}) which sporadically sample the lightcurves between CHP and DUSTiNGS. The observations target 10 galaxies (listed in Table \ref{table:galaxies}) that span a range in size ($-$8.1\,>\,$M$\textsubscript{V}\,>\,$-$15.2 mag), distance (0.62\,--\,1.43\,Mpc), and most notably, metal content ($-$1.27\,>\,[Fe/H]\,>\,$-$1.85). We will use this range in metallicity to investigate its effect on the pulsation properties of evolved stars.

\section{Methods}
\subsection{PSF Photometry}
We have performed point-spread function (PSF) photometry on all of the DUSTiNGS sources and archival data using the DUSTiNGS pipeline (described in Paper I). We performed PSF photometry using DAOphot II and ALLSTAR \citep{Stetson1987} on the co-added frames for the fainter sources ([3.6]\,>\,16 mag) and individual frames for the brighter sources. For the fainter sources, mosaicked images were used to reduce Eddington bias \citep{Eddington1913} in sources near the detection threshold. A mosaicked and subsampled image can smear the PSF, for example, if it includes a rotation between frames. As a result, single frames were used for the brighter sources, which are more sensitive to changes in the PSF. For the photometry of our sample in IC 10, we have adjusted the magnitudes by 0.2 mag to account for foreground interstellar extinction (described further in $\S$\ref{sec:IR_PL_relation}). Paper I provides additional details on the photometry, saturation limits, the photometric correction, and the photometric completeness.

\subsection{Identifying LPVs}
\label{sect: identifying LPVs}

The non-uniformity of the observing programs has resulted in varying depths and spatial coverage for each epoch. Therefore, many lightcurves are sparsely sampled. We have implemented the Lomb-Scargle algorithm \citep{Lomb1976,Scargle1982}, to determine the pulsation periods and pulsation amplitudes. The method fits a simple single-term sinusoidal lightcurve to different frequencies and then normalizes the results using the residuals. This method reduces the effects of unevenly-spaced data using a more appropriate means of weighting within the Fourier transform \citep[see review by ][]{VanderPlas2018}.

\begin{deluxetable*}{crcccccccccccr}
\tablewidth{\linewidth}
\tabletypesize{\normalsize}
\tablecolumns{14}
\tablecaption{The results of 3.6 and 4.5\,$\mu$m lightcurve fitting \label{table:fitting_results}}

\tablehead{
\colhead{{\tiny(1)}}&
\colhead{{\tiny(2)}}&
\colhead{{\tiny(3)}}&
\colhead{{\tiny(4)}}&
\colhead{{\tiny(5)}}&
\colhead{{\tiny(6)}}&
\colhead{{\tiny(7)}}&
\colhead{{\tiny(8)}}&
\colhead{{\tiny(9)}}&
\colhead{{\tiny(10)}}&
\colhead{{\tiny(11)}}&
\colhead{{\tiny(12)}}&
\colhead{{\tiny(13)}}&
\colhead{{\tiny(14)}}
\\
\multirow{2}{*}{Galaxy}& 
\colhead{Target}& 
\colhead{RA}& 
\colhead{Dec}& 
\colhead{\llap{$\langle$}3.6\,$\mu$m\rlap{$\rangle$}}&
\colhead{$\langle$4.5\,$\mu$m$\rangle$}&
\colhead{P}&
\colhead{P[4.5]} &
\colhead{\llap{2}nd fi\rlap{t}}&
\colhead{2nd[4.5]} &
\colhead{amp}& 
\colhead{color}&
\colhead{$\sigma_{[3.6]-[4.5]}$} &
\multirow{2}{*}{Flag} \\
&
\colhead{ID}&
\colhead{(deg)}&
\colhead{(deg)}&
\colhead{(mag)}&
\colhead{(mag)}&
\colhead{(d)}&
\colhead{(d)}&
\colhead{(d)}&
\colhead{(d)}&
\colhead{(mag)}&
\colhead{(mag)}&
\colhead{(\%)}&
} 

\startdata
DDO 216 & 45582 & 352.2400208 & 14.61585045 & 16.65 & 16.46 & 101         & $\ldots$  & 131         &$\ldots$& 0.35 &$\ldots$&$\ldots$& IE \\
DDO 216 & 58694 & 352.2214661 & 14.70353794 & 15.80 & 15.07 & 389         & 389 & 482         & 354         & 1.05 & 0.70 & \llap{1}3.5 & RF \\
DDO 216 & 76368 & 352.1983032 & 14.75455666 & 16.98 & 16.06 & \llap{1}308 & 196 & 223         & 178         & 0.34 & 0.92 & \llap{1}6.2 & IE \\
DDO 216 & 77533 & 352.1968384 & 14.73819065 & 17.04 & 15.91 & 103         & 174 & 110         & \llap{1}008        & 0.23 & 1.14 & \llap{1}3.9 & UF \\
DDO 216 & 83518 & 352.1894836 & 14.68728733 & 16.11 & 15.62 & 212         & 212 & \llap{1}339 & \llap{1}339 & 0.48 & 0.52 & \llap{1}1.8 & UF 

\enddata

\tablenotetext{}{\small{\textbf{Note.} Column 2 lists the DUSTiNGS IDs from Paper II, except for IDs over 5,000,000, which are new in this work. Columns 5 and 6 list the mid-line 3.6 and 4.5\,$\mu$m magnitudes taken from the mid-line values of the best-fit lightcurve, Columns 7 and 9 show the best- and second-best-fit pulsation periods (P, 2nd fit), Columns 8 and 10 list the same values for the 4.5$\mu$m fit, Column 11 lists the fitted peak-to-peak 3.6\,$\mu$m amplitude, Columns 12 and 13 list the median and standard deviation of the [3.6]$-$[4.5] color, and Column 14 lists a quality flag for a high-confidence reliable fit (RF), insufficient epochs (IE), an unreliable fit (UF), or a LPV\,$5000\Plus$ (see Section \ref{section: S-LPV}). The full catalog ($n=261$) is available in the electronic version and on VizieR.}\\}

\end{deluxetable*} 

The nonuniform temporal and spatial coverage of the archival data has resulted in many sources with incomplete lightcurves from which we cannot derive reliable periods. To identify sources with sufficiently sampled lightcurves, we start by calculating the variability index \citep[e.g.\,][]{Gallart2004} that is defined as the ratio of standard deviation of the measurements for a given star to the mean internal photometric uncertainty. A value of 1, 2, and 3 indicates variability at the 1, 2, and 3-sigma levels. We first restrict our lightcurve fitting to stars with variability index >\,1. Second, we exclude sources with <\,6 epochs from our lightcurve analysis, a number that was also concluded as sufficient by \citet{Javadi2015}. Finally, we restrict lightcurve fitting to sources brighter than M$_{[3.6]}$\,=\,$-$7.5 mag, which includes all extreme AGB stars \citep[x-AGB;][]{Blum2006} in the LMC sample from \citet{Riebel2010}. The x-AGB stars are the dustiest AGB stars that are likely in the superwind phase and very close to the end of their evolution. By restricting our sample to stars in the same brightness range as the LMC x-AGB stars, we limit contamination from fainter variable dusty objects, such as young stellar objects (YSOs) and background active galactic nuclei (AGN).

\subsection{Lightcurve Analysis}
\label{section:lightcurve_classification}

For the sources that were included in the lightcurve fitting, frequencies corresponding to 100--2000\,d were fitted to each of the lightcurves using the Lomb-Scargle periodogram. Sources that were considered for further analysis were those which fit the following criteria: 

\begin{enumerate}
\item{A [3.6]$-$[4.5] color of which the standard deviation did not deviate by more than 50\% to eliminate sources that were not clearly dusty\footnote{Sources that lacked 4.5\,$\mu$m measurements were still included in the final categorization and analysis.} across epochs.}
\item{A best-fit solution where the peak frequency within the Fourier power spectrum was more than 6\% higher than any other peak.}
\item{A best-fit solution constrained within a 95\% confidence interval.}
\end{enumerate}

These three criteria determine which stars are included in further analysis. Stars excluded by these criteria are more likely to suffer from aliasing or poor data quality. The quality of the remaining sources was determined visually, with an eye for ensuring that the direction of brightness changes in the lightcurves matched the best-fit lightcurve, especially for short-term changes.  The results of the lightcurve fitting are shown in Table \ref{table:fitting_results}, which includes the mid-line magnitudes, the fit periods and amplitudes for both 3.6 and 4.5 \,$\mu$m data, the color properties and the classification confidence (described further below). An initial inspection of stars that pass these above three criteria indicate that, at these distances, we are only able to reliably determine pulsation periods of the dusty evolved sources using \textit{Spitzer}.

Variations in spatial coverage between epochs caused some stars to be observed in only one of the filters, either [3.6] or [4.5]. Some of these lightcurves can be augmented where [3.6] data does not exist by including the 4.5\,$\mu$m data and using the mean color to derive 3.6\,$\mu$m magnitudes. This was done only if the color was determined to be stable ($\sigma_{[3.6]-[4.5]}$\,<\,20\%) and had at least three epochs with color values. We refer to these photometric points as ``simulated'' photometry and show them in our lightcurves as empty circles (Appendix A, Figure \ref{fig:high_confidence_lightcurves})\footnote{The displayed errors for the simulated photometry include the $\sigma_{[3.6]-[4.5]}$ as well as the 4.5\,$\mu$m photometric uncertainties.}. We have also fitted the lightcurves of any sources with at least six epochs of 4.5\,$\mu$m data similarly using simulated 4.5\,$\mu$m photometric values when possible. Only one source (IC\,10 57276) did not have enough epochs to fulfill this requirement. These data are less sensitive (and therefore noisier), so we include them in Table \ref{table:fitting_results} but use only the [3.6] data for further classification. For 65\% of the full sample, the 3.6 and 4.5$\mu$m periods are the same. For 87\% of the sample, the periods agree to within 10\%. For the remaining 11 sources, the standard deviation of the [3.6]$-$[4.5] color is large, and there is a difference in the number of epochs for all but two of the sources. 

\subsection{Lightcurve Categorization}

The lightcurves that pass the three criteria listed in the previous section are further categorized based on our confidence in the fits. These fall into 4 categories, two each considered ``high-confidence'' and ``low-confidence''. These groups are discussed further in Sections \ref{section:high-confidence variables} and \ref{section:low-confidence variables}; the lightcurves of the high-confidence variables are shown in Appendix A. \\

\noindent High-confidence variables:
\begin{itemize}
\item{Reliable Fit (RF) sources with lightcurves that pass a visual inspection, meant to isolate sources with unique fit solutions.}
\item{LPV\,$5000\Plus$ sources that do not necessarily have reliable fits but are clearly variable on long timescales; described further in Section \ref{section: S-LPV}.}
\end{itemize}

\noindent Low-confidence variables:
\begin{itemize} 
\item{Insufficient epochs (IE) variable sources with reliable fits but where the uniqueness of the fit is unclear.}
\item{Unreliable Fit (UF) variable sources that do not pass visual inspection.}
\end{itemize} 

\begin{figure}
 \centering
 \includegraphics[width=\columnwidth]{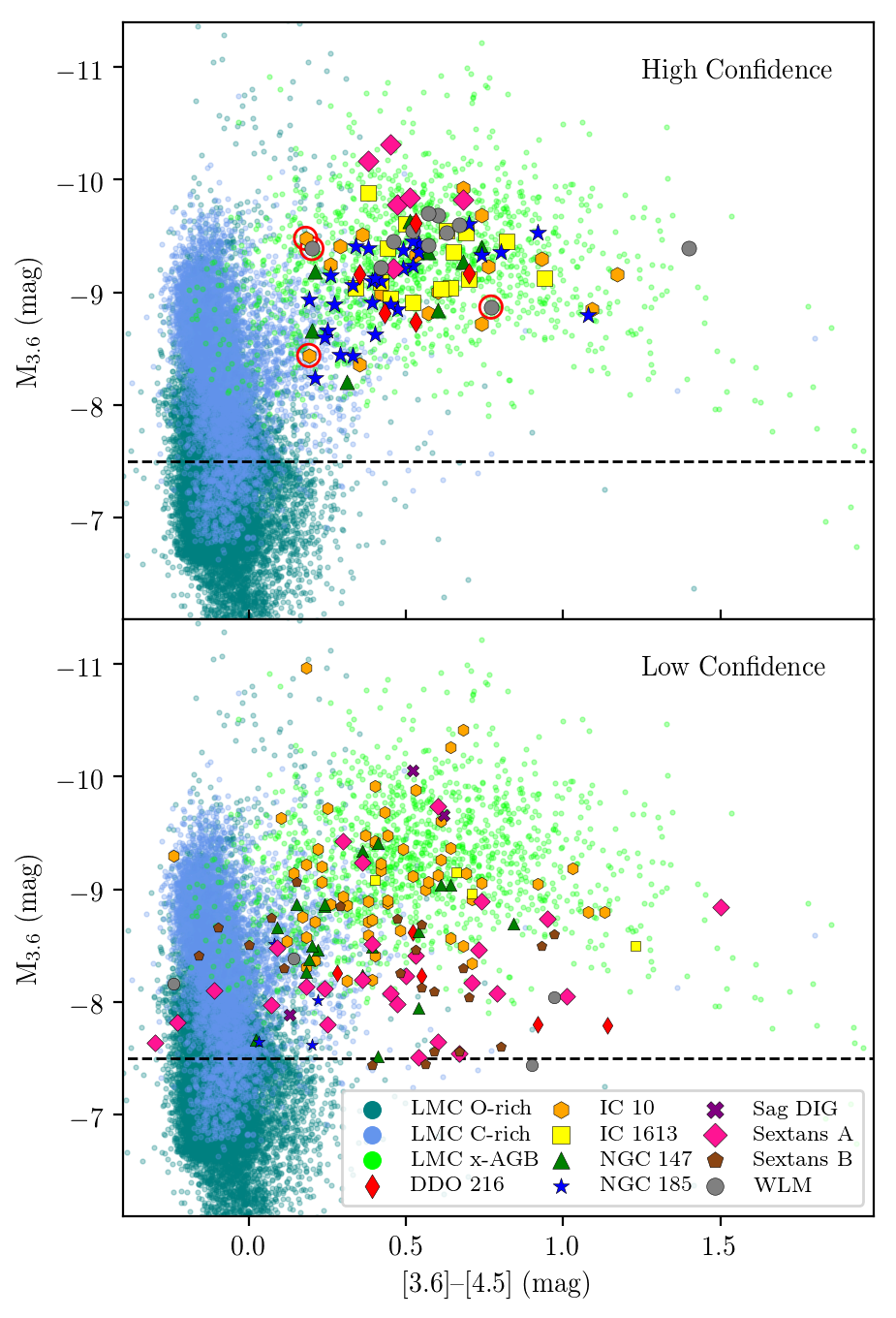}
 \caption{The average \textit{Spitzer} IRAC [3.6]$-$[4.5] vs. 3.6\,$\mu$m absolute magnitude for the high- (top) and low-confidence (bottom) DUSTiNGS sources compared to the LMC sample from \citet{Riebel2010}. M$_{3.6}$ was calculated using the mid-line value of the best-fit lightcurve. The brightness threshold for our lightcurve-fitting analysis is shown with the dotted line. The LPV5000$\Plus$ sources (described in Section \ref{section:lightcurve_classification}) are shown with red circles. Adopted distances to the DUSTiNGS galaxies are shown in Table \ref{table:galaxies}. For the LMC we adopt a distance modulus of $M-m$\,=\,18.52 mag \citep{Kovacs2000}. \\ }
 \label{fig:cmd} 
\end{figure}

\begin{figure}
 \centering
 \includegraphics[width=\columnwidth]{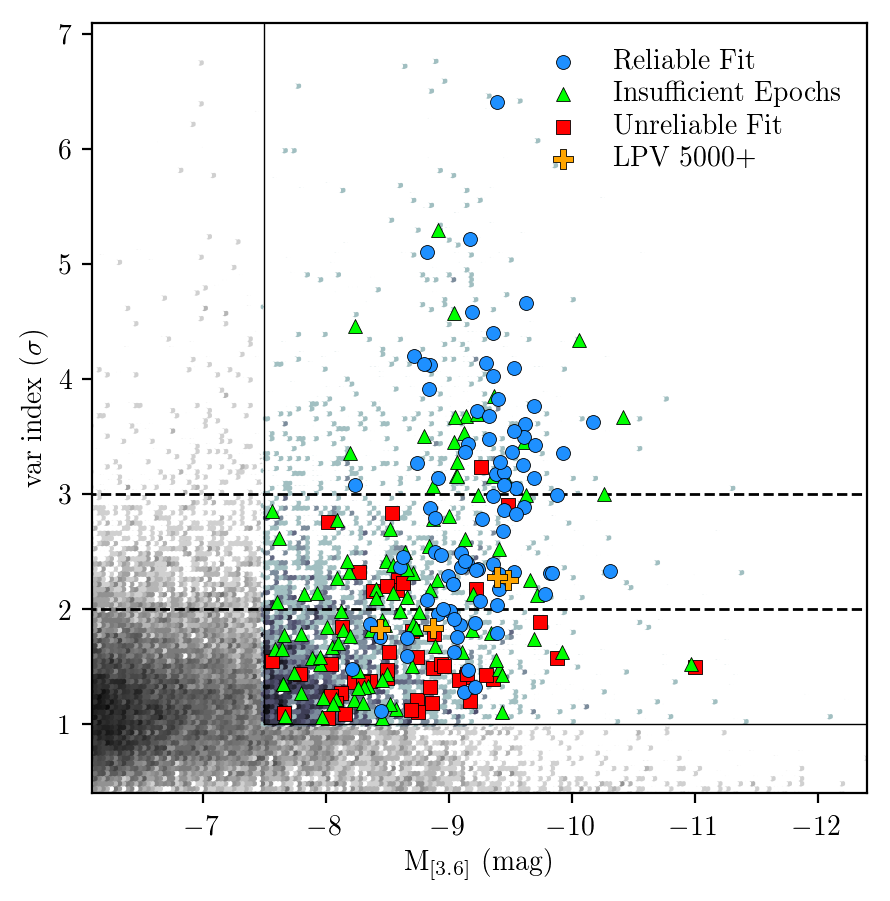}
 \caption{The standard deviation of the 3.6\,$\mu$m magnitude of the lightcurve divided by the average photometric uncertainty (variability index) vs. absolute 3.6\,$\mu$m magnitude for the DUSTiNGS sources with fitted lightcurves. Small points are those which did not meet the requirements of our lightcurve fitting analysis (gray) and those that met the criteria but were not considered credible enough for further analysis (bluish). The 2 and 3\,$\sigma$ intervals have been shown with dotted lines, and the designation of ``LPV\,$5000\Plus$'' is described further in Section \ref{section: S-LPV}.}
 \label{fig:var_index}
\end{figure}

\begin{deluxetable}{lccccc}
\tablewidth{\linewidth}
\tabletypesize{\normalsize}
\tablecolumns{6}
\tablecaption{Results of lightcurve fitting \label{table:gal_stats}}

\tablehead{
\multirow{2}{*}{Galaxy} &
\colhead{Initial} &
\colhead{Var.} &
\colhead{Dusty \&} &
\colhead{Low} &
\colhead{High}  \\
\colhead{} &
\colhead{sample} & 
\colhead{index\,>\,2} &
\colhead{variable} &
\colhead{conf.} &
\colhead{conf.} }
\startdata    
And IX    & 16            & 2            & 1          & 0          & 0 \\
DDO 216   & \llap{24,6}88  & \llap{2,16}6  & \llap{1}5  & 7          & 5 \\
IC 10     & \llap{80,3}51  & \llap{5,55}5  & \llap{22}8 & \llap{5}9  & \llap{1}8 \\
IC 1613   & \llap{3}11    & \llap{9}6    & \llap{2}7  & 5          & \llap{1}5 \\
NGC 147   & \llap{60,4}46  & \llap{3,71}2  & \llap{7}0  & \llap{2}0  & 8 \\
NGC 185   & \llap{53,0}11  & \llap{3,50}6  & \llap{5}8  & 4          & \llap{2}9 \\
Sag DIG   & 51            & \llap{1}0    & 8          & 3          & 0 \\
Sextans A & \llap{22,9}46  & \llap{2,95}9  & \llap{8}3  & \llap{4}0  & 6 \\
Sextans B & \llap{22,6}29  & \llap{2,28}1  & \llap{8}9  & \llap{2}6  & 0 \\
WLM       & \llap{33,1}84  & \llap{2,86}0  & \llap{3}4  & 5          & \llap{1}1 \\
\textit{Total} & \llap{308,0}83 & \llap{24,16}0 & \llap{61}3 & \llap{16}9 & \llap{9}2 
\enddata 
\tablenotetext{}{\small{\textbf{Note.} Table lists the number of high- and low-confidence variables per galaxy. Also shown is the number of sources, the number of sources that we determined variable (variability index\,>\,2), and dusty and variable with a [3.6]$-$[4.5]\,>\,0.2 mag and a variation in the color of less than 20\%.}\\} 
\end{deluxetable}

Recall that both the high- and low-confidence variables have a $\chi^2$ of at least 95\% (Section \ref{section:lightcurve_classification}), yet additional information about the uniqueness or quality of the fit is taken into account. The sources categorized as insufficient epochs are sources where a unique fit solution to the unphased lightcurve is not visibly clear. This may include a lightcurve lacking temporal coverage towards the maximum or minimum of the lightcurve, or where a shorter pulsation period could plausibly fit the source. A source designated as an unreliable fit is typically one that has a change in brightness in several epochs that is in the opposite direction of the change in the best-fit unphased lightcurve, outside the value of the uncertainty. Figure \ref{fig:low_confidence_lightcurves} in Appendix B shows examples of an insufficient epochs source and unreliable-fit source. 

We show the average [3.6]$-$[4.5] vs. absolute 3.6\,$\mu$m magnitude color-magnitude diagram (CMD) of both high- and low-confidence sources in Figure \ref{fig:cmd}. It is clear that we are only sensitive to the highly evolved and dusty x-AGB stars here as relatively dust-free C-AGB and O-AGB stars have smaller amplitude pulsations closer to the level of our photometric uncertainty \citep{Riebel2015}. We are also less sensitive to shorter period variables due to our observing cadence. As a result, we do not obtain any high-confidence variables with median [3.6]$-$[4.5] colors less than 0.1 mag and most of our measured pulsation periods are longer than 200 days. 

Our final results include 92 high-confidence variables with 4 LPV\,$5000\Plus$ and 88 reliable-fit sources with a median period of 437\,d. Figure \ref{fig:var_index} shows the variability of our low- and high-confidence variables. Note that unreliable-fit sources cluster towards low variability index. We categorize the LPV\,$5000\Plus$ sources as high-confidence variables but with a limited temporal baseline, we cannot definitively confirm their periodicity. The lightcurves of the low- and high-confidence variables are available for download in the electronic version and on VizieR. \\

\begin{deluxetable*}{lcccccccc}
\tablewidth{0.85\linewidth}
\tabletypesize{\normalsize}
\tablecolumns{9}
\tablecaption{Comparison to literature periods \label{table:previously_detected_matches}}

\tablehead{
\multirow{2}{*}{Galaxy} & 
\multirow{2}{*}{ID}& 
\multirow{2}{*}{ID\textsubscript{Lit.}} & 
\colhead{RA} & 
\colhead{Dec} & 
\colhead{P} & 
\colhead{P\textsubscript{Lit.}} &
\multirow{2}{*}{\llap{Ty}\rlap{pe}} &
\multirow{2}{*}{Flag} \\ 
& 
& 
& 
\colhead{(deg)} & 
\colhead{(deg)} & 
\colhead{(d)} & 
\colhead{(d)} &
&
 } 

\startdata
IC 1613 & 95038         & \llap{1}093 & \llap{1}6.24090 & 2.15469         & 318 & \llap{3}05 & C & RF \\
IC 1613 & \llap{1}42830 & \llap{3}198 & \llap{1}6.18219 & 2.05673         & 395 & \llap{3}70 & C & RF \\
NGC 147 & 68407         & 112         & 8.35656         & \llap{4}8.55440 & 449 & \llap{4}06 & $\ldots$ & RF \\
NGC 147 & \llap{1}12918 & 171         & 8.27302         & \llap{4}8.47706 & 385 & $\ldots$ & C & RF \\
NGC 147 & \llap{1}13288 & 158         & 8.27241         & \llap{4}8.50505 & 317 & \llap{2}26 & C & IE \\
NGC 147 & \llap{1}23715 & 161         & 8.25264         & \llap{4}8.46111 & 335 & \llap{3}71 & $\ldots$ & UF \\
NGC 185 & 70862         & 009         & 9.79879         & \llap{4}8.32867 & 196 & \llap{2}77 & C & RF \\
NGC 185 & 77053         & 313         & 9.78671         & \llap{4}8.35519 & 418 & \llap{5}19 & $\ldots$ & RF \\
NGC 185 & 83286         & 049         & 9.77481         & \llap{4}8.38233 & 416 & \llap{3}99 & M & RF \\
NGC 185 & 87065         & 062         & 9.76783         & \llap{4}8.36058 & 226 & \llap{2}19 & $\ldots$ & RF \\
NGC 185 & 87213         & 398         & 9.76754         & \llap{4}8.35008 & 837 & \llap{4}27 & $\ldots$ & RF \\
NGC 185 & 89650         & 384         & 9.76315         & \llap{4}8.32895 & 637 & 82 & S & RF \\
NGC 185 & 91361         & 076         & 9.76007         & \llap{4}8.31672 & 360 & \llap{3}58 & C & RF \\
NGC 185 & 92015         & 078         & 9.75888         & \llap{4}8.32191 & 416 & \llap{4}20 & S & RF \\
NGC 185 & 95982         & 404         & 9.75180         & \llap{4}8.32625 & 381 & \llap{3}67 & $\ldots$ & RF \\
NGC 185 & 96014         & 099         & 9.75173         & \llap{4}8.32416 & 327 & \llap{2}87 & C & RF \\
NGC 185 & \llap{1}31142 & 249         & 9.68842         & \llap{4}8.33260 & 227 & \llap{2}31 &  $\ldots$& RF \\
NGC 185 & \llap{1}36723 & 160         & 9.67694         & \llap{4}8.31031 & 347 & \llap{3}61 & C & RF \\
NGC 185 & \llap{1}32331 & 396         & 9.68612         & \llap{4}8.35306 & 227 & \llap{2}34 & C & UF 
\enddata

\tablenotetext{}{\small{\textbf{Note.} Literature values (Lit.) are from \citet{Menzies2015} and \citet{Lorenz2011}, where pulsation periods were derived using near-IR photometry ($JHK$). \\}}
\end{deluxetable*}

\section{Results}
\subsection{High-confidence variables}
\label{section:high-confidence variables}
Figure \ref{fig:cmd} shows that the high-confidence variables occupy the same space as the x-AGB sources found in the SAGE program \citep{Riebel2010}. The reliable-fit sample has a median 3.6\,$\mu$m absolute magnitude of $-$9.2\,$\pm$\,0.26\,(1$\sigma$). Four high-confidence variables have IR colors much redder than the rest of the sample ([3.6]$-$[4.5]\,>\,1): IC\,10 98211, IC\,10 105991, NGC\,185 90369, WLM 84699; their lightcurves, as well as the rest of the reliable-fit sources, are shown in Figure \ref{fig:high_confidence_lightcurves}. We were not able to detect any reliable fits in And IX, Sextans B, or Sag DIG. This is not surprising given And IX and Sag DIG have limited temporal coverage and a small AGB population.

\subsection{Low-confidence variable}
\label{section:low-confidence variables}
The low-confidence sample is composed of 113 sources with insufficient epochs and 56 sources with unreliable fits. The low-confidence variables have a median absolute magnitude of $-$8.55\,$\pm$\,0.17\,(1$\sigma$), which is low compared to most of the LMC x-AGB sample. Sextans B has 14 sources with insufficient epochs and 12 sources with unreliable fits. This may be the result of stochastic sampling. At fainter magnitudes the photometric uncertainty is higher, making it more difficult to detect changes in brightness. This likely indicates that some sources in the low-confidence sample are not LPVs but in fact YSOs and AGN, which sometimes show irregular variability in the IR. We expect that low-confidence sources that cluster together may be YSOs in a star formation region. Sources far from the galaxy center may be AGN. However, given the non-uniform positioning of the detectors and our sporadic temporal coverage, these sources are hard to disentangle (Paper I).

\begin{figure*}
% \begin{flushright}
\includegraphics[width=\linewidth]{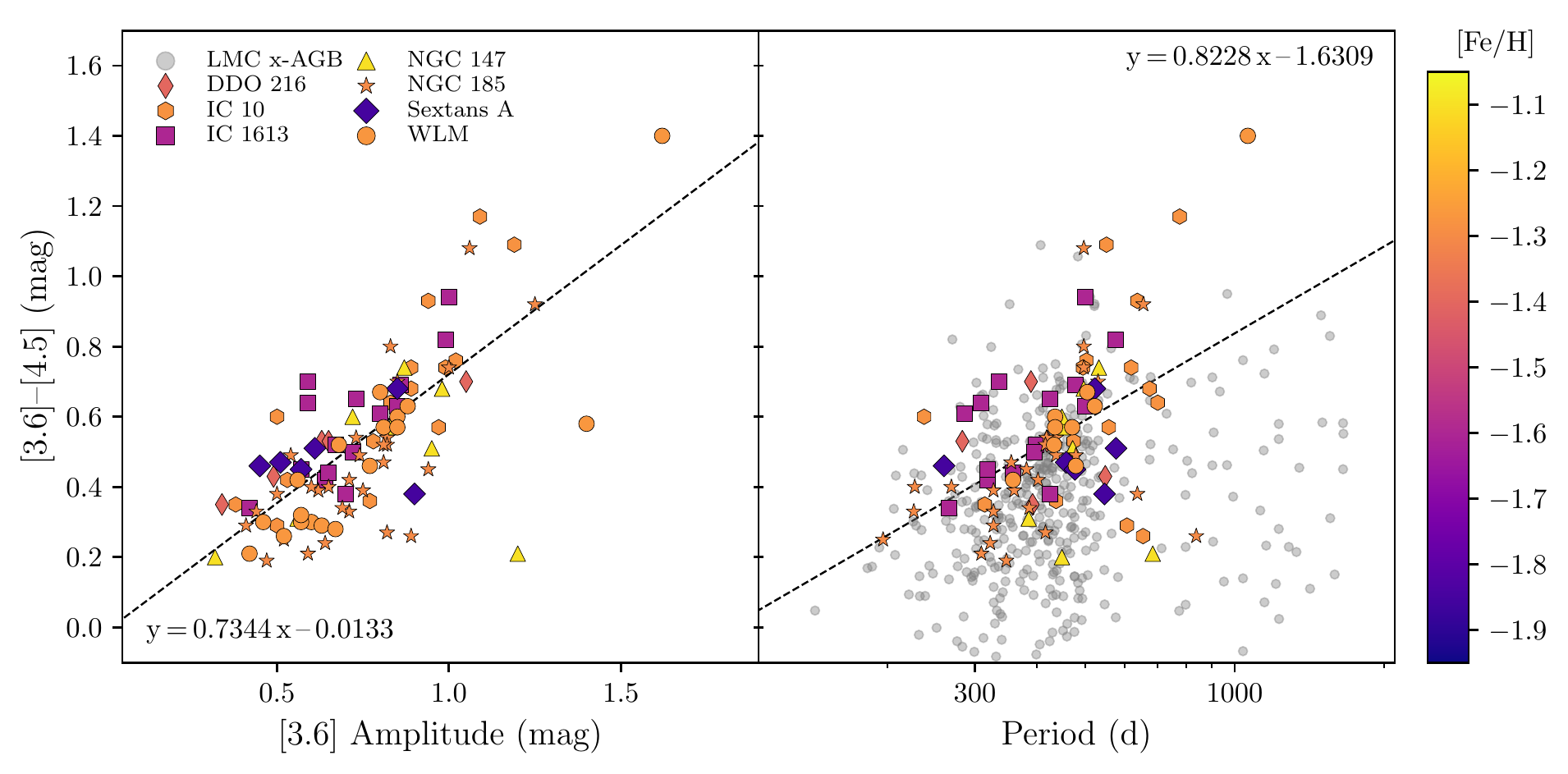}
% \end{flushright}
\caption{The pulsation amplitude and period of our reliable-fit sources vs. their \textit{Spitzer} [3.6]$-$[4.5] color (indicative of dust content) with metallicity shown in color. The best-fit results and lines are shown in the figure. Also shown are pulsation periods for the x-AGB sample from the LMC \citep{Riebel2010}; 3.6\,$\mu$m amplitudes were not measured. The low-confidence variables are shown in Figure \ref{fig:color_subplot_low_confidence}. The cadence and number of epochs in the SAGE-Var survey \citep{Riebel2015} are insufficient for getting reliable amplitudes which may contribute to the scatter for the LMC sample. \\}
\label{fig:color_subplot}
\end{figure*}

\subsection{Detection statistics}

Of sources that were included in our lightcurve analysis, we have isolated a small subset of them as potential AGB stars from their [3.6]$-$[4.5] color and variability alone. These sources have already been limited to those which have at least 6 epochs and are bright in the IR (M$_{[3.6]}$\,<\,$-$7.5 mag). Table \ref{table:gal_stats} shows the number of variable sources (Var. index > 2) in this sample, and the number of sources that were dusty ([3.6]$-$[4.5]\,>\,0.2) and also variable. Many of the sources that we were not able to confirm as AGB stars may be less luminous and less dusty. These may have been confirmed as AGB stars given more observations.

In Paper II, we showed that our photometry is sensitive enough to detect variability down to peak-to-peak amplitudes of $\sim$\,0.15 mag. We have now compared the number of sources with high-confidence reliable fits that were covered in each epoch of the Cycle 11 observations with the number of 3$\sigma$ x-AGB stars from Paper II also found in that region (Appendix C). Of the 3$\sigma$ variables originally detected in Paper II and categorized as x-AGB stars, we have confirmed 19\% as reliable fits and likely TP-AGB stars. The remaining Paper II variables are outside of our spatial footprints and/or have temporal coverage that is too sporadic to measure a reliable lightcurve. The variable star catalogs presented here should therefore be considered a representative subset of the total variable population in each galaxy.

% \subsection{Previously detected LPVs}
LPVs have been previously detected in four of our galaxies: IC 1613 \citep{Menzies2015}, NGC 147, NGC 185 \citep{Lorenz2011}, and Sag DIG \citep{Whitelock2018}. We have classified nineteen of these previously detected sources as high-confidence LPVs. The pulsation periods measured in these works are comparable to those measured in this work (Table \ref{table:previously_detected_matches}).

\section{discussion}

Our new sample of LPVs allows us to study the pulsation properties of evolved stars in metal-poor environments and how they are affected by other observable parameters. These results along with previous observations suggest that dust production is unaffected by metallicity. 

\subsection{Dust \& Pulsation}

\begin{figure}
\includegraphics[width=\linewidth]{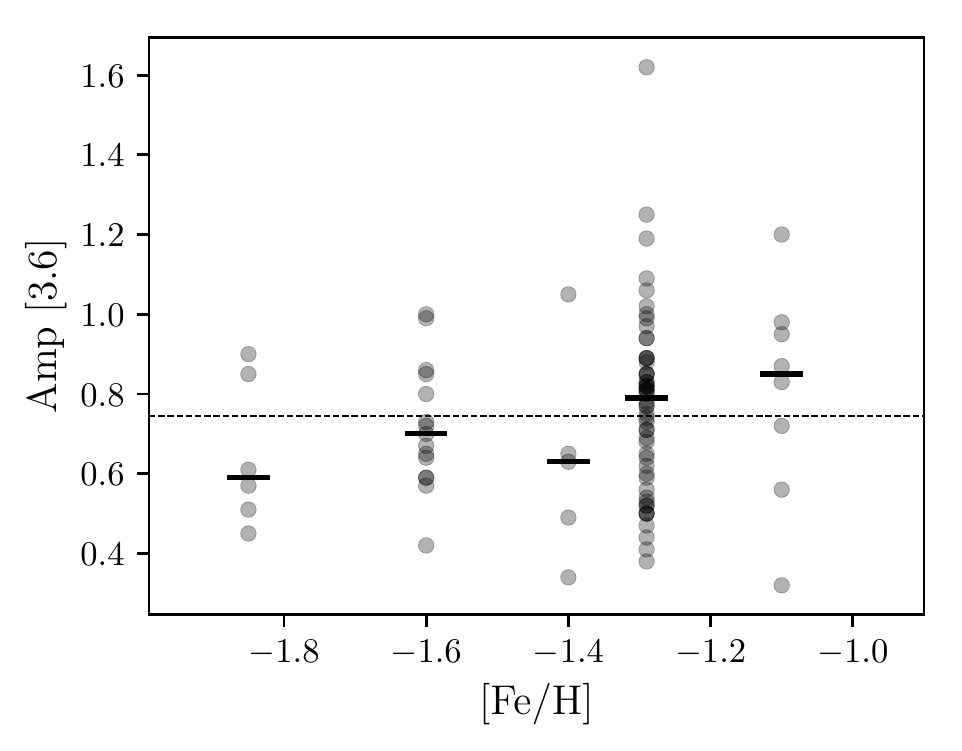}
\caption{The 3.6\,$\mu$m best-fit amplitudes with respect to [Fe/H]. The dotted line shows the median amplitude of the reliable-fit sample. Black lines show the median amplitude in five metallicity bins. There is an apparent increase of the median amplitudes towards higher metallicity. However, this trend is dominated by stochastic sampling.  \\}
\label{fig:amp_vs_z}
\end{figure}

The [3.6]$-$[4.5] color has been shown to scale approximately with the dust content \citep[Paper II;][]{Riebel2015}. Within our sample, both pulsation period and (especially) pulsation amplitude also correlate well with [3.6]$-$[4.5] color\footnote{Pulsation amplitude and period have also been seen to scale with mass-loss rate \citep{Javadi2013,Goldman2017}.} (Figure \ref{fig:color_subplot}). This correlation has been seen in more metal-rich samples in the galaxy, Magellanic Clouds, M33, and Sgr dSph \citep{Whitelock2006,McQuinn2007,Javadi2011a,McDonald2014,Riebel2015}. Figure \ref{fig:amp_vs_z} shows our amplitudes with respect to metallicity. We see an apparent slight increase in median amplitude towards higher metallicity. This is however dominated by stochastic sampling especially at the metallicity extremes, and if real, would only result in an increase of $\sim$\,0.1 mag over the entire metallicity range of our sample. Based on these relationships, it follows that the dust production should also be unaffected by metallicity.

\paragraph{Uncertainties}
There are uncertainties underlying our assumptions of dust production and metallicity. Given that most of our LPVs are expected to be carbon stars, we can only claim that dust production is unaffected by metallicity for carbon stars. The [3.6]$-$[4.5] color, a key metric in this analysis, will also depend on the dust temperature and wind speed, and the opacity of the dust may also differ at lower metallicities \citep{McDonald2011,McDonald2019a}. In determining the impact of metallicity on the dust production, we have assumed metallicities for our sample that were derived primarily from samples of RGB stars. These stars represent populations older and more metal-poor than our intermediate-mass LPVs. We expect the true metallicities of our LPVs to lie between these metallicities and ISM gas-phase oxygen abundances (shown in Table \ref{table:galaxies}), yet neither show a correlation with amplitude. 

\subsection{Infrared $P$--$L$ relation}
\label{sec:IR_PL_relation}
\citet{Wood2015} reviewed what is known about the $P$--$L$ sequences of variable stars and suggests an evolutionary scenario with current mass decreasing towards longer period at a given luminosity. This allows us to follow the amount of mass that has been lost as a star moves towards the latest stages of its evolution.

\begin{figure*}
 \centering
 \includegraphics[width=\linewidth]{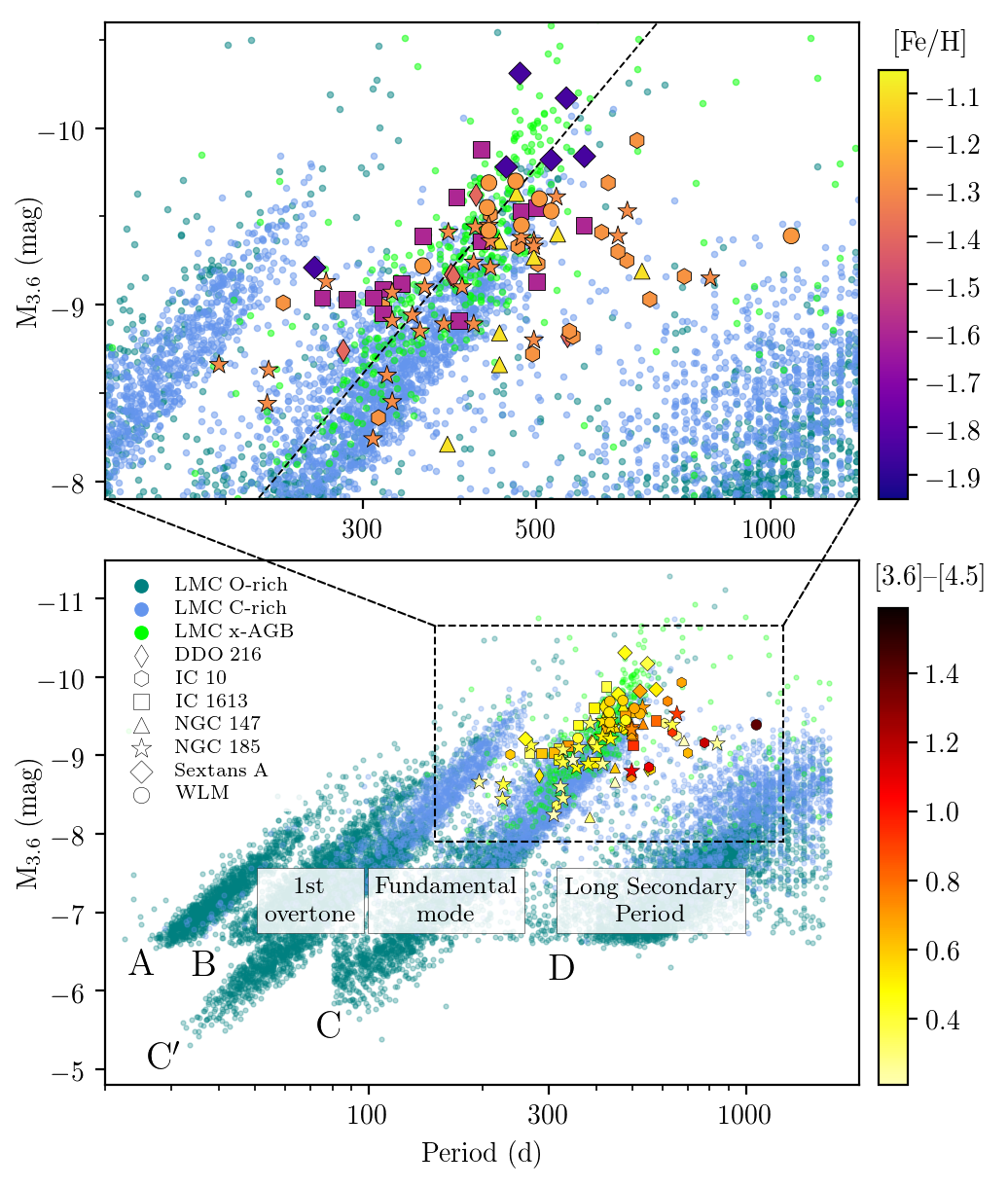}\
 \caption{The $P$--$L$ relation of the high-confidence DUSTiNGS sample with the color of the symbols showing the metallicity (Top) and [3.6]$-$[4.5] color (Bottom) as well as galaxy membership indicated by shapes. We show the MACHO-SAGE sample from \citet{Riebel2010} containing oxygen- and carbon-rich AGB stars as well as more evolved and dusty extreme x-AGB stars of both spectral types from the DUSTiNGS sample and MACHO-SAGE samples. Also shown is the best fit of the LMC x-AGB sources that are clearly fundamental mode pulsators. M$_{3.6}$ was calculated using the mid-line value of the best-fit lightcurve for the DUSTiNGS sample. \\}
 \label{fig:PL_relation} 
\end{figure*}

Figure \ref{fig:PL_relation} shows period with respect to luminosity for our LPVs, with [3.6]$-$[4.5] color and metallicity in color. Our sample spans $\gtrsim$\, 1 dex in metallicity, providing a first look at how the IR $P$--$L$ relation behaves at very low metallicity. Compared to the SAGE+MACHO sample \citep{Riebel2010}, most of our LPVs follow the fundamental-mode sequence. We find that the reddest objects fall below the fundamental mode at 3.6\,$\mu$m, a phenomenon that has also been seen in LPVs in the Magellanic Clouds \citep{Ita2004a,Ita2011} and IC 1613 \citep{Whitelock2017}. These sources, likely obscured by circumstellar extinction, have a decreased 3.6\,$\mu$m flux. IC 10, in particular, has a high number of reddened sources, which may be a combination of circumstellar and, to a lesser degree, interstellar extinction. IC 10 lies near the Galactic Plane and has an estimated interstellar extinction of A\textsubscript{V}\,$\sim$\,2.33 mag which has been measured using a CMD analysis \citep{Weisz2014}. This should cause a mean approximate shift of $\sim$\,0.2 mag in the IRAC magnitudes for which we have corrected for all sources in IC 10. 

\begin{figure*}
 \centering
 \includegraphics[width=\linewidth]{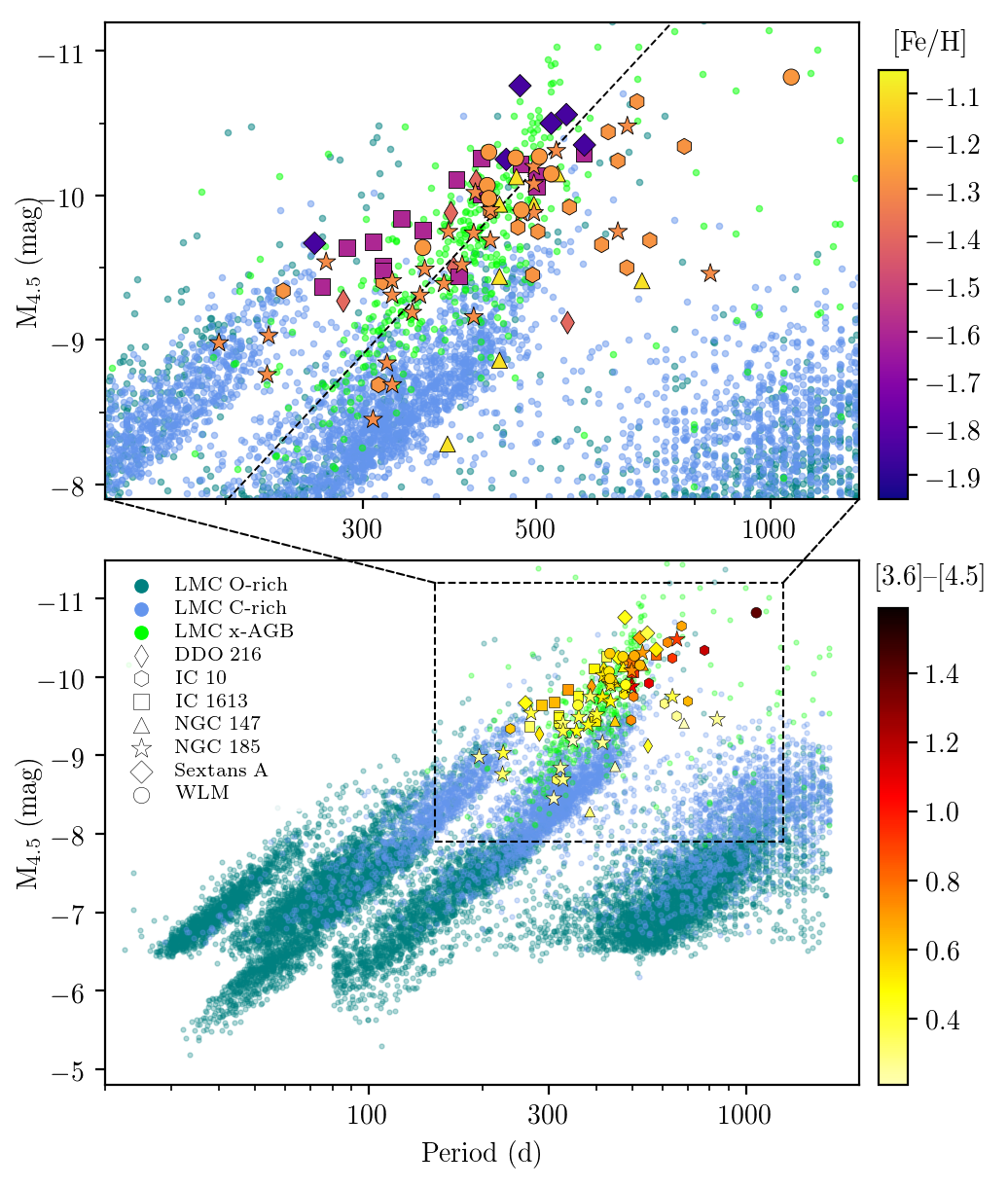}
 \caption{The same as Figure \ref{fig:PL_relation} but showing 4.5\,$\mu$m magnitudes. Dusty stars get brighter at 4.5\,$\mu$m, creating a sequence more in-line with the other sources. We see a tighter sequence than the sequence at 3.6\,$\mu$m, which is affected by the [3.6]$-$[4.5] color.}
 \label{fig:PL_relation45} 
\end{figure*}

\begin{figure*}
  \begin{center}
    %\vspace{-0.15cm}
	\includegraphics[width=6cm]{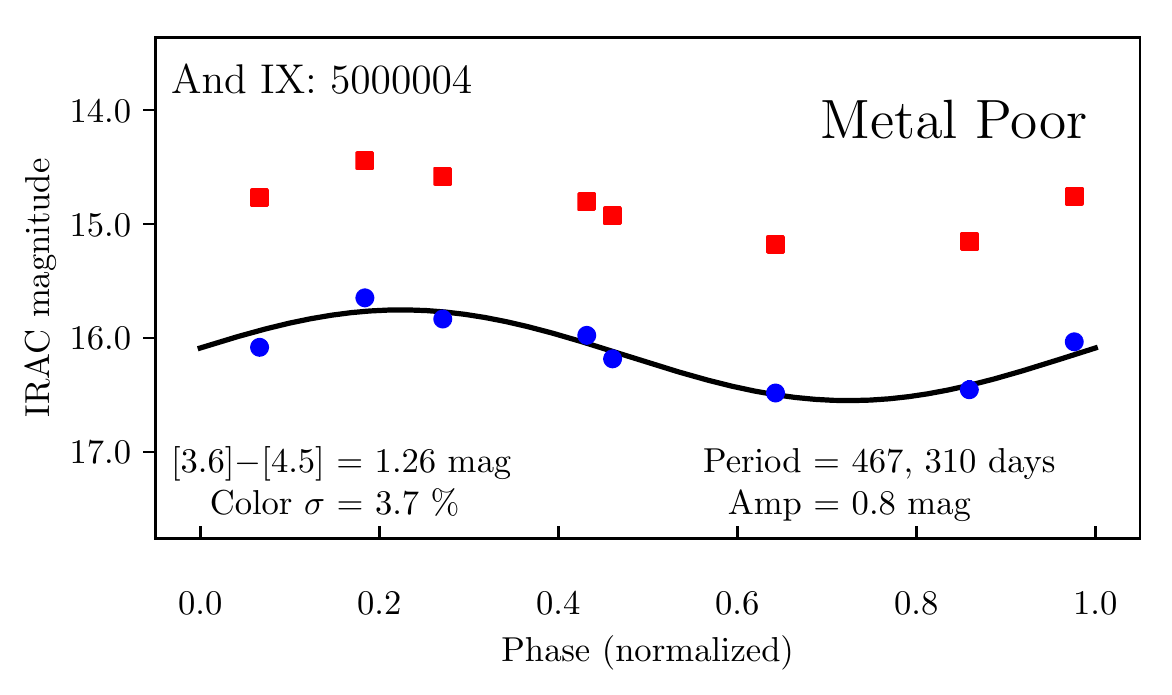} \hspace{-0.32cm}
	\includegraphics[width=6cm]{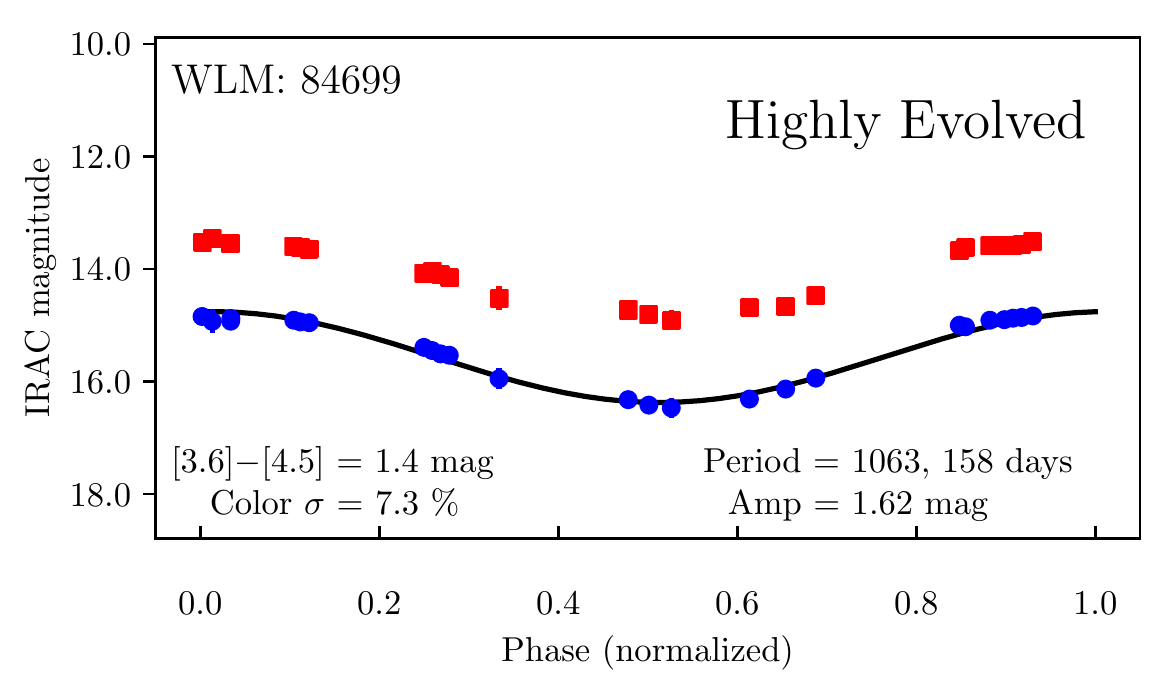} \hspace{-0.32cm}
	\includegraphics[width=6cm]{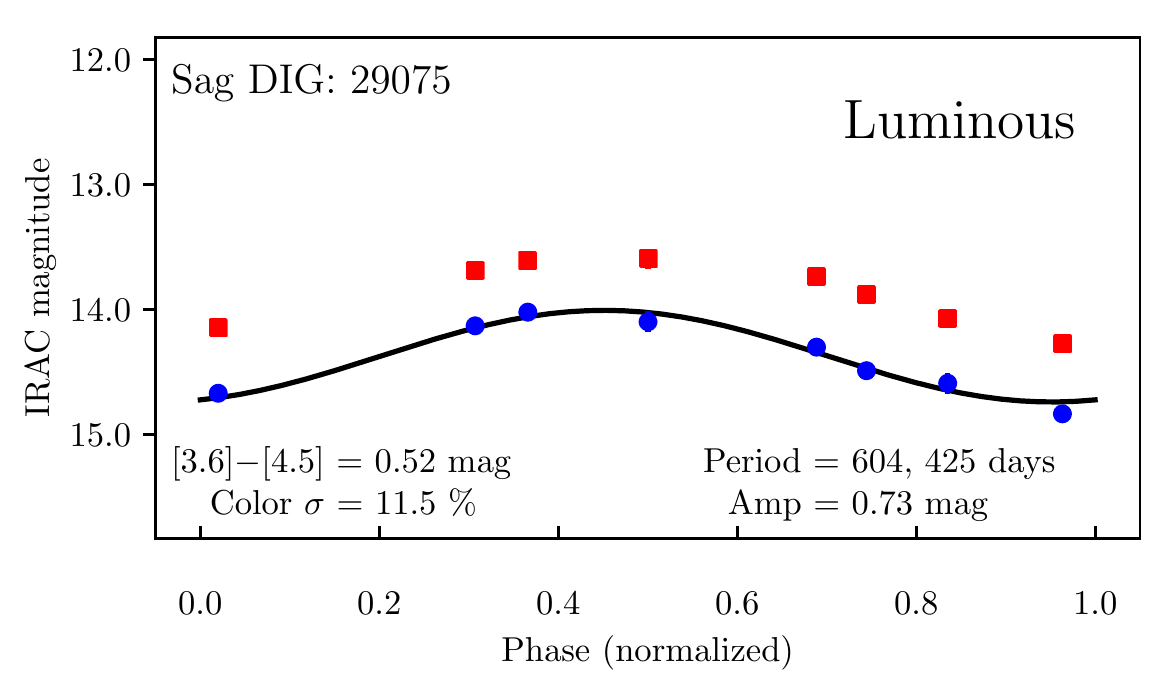} \vspace{-0.75cm} \\ 

	\includegraphics[width=6cm]{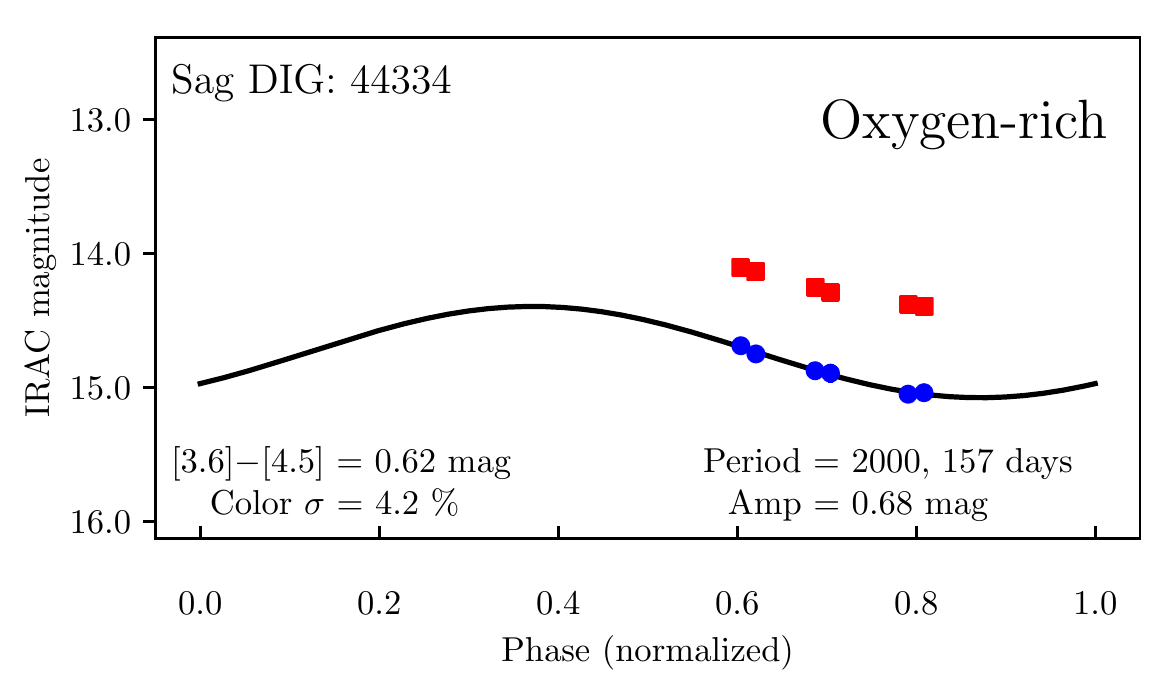} \hspace{-0.32cm}
	\includegraphics[width=6cm]{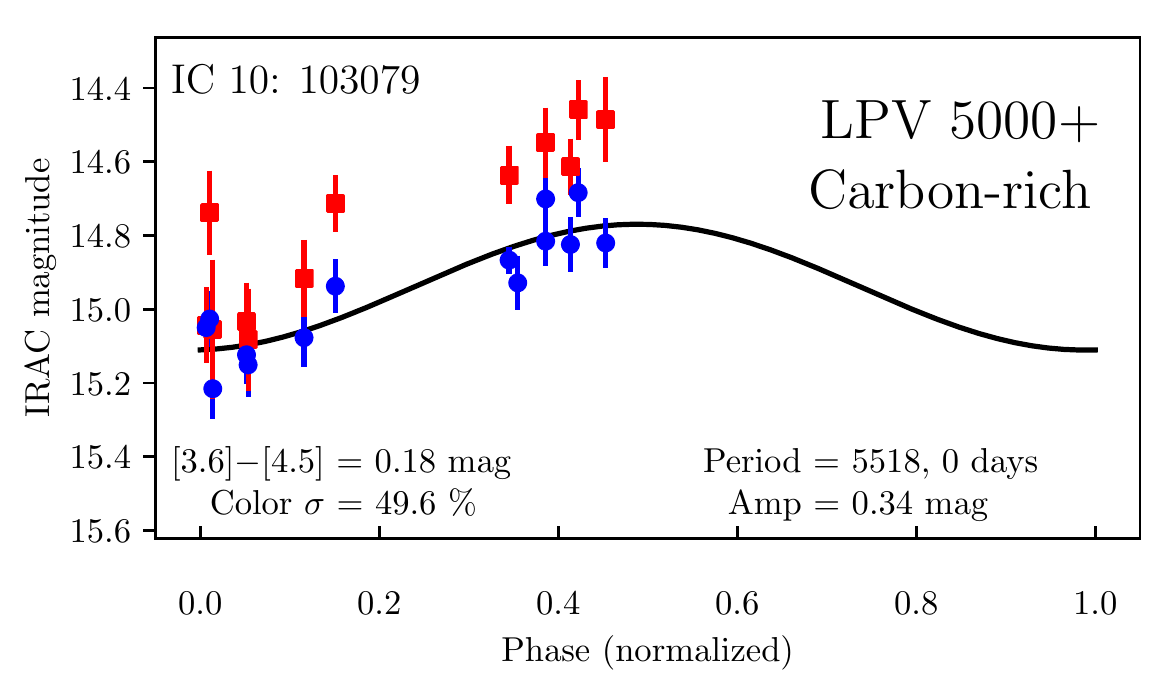} \hspace{-0.32cm}
	\includegraphics[width=6cm]{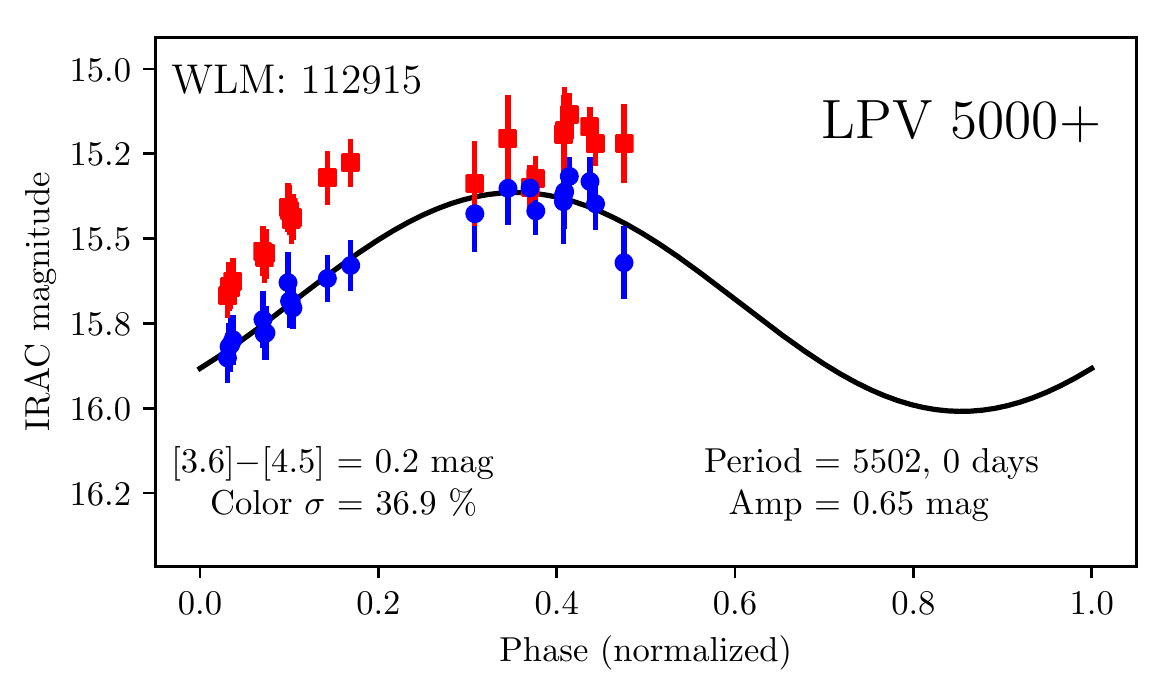} \vspace{-0.75cm} \\

	\includegraphics[width=6cm]{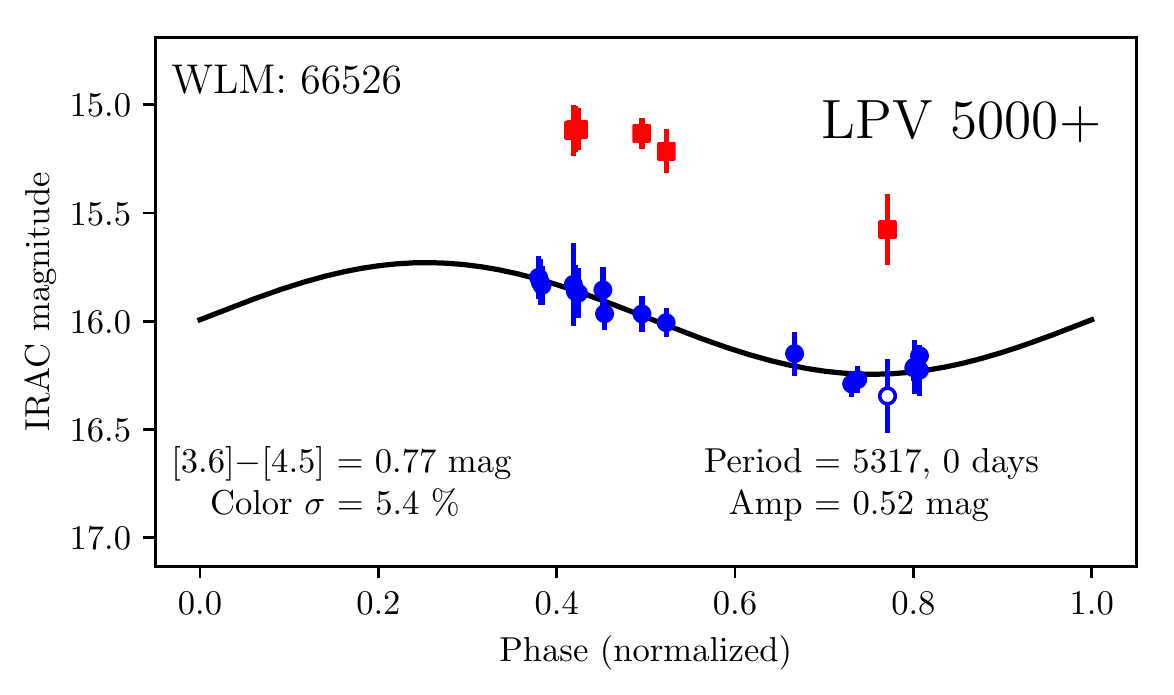} \hspace{-0.32cm}
	\includegraphics[width=6cm]{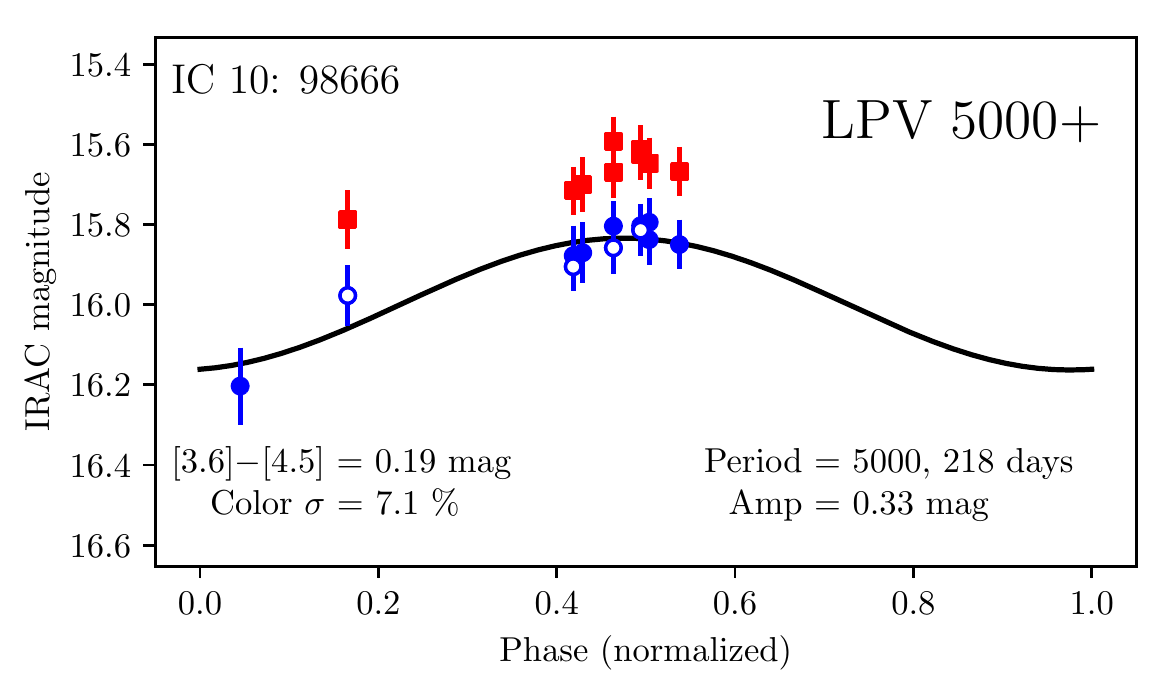} \hspace{-0.32cm}
	\transparent{0.01}\includegraphics[width=6cm]{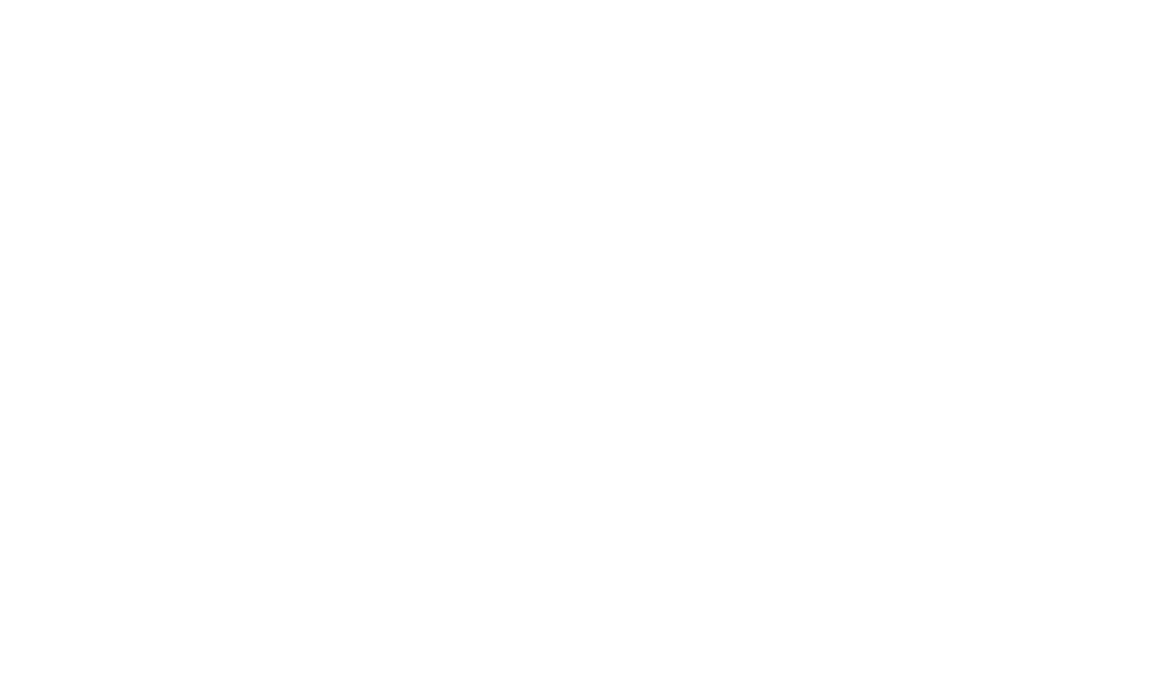}
	\vspace{-0.3cm}
	\end{center}
	\caption{ The lightcurves of particularly interesting sources with the IRAC 3.6\,$\mu$m (blue) and IRAC 4.5\,$\mu$m (red) photometry. Unless indicated, the error bars are smaller than the plotting symbols. Shown from Top left to Bottom right are examples of sources that are metal-poor, highly evolved and dusty, bright in the IR, categorized as oxygen-rich, and the four LPV\,$5000\Plus$ sources. The LPV\,$5000\Plus$ source IC\,10 103079 also has a confirmed carbon-rich chemistry.}
 \label{fig:interesting_lightcurves}
\end{figure*}

Our data suggest that, down to a [Fe/H]\,=\,$-$1.85, the $P$--$L$ relation of the fundamental mode (at 3.6\,$\mu$m) is largely unaffected by metallicity. This suggests that the fundamental mode may be a robust tool for measuring distances to galaxies in the IR \citep[e.g.][]{Yuan2018}. Since dusty AGB stars are among the brightest IR sources in galaxies, this technique can reach more distant galaxies than what can currently be measured with TRGB distance estimates. However, there is a $\sim$\,1 mag uncertainty in the absolute magnitude stemming from the width of the fundamental mode sequence at 3.6\,$\mu$m. The spread of the sequence may be a result of differences in the current mass, as a result of mass loss. For the DUSTiNGS reliable-fit sources and x-AGB stars in the LMC that are firmly on the fundamental mode, we calculate standard errors of the best fit of the x-AGB sample of 0.066 and 0.022, respectively. This excludes several shorter period and fainter LPVs, and several strongly affected by the circumstellar dust. The smaller standard error we calculate for the LMC sample is expected given the galaxy's larger and more complete sample, and more accurately known interstellar extinction and distance. We modeled the LMC x-AGB $P$--$L$ relation with 3 parameters: the slope, the intercept, and the intrinsic scatter. We used a 1st order Student t likelihood function, which has more weight in the tails than a Gaussian, and is therefore less vulnerable to the effects of outliers \citep{Galliano2018}. We sampled this likelihood function using \textsc{PyMC3}, a Monte Carlo Markov Chain package for Python \citep{Salvatier2016}, with 5000 steps along 15 independent chains. Equation 1 shows the fit of the fundamental mode x-AGB sample, calculated with a scatter of $0.25\pm0.01$; the fit is also shown in Figure \ref{fig:PL_relation}.

\begin{equation}
\textrm{M}_{3.6}\,=\,-5.26^{+0.15}_{-0.14}\,\textrm{log}\,P +4.42^{+0.38}_{-0.38}
\end{equation}

Using the intercept error, this translates to a 4\% uncertainty on the distance to a galaxy measured using the LMC $P$--$L$ relationship. This is smaller than the $\sim$\,8$\%$ uncertainty typically measured using the TRGB method \citep[e.g.][]{McQuinn2013}, but uses only sources firmly on the fundamental mode. This may provide a more accurate tool at larger distances.

Using a similar approach as in Eq. 1, we have also fitted the same LMC x-AGB sample at 4.5\,$\mu$m with a scatter of $0.35\pm0.01$:

\begin{equation}
\textrm{M}_{4.5}\,=\,-5.79^{+0.20}_{-0.21}\,\textrm{log}\,P +5.44^{+0.54}_{-0.53}
\end{equation}

At 4.5\,$\mu$m (Figure \ref{fig:PL_relation45}) the reliable-fit LPVs appear to be associated with a shifted fundamental-mode sequence, significantly affected by circumstellar dust. This dust will veil molecular features like CO, known to affect the $P$--$L$ relation in Cepheids \citep{Blum2014,Scowcroft2011}. This shift off of the fundamental mode suggests that it may be more challenging to use this wavelength region for measuring distance using the brightest and dustiest stars. \\

\subsection{Classification of stellar chemistry with the \textit{Hubble Space Telescope}}

Some of our reliable-fit LPV sources were previously observed with the \textit{Hubble Space Telescope} (\textit{HST}; Paper IV); Figure \ref{fig:spitzer_mosaics} in Appendix C shows the footprints of these observations. Medium-band optical photometry in the F127M, F139M, and F153M bands were used to categorize evolved stars by their chemical type (Paper IV). For our sample of reliable-fit LPVs, eight have \textit{HST} counterparts with chemical types, all of which were found to be carbon-rich (Table \ref{table:hubble_matches}). We found counterparts for nine sources with insufficient epochs, which include two oxygen-rich sources, and seven carbon-rich sources with unreliable fits. Theoretical models predict that most of these AGB populations will be dominated by carbon stars (Paper IV), with many fewer higher-mass oxygen-rich sources a result of their metal-poor environments \citep{DellAgli2016,DellAgli2018,DellAgli2019} and star formation histories \citep{Javadi2017,HamedaniGolshan2017,Goldman2018,Hashemi2019}. Additionally, they may be too obscured in the near-IR, lack sufficient temporal coverage, or were not covered in the \textit{HST} observations. The fact that oxygen-rich LPVs have not been detected here is not proof that they do not exist.

\begin{deluxetable*}{lrccccccccccc}
% \tablewidth{\linewidth}
\tabletypesize{\normalsize}
\tablecolumns{13}
\tablecaption{DUSTiNGS sources with both derived pulsation periods and determined chemical type \label{table:hubble_matches}}

\tablehead{
\colhead{{\tiny(1)}}&
\colhead{{\tiny(2)}}&
\colhead{{\tiny(3)}}&
\colhead{{\tiny(4)}}&
\colhead{{\tiny(5)}}&
\colhead{{\tiny(6)}}&
\colhead{{\tiny(7)}}&
\colhead{{\tiny(8)}}&
\colhead{{\tiny(9)}}&
\colhead{{\tiny(10)}}&
\colhead{{\tiny(11)}}&
\colhead{{\tiny(12)}}&
\colhead{{\tiny(13)}}
\\
\multirow{2}{*}{Galaxy}& 
\multirow{2}{*}{ID}& 
\multirow{2}{*}{RA}& 
\multirow{2}{*}{Dec}& 
\multirow{2}{*}{Type}& 
\colhead{Period}& 
\colhead{[3.6] amp}& 
\colhead{F127M} &
\colhead{F139M} &
\colhead{F153M} &
\colhead{[3.6]} &
\colhead{[4.5]} &
\multirow{2}{*}{Flag} \\
&
&
&
&
&
(d) &
(mag) &
(mag) &
(mag) &
(mag) &
(mag) &
(mag) &
}

\startdata
NGC 147 & 68407 & 00:33:25.61 & +48:33:15\rlap{.1} & C & 449 & 0.83 & 20.18 & 19.71 & 19.35 & 14.98 & 14.43 & RF \\
NGC 147 & 112918 & 00:33:05.55 & +48:28:37\rlap{.2} & C & 385 & 0.54 & 19.45 & 19.09 & 18.80 & 16.24 & 15.94 & RF \\
DDO 216 & 96974 & 23:28:41.54 & +14:44:45\rlap{.9} & C & 392 & 0.34 & 19.76 & 19.24 & 18.94 & 15.70 & 15.35 & RF \\
DDO 216 & 98916 & 23:28:40.97 & +14:44:01\rlap{.3} & C & 283 & 0.63 & 20.36 & 19.90 & 19.48 & 16.08 & 15.55 & RF \\
DDO 216 & 101247 & 23:28:40.29 & +14:44:38\rlap{.1} & C & 419 & 0.65 & 19.26 & 18.80 & 18.33 & 15.29 & 14.76 & RF \\
Sextans A & 86434 & 10:11:00.77 & $-$04:41:54\rlap{.0} & C & 458 & 0.51 & 20.18 & 19.71 & 19.54 & 16.01 & 15.54 & RF \\
Sextans A & 90941 & 10:10:59.20 & $-$04:42:23\rlap{.0} & C & 260 & 0.45 & 20.27 & 19.82 & 19.67 & 16.53 & 16.07 & RF \\
Sextans A & 98908 & 10:10:56.45 & $-$04:41:33\rlap{.0} & C & 477 & 0.57 & 20.27 & 19.79 & 19.41 & 15.57 & 15.12 & RF \\
IC 10 & 101088 & 00:20:15.71 & +59:16:00\rlap{.9} & C & 215\rlap{\tablenotemark{a}} & 0.56 & 19.99 & 19.47 & 19.19 & 15.70 & 15.33 & IE \\
IC 10 & 101812 & 00:20:15.24 & +59:15:59\rlap{.3} & C & 210\rlap{\tablenotemark{a}} & 0.41 & 18.90 & 18.33 & 18.08 & 14.89 & 14.68 & IE \\
IC 10 & 109003 & 00:20:10.26 & +59:20:10\rlap{.4} & C & 170\rlap{\tablenotemark{a}} & 0.32 & 18.07 & 17.65 & 17.46 & 14.65 & 14.27 & IE \\
IC 10 & 109003 & 00:20:10.17 & +59:20:11\rlap{.6} & M & 170\rlap{\tablenotemark{a}} & 0.32 & 19.54 & 19.24 & 18.80 & 14.65 & 14.27 & IE \\
IC 10 & 111369 & 00:20:08.84 & +59:16:57\rlap{.6} & C & 170\rlap{\tablenotemark{a}} & 0.69 & 19.22 & 18.70 & 18.47 & 15.58 & 15.03 & IE \\
IC 10 & 117441 & 00:20:04.56 & +59:21:11\rlap{.8} & C & 132\rlap{\tablenotemark{a}} & 0.44 & 18.54 & 18.15 & 17.98 & 15.92 & 15.74 & IE \\
NGC 147 & 55735 & 00:33:32.28 & +48:32:47\rlap{.3} & C & 133\rlap{\tablenotemark{a}} & 0.31 & 18.69 & 18.34 & 18.20 & 15.92 & 15.70 & IE \\
NGC 147 & 113288 & 00:33:05.40 & +48:30:18\rlap{.1} & C & 317\rlap{\tablenotemark{a}} & 0.49 & 18.08 & 17.69 & 17.55 & 15.83 & 15.63 & IE \\
Sag DIG & 44334 & 19:29:57.95 & $-$17:40:17\rlap{.0} & M & \llap{2}000\rlap{\tablenotemark{a}} & 0.68 & 18.03 & 17.72 & 17.26 & 14.88 & 14.26 & IE \\
IC 10 & 65446 & 00:20:39.81 & +59:16:39\rlap{.2} & C & 854\rlap{\tablenotemark{a}} & 0.35 & 19.52 & 19.04 & 18.73 & 15.79 & 15.53 & UF \\
IC 10 & 65548 & 00:20:39.74 & +59:17:25\rlap{.3} & C & 136\rlap{\tablenotemark{a}} & 0.43 & 19.84 & 19.37 & 19.03 & 15.73 & 15.33 & UF \\
IC 10 & 73607 & 00:20:34.08 & +59:15:58\rlap{.1} & C & 695\rlap{\tablenotemark{a}} & 0.41 & 19.40 & 18.89 & 18.53 & 15.32 & 15.10 & UF \\
IC 10 & 107394 & 00:20:11.34 & +59:21:14\rlap{.5} & C & 555\rlap{\tablenotemark{a}} & 0.80 & 20.14 & 19.64 & 19.39 & 16.01 & 15.58 & UF \\
IC 10 & 114178 & 00:20:06.78 & +59:19:57\rlap{.2} & C & 378\rlap{\tablenotemark{a}} & 0.40 & 19.46 & 18.99 & 18.66 & 15.70 & 15.37 & UF \\
Sextans A & 91449 & 10:10:59.05 & $-$04:40:14\rlap{.0} & C & 524\rlap{\tablenotemark{a}} & 0.36 & 20.61 & 20.07 & 19.77 & 16.07 & 15.47 & UF \\
IC 10 & 103079 & 00:20:14.25 & +59:19:07\rlap{.0} & C & $\ldots$ & $\ldots$ & 18.93 & 18.41 & 18.20 & 15.14 & 14.98 & LPV\,$5000\Plus$ 
\enddata 

\tablenotetext{}{\small{\textbf{Note.} The DUSTiNGS LPVs with \textit{HST} photometry used to determine the spectral type (Paper IV). Columns 11 and 12 show the average magnitudes of all of the 3.6 and 4.5\,$\mu$m epochs. Flag is the same as in Table \ref{table:fitting_results}.  \\ \tablenotemark{a}Value unlikely due to the low-confidence of the fit solution \\}}
%lmc distance from kovacs 2010
%$^{*}$Value unlikely due to the low confidence of the fit solution

\end{deluxetable*}

\subsection{Individual sources}
\label{section: S-LPV}

We have discovered several sources with particularly interesting lightcurves (Figure \ref{fig:interesting_lightcurves}). And IX is the most metal-poor galaxy in our sample, and we detected one LPV candidate, And\,IX 5000004, which has a clear variability and is one of the reddest sources in our sample. It lies near the outer regions of M31's disk at 37\,kpc from the galaxy's center and, while unlikely, could be a metal-rich M31 interloper. More epochs are needed to constrain the precise period, and spectroscopy is needed to confirm its membership to And IX. 

WLM 84699 is our most massive source, with a pulsation period of 1063 days, fitted peak-to-peak pulsation amplitude of 1.62 mag at 3.6\,$\mu$m and a [3.6]$-$[4.5] color of 1.6 mag. This is indicative of a very late stage of evolution, a high mass-loss rate, and a high dust-production rate. This source was also analyzed in \citet{Karambelkar2019} and lies between the lower mass population of Mira variables, and what they claim are massive AGB stars, in the $P$--$L$ diagram. These massive AGB stars are also in galaxies with younger populations than WLM, making it a particularly interesting target for spectroscopic follow-up and an analysis of the source's dust composition. 

While not as evolved, Sag\,DIG 29075 is our most luminous source and also quite metal-poor, but more observations are needed to constrain the period. Another source within Sag DIG, 44334, was previously categorized in Paper IV as oxygen-rich, making it one of the most metal-poor and dusty oxygen-rich evolved stars known. This source has a known pulsation period of 950 days  from ground-based observations in the near-IR \citep{Whitelock2018}. We lack enough IR data to further constrain the period. However, the clear variable nature of the source together with its red color strongly suggest that it is producing dust. 

The remaining four sources in Figure \ref{fig:interesting_lightcurves} have been categorized as LPV\,$5000\Plus$, and show a gradual change over the entirety of the lightcurve. While these sources may be shorter-period evolved stars with a coincidental cadence, they may also have dominant pulsation periods that are considerably longer than a typical AGB star (P\,$\gtrsim$\,2000 d), or may just be growing or diminishing in brightness, with no periodicity. Only one of the LPV\,$5000\Plus$ sources (IC\,10 103079) has been confirmed as carbon-rich and thus an AGB star (Paper IV).

\subsection{Long secondary period}
\label{sec: lsp}

Distinct from the LPV\,$5000\Plus$ are sources with long secondary periods (LSPs). The sequence hosts less-evolved stars as well as TP-AGB stars, with periods between 400--1200 days. Recent work by \citet{Wood2015} and \citet{Trabucchi2017,Trabucchi2018} has improved our understanding of how stars evolve along these sequences, but the mechanism that drives the LSP is still unclear. It is now known that stars with primary periods associated with the LSP, sequence D, have secondary periods that lie in the middle of the first overtone sequence made up of B and C$^{\prime}$. The reason for the appearance of the sequence D period in these stars is unknown, but may arise from convection, binarity or changes in the internal chemistry of the star \citep{Nicholls2009,Mathias2018}. The pulsation behavior of our LPV\,$5000\Plus$ sources may be that of an LSP.

\subsection{High-redshift dust}
Our observations provide further evidence of the evolved nature of the dusty sources found in metal poor galaxies, and for significant AGB dust production in these environments. The lowest metallicities of our LPV sample are characteristic of galaxies $\sim$\,12.3 Gyr ago and redshifts of $z$\,$\sim$\,5 \citep{Rafelski2012,Poudel2017}. Paper IV identified both carbon- and oxygen-rich evolved stars at low metallicity. As luminous oxygen-rich sources are more massive than their lower mass carbon-rich counterparts, they are capable of injecting dust into the ISM as early as 30\,Myr after forming (for a 10\,M$_{\odot}$ star), while carbon stars are expected to take longer \citep[as soon as $\sim$\,200--300\,Myr;][]{Sloan2009}. Most dust evolution models ignore dust produced by metal-poor oxygen-rich stars.

While we expect that AGB stars may produce dust in this regime, it is unclear whether they are the dominant dust producers. Supernovae may produce considerable dust; however, their net contribution is still unclear due to the unknown efficiency of dust destruction \citep{Temim2015,Lakicevic2015}. Another alternative is dust produced by grain growth within the ISM \citep{Zhukovska2008}. However, a theory as to how the grains grow and what seeds their nucleation has yet to be identified. While the pulsation properties and 3--5\,$\mu$m observations highlight the important role that AGB stars play in dust production, longer wavelength observations are critical to constraining the amount of cooler dust surrounding these stars. More observations are needed to confirm that AGB stars are capable of producing dust out to $z$\,$\sim$\,6. The soon-to-launch \textit{James Webb Space Telescope} (\textit{JWST}) will allow us to study Local Group AGB samples in much greater detail. In particular, observations with the \textit{JWST} Mid-InfraRed Instrument \citep[MIRI;][]{Rieke2015} will be able to obtain photometry for every star in this sample out to 25\,$\mu$m \citep{Jones2017a}.

\section{Conclusion}
This survey has provided the first infrared lightcurves of dusty evolved stars in metal-poor environments. We surveyed ten metal-poor dwarf galaxies within 1.5\,Mpc at 3.6 and 4.5\,$\mu$m. By combining our multi-epoch observations with archival observations, we identified the dustiest evolved AGB stars within these galaxies, sources that may have been missed in the near-IR or optical surveys due to dust obscuration. We have identified 88 sources in seven of these galaxies as high-confidence LPV candidates, eight of which have been confirmed as carbon-rich. 

We find that metallicity does not seem to have a strong impact on AGB pulsation or dust production. This has implications for the dust seen at high redshift and the origin of dust in the early Universe. We also find that the fundamental mode of the IR $P$--$L$ relationship seems unaffected by metallicity, at least between one half and one hundredth solar. This suggests that the $P$--$L$ relation can be a useful tool in measuring distance. With IR observations with \textit{JWST}, the $P$--$L$ relation can be used to confirm distances to Type Ia supernovae in distant galaxies, providing additional constraints on the Hubble constant ($H_0$). 

\acknowledgements We would like to thank the SPIRITS team (PI Mansi Kasliwal) for assistance with the inclusion of the SPIRITS data in four of our galaxies, and Chris Clark for his help with the $P$--$L$ distance uncertainty calculation. This work is supported by \textit{Spitzer} via grant GO11041 and by the NASA Astrophysics Data Analysis Program grant number NNX16AT56G. RDG was supported by NASA and the United States Air Force. OCJ has  received  funding from the European Union's Horizon 2020 research and innovation programme under the Marie Sklodowska-Curie grant agreement No. 665593 awarded to the Science and Technology Facilities Council.

\bibliographystyle{mn2e_new}
\bibliography{references_2018}

\begin{thebibliography}{113}
\expandafter\ifx\csname natexlab\endcsname\relax\def\natexlab#1{#1}\fi

\bibitem[{Alcock {et~al}\mbox{.}(1997)Alcock, Allsman, Alves, Axelrod, Becker,
  Bennett, Cook, Freeman, Griest, Guern, Lehner, Marshall, Peterson, Pratt,
  Quinn, Rodgers, Stubbs, Sutherland, \& Welch}]{Alcock1997}
Alcock C. {et~al.}, 1997, \apj, 486, 697

\bibitem[{{Bellazzini} {et~al}\mbox{.}(2014){Bellazzini}, {Beccari},
  {Fraternali}, {Oosterloo}, {Sollima}, {Testa}, {Galleti}, {Perina},
  {Faccini}, \& {Cusano}}]{Bellazzini2014}
{Bellazzini} M. {et~al.}, 2014, \aap, 566, A44

\bibitem[{Bladh {et~al}\mbox{.}(2015)Bladh, H\"{o}fner, Aringer, \&
  Eriksson}]{Bladh2015}
Bladh S., H\"{o}fner S., Aringer B., Eriksson K., 2015, A\&A, 575, A105

\bibitem[{Blum {et~al}\mbox{.}(2006)Blum, Mould, Olsen, Frogel, Werner,
  Meixner, Markwick-Kemper, Indebetouw, Whitney, Meade, Babler, Churchwell,
  Gordon, Engelbracht, For, Misselt, Vijh, Leitherer, Volk, Points, Reach,
  Hora, Bernard, Boulanger, Bracker, Cohen, Fukui, Gallagher, Gorjian, Harris,
  Kelly, Kawamura, Latter, Madden, Mizuno, Mizuno, Nota, Oey, Onishi, Paladini,
  Panagia, Perez-Gonzalez, Shibai, Sato, Smith, Staveley-Smith, Tielens, Ueta,
  Van~Dyk, \& Zaritsky}]{Blum2006}
Blum R.~D. {et~al.}, 2006, \aj, 132, 2034

\bibitem[{Blum {et~al}\mbox{.}(2014)Blum, Srinivasan, Kemper, Ling, \&
  Volk}]{Blum2014}
Blum R.~D., Srinivasan S., Kemper F., Ling B., Volk K., 2014, \aj, 148, 86

\bibitem[{{Boyer} {et~al}\mbox{.}(2015{\natexlab{a}}){Boyer}, {McQuinn},
  {Barmby}, {Bonanos}, {Gehrz}, {Gordon}, {Groenewegen}, {Lagadec}, {Lennon},
  {Marengo}, {McDonald}, {Meixner}, {Skillman}, {Sloan}, {Sonneborn}, {van
  Loon}, \& {Zijlstra}}]{Boyer2015}
{Boyer} M.~L. {et~al.}, 2015{\natexlab{a}}, \apj, 800, 51

\bibitem[{{Boyer} {et~al}\mbox{.}(2015{\natexlab{b}}){Boyer}, {McQuinn},
  {Barmby}, {Bonanos}, {Gehrz}, {Gordon}, {Groenewegen}, {Lagadec}, {Lennon},
  {Marengo}, {Meixner}, {Skillman}, {Sloan}, {Sonneborn}, {van Loon}, \&
  {Zijlstra}}]{Boyer2015a}
{Boyer} M.~L. {et~al.}, 2015{\natexlab{b}}, \apjs, 216, 10

\bibitem[{Boyer {et~al}\mbox{.}(2017)Boyer, McQuinn, Groenewegen, Zijlstra,
  Whitelock, van Loon, Sonneborn, Sloan, Skillman, Meixner, McDonald, Jones,
  Javadi, Gehrz, Britavskiy, \& Bonanos}]{Boyer2017}
Boyer M.~L. {et~al.}, 2017, \apj, 851, 152

\bibitem[{Clement {et~al}\mbox{.}(2001)Clement, Muzzin, Dufton, Ponnampalam,
  Wang, Burford, Richardson, Rosebery, Rowe, \& Hogg}]{Clement2001}
Clement C.~M. {et~al.}, 2001, \aj, 122, 2587

\bibitem[{Dell'Agli {et~al}\mbox{.}(2016)Dell'Agli, Di~Criscienzo, Boyer, \&
  Garc{\'{\i}}a-Hern{\'a}ndez}]{DellAgli2016}
Dell'Agli F., Di~Criscienzo M., Boyer M.~L., Garc{\'{\i}}a-Hern{\'a}ndez D.~A.,
  2016, \mnras, 460, 4230

\bibitem[{Dell'Agli {et~al}\mbox{.}(2019)Dell'Agli, Di~Criscienzo,
  Garc{\'{\i}}a-Hern{\'a}ndez, Ventura, Limongi, Marini, \&
  Jones}]{DellAgli2019}
Dell'Agli F., Di~Criscienzo M., Garc{\'{\i}}a-Hern{\'a}ndez D.~A., Ventura P.,
  Limongi M., Marini E., Jones O.~C., 2019, \mnras, 482, 4733

\bibitem[{Dell'Agli {et~al}\mbox{.}(2018)Dell'Agli, Di~Criscienzo, Ventura,
  Limongi, Garc{\'{\i}}a-Hern{\'a}ndez, Marini, \& Rossi}]{DellAgli2018}
Dell'Agli F., Di~Criscienzo M., Ventura P., Limongi M.,
  Garc{\'{\i}}a-Hern{\'a}ndez D.~A., Marini E., Rossi C., 2018, \mnras, 479,
  5035

\bibitem[{Doherty {et~al}\mbox{.}(2015)Doherty, Gil-Pons, Siess, Lattanzio, \&
  Lau}]{Doherty2015}
Doherty C.~L., Gil-Pons P., Siess L., Lattanzio J.~C., Lau H. H.~B., 2015,
  \mnras, 446, 2599

\bibitem[{Eddington(1913)}]{Eddington1913}
Eddington A.~S., 1913, \mnras, 73, 359

\bibitem[{Evans \& Gehrz(2012)}]{Evans2012}
Evans A., Gehrz R.~D., 2012, Bulletin of the Astronomical Society of India, 40,
  213

\bibitem[{{Fazio} {et~al}\mbox{.}(2004){Fazio}, {Hora}, {Allen}, {Ashby},
  {Barmby}, {Deutsch}, {Huang}, {Kleiner}, {Marengo}, {Megeath}, {Melnick},
  {Pahre}, {Patten}, {Polizotti}, {Smith}, {Taylor}, {Wang}, {Willner},
  {Hoffmann}, {Pipher}, {Forrest}, {McMurty}, {McCreight}, {McKelvey},
  {McMurray}, {Koch}, {Moseley}, {Arendt}, {Mentzell}, {Marx}, {Losch},
  {Mayman}, {Eichhorn}, {Krebs}, {Jhabvala}, {Gezari}, {Fixsen}, {Flores},
  {Shakoorzadeh}, {Jungo}, {Hakun}, {Workman}, {Karpati}, {Kichak}, {Whitley},
  {Mann}, {Tollestrup}, {Eisenhardt}, {Stern}, {Gorjian}, {Bhattacharya},
  {Carey}, {Nelson}, {Glaccum}, {Lacy}, {Lowrance}, {Laine}, {Reach},
  {Stauffer}, {Surace}, {Wilson}, {Wright}, {Hoffman}, {Domingo}, \&
  {Cohen}}]{Fazio2004}
{Fazio} G.~G. {et~al.}, 2004, \apjs, 154, 10

\bibitem[{Feast, Whitelock \& Menzies(2002)Feast, Whitelock, \&
  Menzies}]{Feast2002}
Feast M., Whitelock P., Menzies J., 2002, \mnras, 329, L7

\bibitem[{{Feast} {et~al}\mbox{.}(1989){Feast}, {Glass}, {Whitelock}, \&
  {Catchpole}}]{Feast1989}
{Feast} M.~W., {Glass} I.~S., {Whitelock} P.~A., {Catchpole} R.~M., 1989,
  \mnras, 241, 375

\bibitem[{Ferrarotti \& Gail(2006)}]{Ferrarotti2006}
Ferrarotti A.~S., Gail H.-P., 2006, A {\&}A, 447, 553

\bibitem[{Freedman {et~al}\mbox{.}(2011)Freedman, Madore, Scowcroft, Monson,
  Persson, Seibert, Rigby, Sturch, \& Stetson}]{Freedman2011}
Freedman W.~L. {et~al.}, 2011, \aj, 142, 192

\bibitem[{Gallart {et~al}\mbox{.}(2004)Gallart, Aparicio, Freedman, Madore,
  Mart{\'{\i}}nez-Delgado, \& Stetson}]{Gallart2004}
Gallart C., Aparicio A., Freedman W.~L., Madore B.~F., Mart{\'{\i}}nez-Delgado
  D., Stetson P.~B., 2004, \aj, 127, 1486

\bibitem[{Galliano(2018)}]{Galliano2018}
Galliano F., 2018, \mnras, 476, 1445

\bibitem[{{Gehrz} {et~al}\mbox{.}(2007){Gehrz}, {Roellig}, {Werner}, {Fazio},
  {Houck}, {Low}, {Rieke}, {Soifer}, {Levine}, \& {Romana}}]{Gehrz2007}
{Gehrz} R.~D. {et~al.}, 2007, Review of Scientific Instruments, 78, 011302

\bibitem[{{Gerasimovi\v{c}}(1928)}]{Gerasimovic1928}
{Gerasimovi\v{c}} B.~P., 1928, PNAS, 14, 963

\bibitem[{{Glass} {et~al}\mbox{.}(2009){Glass}, {Schultheis}, {Blommaert},
  {Sahai}, {Stute}, \& {Uttenthaler}}]{Glass2009}
{Glass} I.~S., {Schultheis} M., {Blommaert} J.~A.~D.~L., {Sahai} R., {Stute}
  M., {Uttenthaler} S., 2009, \mnras, 395, L11

\bibitem[{{Goldman} {et~al}\mbox{.}(2018){Goldman}, {van Loon}, {G{\'o}mez},
  {Green}, {Zijlstra}, {Nanni}, {Imai}, {Whitelock}, {Groenewegen}, \&
  {Oliveira}}]{Goldman2018}
{Goldman} S.~R. {et~al.}, 2018, \mnras, 473, 3835

\bibitem[{{Goldman} {et~al}\mbox{.}(2017){Goldman}, {van Loon}, {Zijlstra},
  {Green}, {Wood}, {Nanni}, {Imai}, {Whitelock}, {Matsuura}, {Groenewegen}, \&
  {G{\'o}mez}}]{Goldman2017}
{Goldman} S.~R. {et~al.}, 2017, \mnras, 465, 403

\bibitem[{Gordon {et~al}\mbox{.}(2011)Gordon, Meixner, Meade, Whitney,
  Engelbracht, Bot, Boyer, Lawton, Sewi{\l}o, Babler, Bernard, Bracker, Block,
  Blum, Bolatto, Bonanos, Harris, Hora, Indebetouw, Misselt, Reach, Shiao,
  Tielens, Carlson, Churchwell, Clayton, Chen, Cohen, Fukui, Gorjian, Hony,
  Israel, Kawamura, Kemper, Leroy, Li, Madden, Marble, McDonald, Mizuno,
  Mizuno, Muller, Oliveira, Olsen, Onishi, Paladini, Paradis, Points,
  Robitaille, Rubin, Sandstrom, Sato, Shibai, Simon, Smith, Srinivasan, Vijh,
  Van~Dyk, van Loon, \& Zaritsky}]{Gordon2011}
Gordon K.~D. {et~al.}, 2011, \aj, 142, 102

\bibitem[{Hamedani~Golshan {et~al}\mbox{.}(2017)Hamedani~Golshan, Javadi, van
  Loon, Khosroshahi, \& Saremi}]{HamedaniGolshan2017}
Hamedani~Golshan R., Javadi A., van Loon J.~T., Khosroshahi H., Saremi E.,
  2017, \mnras, 466, 1764

\bibitem[{Harris(1996)}]{Harris1996}
Harris W.~E., 1996, \aj, 112, 1487

\bibitem[{Hashemi, Javadi \& van Loon(2019)Hashemi, Javadi, \& van
  Loon}]{Hashemi2019}
Hashemi S.~A., Javadi A., van Loon J.~T., 2019, \mnras, 483, 4751

\bibitem[{{Herwig}(2005)}]{Herwig2005}
{Herwig} F., 2005, \araa, 43, 435

\bibitem[{{H{\"o}fner} \& {Olofsson}(2018)}]{Hoefner2018}
{H{\"o}fner} S., {Olofsson} H., 2018, \aapr, 26

\bibitem[{Huang {et~al}\mbox{.}(2018)Huang, Riess, Hoffmann, Klein, Bloom,
  Yuan, Macri, Jones, Whitelock, Casertano, \& Anderson}]{Huang2018}
Huang C.~D. {et~al.}, 2018, \apj, 857, 67

\bibitem[{{Hughes} \& {Wood}(1990)}]{Hughes1990}
{Hughes} S. M.~G., {Wood} P.~R., 1990, \aj, 99, 784

\bibitem[{Ita \& Matsunaga(2011)}]{Ita2011}
Ita Y., Matsunaga N., 2011, \mnras, 412, 2345

\bibitem[{Ita {et~al}\mbox{.}(2004)Ita, Tanab{\'e}, Matsunaga, Nakajima,
  Nagashima, Nagayama, Kato, Kurita, Nagata, Sato, Tamura, Nakaya, \&
  Nakada}]{Ita2004a}
Ita Y. {et~al.}, 2004, \mnras, 353, 705

\bibitem[{{Ita} {et~al}\mbox{.}(2004){Ita}, {Tanab{\'e}}, {Matsunaga},
  {Nakajima}, {Nagashima}, {Nagayama}, {Kato}, {Kurita}, {Nagata}, {Sato},
  {Tamura}, {Nakaya}, \& {Nakada}}]{Ita2004}
{Ita} Y. {et~al.}, 2004, \mnras, 347, 720

\bibitem[{Javadi {et~al}\mbox{.}(2015)Javadi, Saberi, van Loon, Khosroshahi,
  Golabatooni, \& Mirtorabi}]{Javadi2015}
Javadi A., Saberi M., van Loon J.~T., Khosroshahi H., Golabatooni N., Mirtorabi
  M.~T., 2015, \mnras, 447, 3973

\bibitem[{Javadi {et~al}\mbox{.}(2013)Javadi, van Loon, Khosroshahi, \&
  Mirtorabi}]{Javadi2013}
Javadi A., van Loon J.~T., Khosroshahi H., Mirtorabi M.~T., 2013, \mnras, 432,
  2824

\bibitem[{Javadi {et~al}\mbox{.}(2017)Javadi, van Loon, Khosroshahi,
  Tabatabaei, Hamedani~Golshan, \& Rashidi}]{Javadi2017}
Javadi A., van Loon J.~T., Khosroshahi H.~G., Tabatabaei F., Hamedani~Golshan
  R., Rashidi M., 2017, \mnras, 464, 2103

\bibitem[{Javadi, van Loon \& Mirtorabi(2011{\natexlab{a}})Javadi, van Loon, \&
  Mirtorabi}]{Javadi2011a}
Javadi A., van Loon J.~T., Mirtorabi M.~T., 2011{\natexlab{a}}, \mnras, 411,
  263

\bibitem[{Javadi, van Loon \& Mirtorabi(2011{\natexlab{b}})Javadi, van Loon, \&
  Mirtorabi}]{Javadi2011b}
Javadi A., van Loon J.~T., Mirtorabi M.~T., 2011{\natexlab{b}}, \mnras, 414,
  3394

\bibitem[{{Jones} {et~al}\mbox{.}(2017){Jones}, {Meixner}, {Justtanont}, \&
  {Glasse}}]{Jones2017a}
{Jones} O.~C., {Meixner} M., {Justtanont} K., {Glasse} A., 2017, \apj, 841, 15

\bibitem[{Karakas \& Lattanzio(2014)}]{Karakas2014a}
Karakas A.~I., Lattanzio J.~C., 2014, PASA, 31, e030

\bibitem[{Karambelkar {et~al}\mbox{.}(2019)Karambelkar, Adams, Whitelock,
  Kasliwal, Jencson, Boyer, Goldman, Masci, Cody, Bally, Bond, Gehrz, \&
  Parthasarathy}]{Karambelkar2019}
Karambelkar V.~R. {et~al.}, 2019, arXiv:1901.07179

\bibitem[{{Kasliwal} {et~al}\mbox{.}(2017){Kasliwal}, {Bally}, {Masci}, {Cody},
  {Bond}, {Jencson}, {Tinyanont}, {Cao}, {Contreras}, {Dykhoff}, {Amodeo},
  {Armus}, {Boyer}, {Cantiello}, {Carlon}, {Cass}, {Cook}, {Corgan}, {Faella},
  {Fox}, {Green}, {Gehrz}, {Helou}, {Hsiao}, {Johansson}, {Khan}, {Lau},
  {Langer}, {Levesque}, {Milne}, {Mohamed}, {Morrell}, {Monson}, {Moore},
  {Ofek}, {O' Sullivan}, {Parthasarathy}, {Perez}, {Perley}, {Phillips},
  {Prince}, {Shenoy}, {Smith}, {Surace}, {Van Dyk}, {Whitelock}, \&
  {Williams}}]{Kasliwal2017}
{Kasliwal} M.~M. {et~al.}, 2017, \apj, 839, 88

\bibitem[{{Kirby} {et~al}\mbox{.}(2017){Kirby}, {Rizzi}, {Held}, {Cohen},
  {Cole}, {Manning}, {Skillman}, \& {Weisz}}]{Kirby2017}
{Kirby} E.~N., {Rizzi} L., {Held} E.~V., {Cohen} J.~G., {Cole} A.~A., {Manning}
  E.~M., {Skillman} E.~D., {Weisz} D.~R., 2017, \apj, 834, 9

\bibitem[{Kov{\'a}cs(2000)}]{Kovacs2000}
Kov{\'a}cs G., 2000, \aap, 363, 1

\bibitem[{Lagadec \& Zijlstra(2008)}]{Lagadec2008}
Lagadec E., Zijlstra A.~A., 2008, \mnras, 390, L59

\bibitem[{{Laki{\'c}evi{\'c}} {et~al}\mbox{.}(2015){Laki{\'c}evi{\'c}}, {van
  Loon}, {Meixner}, {Gordon}, {Bot}, {Roman- Duval}, {Babler}, {Bolatto},
  {Engelbracht}, {Filipovi{\'c}}, {Hony}, {Indebetouw}, {Misselt}, {Montiel},
  {Okumura}, {Panuzzo}, {Patat}, {Sauvage}, {Seale}, {Sonneborn}, {Temim},
  {Uro{\v{s}}evi{\'c}}, \& {Zanardo}}]{Lakicevic2015}
{Laki{\'c}evi{\'c}} M. {et~al.}, 2015, \apj, 799, 50

\bibitem[{Lebzelter \& Wood(2005)}]{Lebzelter2005}
Lebzelter T., Wood P.~R., 2005, \aap, 441, 1117

\bibitem[{Lee {et~al}\mbox{.}(2006)Lee, Skillman, Cannon, Jackson, Gehrz,
  Polomski, \& Woodward}]{Lee2006}
Lee H., Skillman E.~D., Cannon J.~M., Jackson D.~C., Gehrz R.~D., Polomski
  E.~F., Woodward C.~E., 2006, \apj, 647, 970

\bibitem[{Liljegren {et~al}\mbox{.}(2018)Liljegren, {o}fner, Freytag, \&
  Bladh}]{Liljegren2018}
Liljegren S., {o}fner S.~H., Freytag B., Bladh S., 2018, A\&A, 619, A47

\bibitem[{Lomb(1976)}]{Lomb1976}
Lomb N.~R., 1976, Astrophysics and Space Science, 39, 447

\bibitem[{{Lorenz} {et~al}\mbox{.}(2011){Lorenz}, {Lebzelter}, {Nowotny},
  {Telting}, {Kerschbaum}, {Olofsson}, \& {Schwarz}}]{Lorenz2011}
{Lorenz} D., {Lebzelter} T., {Nowotny} W., {Telting} J., {Kerschbaum} F.,
  {Olofsson} H., {Schwarz} H.~E., 2011, \aap, 532, 78

\bibitem[{Marigo {et~al}\mbox{.}(2008)Marigo, Girardi, Bressan, Groenewegen,
  Silva, \& Granato}]{Marigo2008}
Marigo P., Girardi L., Bressan A., Groenewegen M. A.~T., Silva L., Granato
  G.~L., 2008, \aap, 482, 883

\bibitem[{Marigo {et~al}\mbox{.}(2017)Marigo, Girardi, Bressan, Rosenfield,
  Aringer, Chen, Dussin, Nanni, Pastorelli, Rodrigues, Trabucchi, Bladh,
  Dalcanton, Groenewegen, Montalb{\'{a}}n, \& Wood}]{Marigo2017}
Marigo P. {et~al.}, 2017, \apj, 835, 77

\bibitem[{Mateo(1998)}]{Mateo1998}
Mateo M., 1998, \araa, 36, 435

\bibitem[{Mathias {et~al}\mbox{.}(2018)Mathias, Auri{\`{e}}re, Ariste, Petit,
  Tessore, Josselin, L{\`{e}}bre, Morin, Wade, Herpin, Chiavassa,
  Montarg{\`{e}}s, Konstantinova-Antova, Kervella, Perrin, Donati, \&
  Grunhut}]{Mathias2018}
Mathias P. {et~al.}, 2018, {A\&A}, 615, A116

\bibitem[{{McConnachie}(2012)}]{McConnachie2012}
{McConnachie} A.~W., 2012, \aj, 144, 4

\bibitem[{McDonald {et~al}\mbox{.}(2019)McDonald, Boyer, Groenewegen, Lagadec,
  Richards, Sloan, \& Zijlstra}]{McDonald2019a}
McDonald I., Boyer M.~L., Groenewegen M. A.~T., Lagadec E., Richards A. M.~S.,
  Sloan G.~C., Zijlstra A.~A., 2019, \mnras, 484, L85

\bibitem[{McDonald {et~al}\mbox{.}(2011)McDonald, Boyer, van Loon, Zijlstra,
  Hora, Babler, Block, Gordon, Meade, Meixner, Misselt, Robitaille, Sewi{\l}o,
  Shiao, \& Whitney}]{McDonald2011}
McDonald I. {et~al.}, 2011, \apjs, 193, 23

\bibitem[{McDonald {et~al}\mbox{.}(2018)McDonald, De~Beck, Zijlstra, \&
  Lagadec}]{McDonald2018}
McDonald I., De~Beck E., Zijlstra A.~A., Lagadec E., 2018, \mnras, 481, 4984

\bibitem[{McDonald \& Trabucchi(2019)}]{McDonald2019}
McDonald I., Trabucchi M., 2019, arXiv:1901.06325

\bibitem[{McDonald {et~al}\mbox{.}(2010)McDonald, van Loon, Dupree, \&
  Boyer}]{McDonald2010}
McDonald I., van Loon J.~T., Dupree A.~K., Boyer M.~L., 2010, \mnras, 405, 1711

\bibitem[{McDonald {et~al}\mbox{.}(2014)McDonald, Zijlstra, Sloan, Kerins,
  Lagadec, \& Minniti}]{McDonald2014}
McDonald I., Zijlstra A.~A., Sloan G.~C., Kerins E., Lagadec E., Minniti D.,
  2014, \mnras, 439, 2618

\bibitem[{McQuinn {et~al}\mbox{.}(2013)McQuinn, Skillman, Berg, Cannon, Salzer,
  Adams, Dolphin, Giovanelli, Haynes, \& Rhode}]{McQuinn2013}
McQuinn K. B.~W. {et~al.}, 2013, \aj, 146, 145

\bibitem[{McQuinn {et~al}\mbox{.}(2007)McQuinn, Woodward, Willner, Polomski,
  Gehrz, Humphreys, van Loon, Ashby, Eicher, \& Fazio}]{McQuinn2007}
McQuinn K. B.~W. {et~al.}, 2007, \apj, 664, 850

\bibitem[{{Meixner} {et~al}\mbox{.}(2006){Meixner}, {Gordon}, {Indebetouw},
  {Hora}, {Whitney}, {Blum}, {Reach}, {Bernard}, {Meade}, {Babler},
  {Engelbracht}, {For}, {Misselt}, {Vijh}, {Leitherer}, {Cohen}, {Churchwell},
  {Boulanger}, {Frogel}, {Fukui}, {Gallagher}, {Gorjian}, {Harris}, {Kelly},
  {Kawamura}, {Kim}, {Latter}, {Madden}, {Markwick-Kemper}, {Mizuno}, {Mizuno},
  {Mould}, {Nota}, {Oey}, {Olsen}, {Onishi}, {Paladini}, {Panagia},
  {Perez-Gonzalez}, {Shibai}, {Sato}, {Smith}, {Staveley-Smith}, {Tielens},
  {Ueta}, {van Dyk}, {Volk}, {Werner}, \& {Zaritsky}}]{Meixner2006}
{Meixner} M. {et~al.}, 2006, \aj, 132, 2268

\bibitem[{{Menzies} {et~al}\mbox{.}(2008){Menzies}, {Feast}, {Whitelock},
  {Olivier}, {Matsunaga}, \& {da Costa}}]{Menzies2008}
{Menzies} J., {Feast} M., {Whitelock} P., {Olivier} E., {Matsunaga} N., {da
  Costa} G., 2008, \mnras, 385, 1045

\bibitem[{{Menzies} {et~al}\mbox{.}(2011){Menzies}, {Feast}, {Whitelock}, \&
  {Matsunaga}}]{Menzies2011}
{Menzies} J.~W., {Feast} M.~W., {Whitelock} P.~A., {Matsunaga} N., 2011,
  \mnras, 414, 3492

\bibitem[{{Menzies}, {Whitelock} \& {Feast}(2015){Menzies}, {Whitelock}, \&
  {Feast}}]{Menzies2015}
{Menzies} J.~W., {Whitelock} P.~A., {Feast} M.~W., 2015, \mnras, 452, 910

\bibitem[{{Menzies} {et~al}\mbox{.}(2010){Menzies}, {Whitelock}, {Feast}, \&
  {Matsunaga}}]{Menzies2010}
{Menzies} J.~W., {Whitelock} P.~A., {Feast} M.~W., {Matsunaga} N., 2010,
  \mnras, 406, 86

\bibitem[{Momany {et~al}\mbox{.}(2002)Momany, Held, Saviane, \&
  Rizzi}]{Momany2002}
Momany Y., Held E.~V., Saviane I., Rizzi L., 2002, \aap, 384, 393

\bibitem[{Nanni {et~al}\mbox{.}(2013)Nanni, Bressan, Marigo, \&
  Girardi}]{Nanni2013}
Nanni A., Bressan A., Marigo P., Girardi L., 2013, \mnras, 434, 2390

\bibitem[{Nanni {et~al}\mbox{.}(2014)Nanni, Bressan, Marigo, \&
  Girardi}]{Nanni2014}
Nanni A., Bressan A., Marigo P., Girardi L., 2014, \mnras, 438, 2328

\bibitem[{{Nicholls} {et~al}\mbox{.}(2009){Nicholls}, {Wood}, {Cioni}, \&
  {Soszy{\'n}ski}}]{Nicholls2009}
{Nicholls} C.~P., {Wood} P.~R., {Cioni} M. R.~L., {Soszy{\'n}ski} I., 2009,
  \mnras, 399, 2063

\bibitem[{Poudel {et~al}\mbox{.}(2017)Poudel, Kulkarni, Morrison, P{\'{e}}roux,
  Som, Rahmani, \& Quiret}]{Poudel2017}
Poudel S., Kulkarni V.~P., Morrison S., P{\'{e}}roux C., Som D., Rahmani H.,
  Quiret S., 2017, \mnras, 473, 3559

\bibitem[{Rafelski {et~al}\mbox{.}(2012)Rafelski, Wolfe, Prochaska, Neeleman,
  \& Mendez}]{Rafelski2012}
Rafelski M., Wolfe A.~M., Prochaska J.~X., Neeleman M., Mendez A.~J., 2012,
  \apj, 755, 89

\bibitem[{{Riebel} {et~al}\mbox{.}(2015){Riebel}, {Boyer}, {Srinivasan},
  {Whitelock}, {Meixner}, {Babler}, {Feast}, {Groenewegen}, {Ita}, {Meade},
  {Shiao}, \& {Whitney}}]{Riebel2015}
{Riebel} D. {et~al.}, 2015, \apj, 807, 1

\bibitem[{{Riebel} {et~al}\mbox{.}(2010){Riebel}, {Meixner}, {Fraser},
  {Srinivasan}, {Cook}, \& {Vijh}}]{Riebel2010}
{Riebel} D., {Meixner} M., {Fraser} O., {Srinivasan} S., {Cook} K., {Vijh} U.,
  2010, \apj, 723, 1195

\bibitem[{Rieke {et~al}\mbox{.}(2015)Rieke, Wright, Böker, Bouwman, Colina,
  Glasse, Gordon, Greene, Güdel, Henning, Justtanont, Lagage, Meixner,
  N{\o}rgaard-Nielsen, Ray, Ressler, van Dishoeck, \& Waelkens}]{Rieke2015}
Rieke G.~H. {et~al.}, 2015, \pasp, 127, 584

\bibitem[{Salvatier, Wiecki{\^a} \& Fonnesbeck(2016)Salvatier, Wiecki{\^a}, \&
  Fonnesbeck}]{Salvatier2016}
Salvatier J., Wiecki{\^a} T.~V., Fonnesbeck C., 2016, Pymc3: Python
  probabilistic programming framework. Astrophysics Source Code Library

\bibitem[{Saviane {et~al}\mbox{.}(2002)Saviane, Rizzi, Held, Bresolin, \&
  Momany}]{Saviane2002}
Saviane I., Rizzi L., Held E.~V., Bresolin F., Momany Y., 2002, \aap, 390, 59

\bibitem[{Scargle(1982)}]{Scargle1982}
Scargle J.~D., 1982, \apj, 263, 835

\bibitem[{Scowcroft {et~al}\mbox{.}(2011)Scowcroft, Freedman, Madore, Monson,
  Persson, Seibert, Rigby, \& Sturch}]{Scowcroft2011}
Scowcroft V., Freedman W.~L., Madore B.~F., Monson A.~J., Persson S.~E.,
  Seibert M., Rigby J.~R., Sturch L., 2011, \apj, 743, 76

\bibitem[{Skillman, Terlevich \& Melnick(1989)Skillman, Terlevich, \&
  Melnick}]{Skillman1989}
Skillman E.~D., Terlevich R., Melnick J., 1989, \mnras, 240, 563

\bibitem[{{Sloan} {et~al}\mbox{.}(2016){Sloan}, {Kraemer}, {McDonald},
  {Groenewegen}, {Wood}, {Zijlstra}, {Lagadec}, {Boyer}, {Kemper}, {Matsuura},
  {Sahai}, {Sargent}, {Srinivasan}, {van Loon}, \& {Volk}}]{Sloan2016}
{Sloan} G.~C. {et~al.}, 2016, \apj, 826, 44

\bibitem[{Sloan {et~al}\mbox{.}(2010)Sloan, Matsunaga, Matsuura, Zijlstra,
  Kraemer, Wood, Nieusma, Bernard-Salas, Devost, \& Houck}]{Sloan2010}
Sloan G.~C. {et~al.}, 2010, \apj, 719, 1274

\bibitem[{Sloan {et~al}\mbox{.}(2012)Sloan, Matsuura, Lagadec, van Loon,
  Kraemer, McDonald, Groenewegen, Wood, Bernard-Salas, \& Zijlstra}]{Sloan2012}
Sloan G.~C. {et~al.}, 2012, \apj, 752, 140

\bibitem[{Sloan {et~al}\mbox{.}(2009)Sloan, Matsuura, Zijlstra, Lagadec,
  Groenewegen, Wood, Szyszka, Bernard-Salas, \& van Loon}]{Sloan2009}
Sloan G.~C. {et~al.}, 2009, Science, 323, 353

\bibitem[{Soszy\'nski {et~al}\mbox{.}(2009)Soszy\'nski, Udalski, Szyma\'{n}ski,
  Kubiak, Pietrzynski, Wyrzykowski, Szewczyk, Ulaczyk, \&
  Poleski}]{Soszynski2009}
Soszy\'nski I. {et~al.}, 2009, AcA, 59, 239

\bibitem[{Stetson(1987)}]{Stetson1987}
Stetson P.~B., 1987, \pasp, 99, 191

\bibitem[{{Temim} {et~al}\mbox{.}(2015){Temim}, {Dwek}, {Tchernyshyov},
  {Boyer}, {Meixner}, {Gall}, \& {Roman-Duval}}]{Temim2015}
{Temim} T., {Dwek} E., {Tchernyshyov} K., {Boyer} M.~L., {Meixner} M., {Gall}
  C., {Roman-Duval} J., 2015, \apj, 799, 158

\bibitem[{Trabucchi {et~al}\mbox{.}(2017)Trabucchi, Wood, Montalb{\'{a}}n,
  Marigo, Pastorelli, \& Girardi}]{Trabucchi2017}
Trabucchi M., Wood P.~R., Montalb{\'{a}}n J., Marigo P., Pastorelli G., Girardi
  L., 2017, \apj, 847, 139

\bibitem[{Trabucchi {et~al}\mbox{.}(2018)Trabucchi, Wood, Montalb{\'a}n,
  Marigo, Pastorelli, \& Girardi}]{Trabucchi2018}
Trabucchi M., Wood P.~R., Montalb{\'a}n J., Marigo P., Pastorelli G., Girardi
  L., 2018, \mnras, 482, 929

\bibitem[{Udalski, Kubiak \& Szyma\'{n}ski(1997)Udalski, Kubiak, \&
  Szyma\'{n}ski}]{Udalski1997}
Udalski A., Kubiak M., Szyma\'{n}ski M., 1997, AcA, 47, 319

\bibitem[{van Loon(2000)}]{Loon2000}
van Loon J.~T., 2000, \aap, 354, 125

\bibitem[{van Loon(2006)}]{Loon2006}
van Loon J.~T., 2006, in Astronomical Society of the Pacific Conference Series,
  Vol. 353, Stellar Evolution at Low Metallicity: Mass Loss, Explosions,
  Cosmology, {Lamers} H.~J.~G.~L.~M., {Langer} N., {Nugis} T., {Annuk} K.,
  eds., p. 211

\bibitem[{{van Loon}, {Marshall} \& {Zijlstra}(2005){van Loon}, {Marshall}, \&
  {Zijlstra}}]{vanLoon2005}
{van Loon} J.~T., {Marshall} J.~R., {Zijlstra} A.~A., 2005, \aap, 442, 597

\bibitem[{VanderPlas(2018)}]{VanderPlas2018}
VanderPlas J.~T., 2018, \apjs, 236, 16

\bibitem[{Weisz {et~al}\mbox{.}(2014)Weisz, Dolphin, Skillman, Holtzman,
  Gilbert, Dalcanton, \& Williams}]{Weisz2014}
Weisz D.~R., Dolphin A.~E., Skillman E.~D., Holtzman J., Gilbert K.~M.,
  Dalcanton J.~J., Williams B.~F., 2014, \apj, 789, 147

\bibitem[{{Werner} {et~al}\mbox{.}(2004){Werner}, {Roellig}, {Low}, {Rieke},
  {Rieke}, {Hoffmann}, {Young}, {Houck}, {Brandl}, {Fazio}, {Hora}, {Gehrz},
  {Helou}, {Soifer}, {Stauffer}, {Keene}, {Eisenhardt}, {Gallagher}, {Gautier},
  {Irace}, {Lawrence}, {Simmons}, {Van Cleve}, {Jura}, {Wright}, \&
  {Cruikshank}}]{Werner2004}
{Werner} M.~W. {et~al.}, 2004, \apjs, 154, 1

\bibitem[{{Whitelock}(2012)}]{Whitelock2012}
{Whitelock} P.~A., 2012, \apss, 341, 123

\bibitem[{Whitelock {et~al}\mbox{.}(2006)Whitelock, Feast, Marang, \&
  Groenewegen}]{Whitelock2006}
Whitelock P.~A., Feast M.~W., Marang F., Groenewegen M. A.~T., 2006, \mnras,
  369, 751

\bibitem[{{Whitelock}, {Kasliwal} \& {Boyer}(2017){Whitelock}, {Kasliwal}, \&
  {Boyer}}]{Whitelock2017}
{Whitelock} P.~A., {Kasliwal} M., {Boyer} M., 2017, in Catalan M., Gieren W.,
  eds, European Physical Journal Web of Conferences Vol. 152, Wide-Field
  Variability Surveys: A 21st Century Perspective, 1009 (arXiv:1702.06797),
  Vol. 152

\bibitem[{{Whitelock} {et~al}\mbox{.}(2018){Whitelock}, {Menzies}, {Feast}, \&
  {Marigo}}]{Whitelock2018}
{Whitelock} P.~A., {Menzies} J.~W., {Feast} M.~W., {Marigo} P., 2018, \mnras,
  473, 173

\bibitem[{{Whitelock} {et~al}\mbox{.}(2009){Whitelock}, {Menzies}, {Feast},
  {Matsunaga}, {Tanab{\'e}}, \& {Ita}}]{Whitelock2009}
{Whitelock} P.~A., {Menzies} J.~W., {Feast} M.~W., {Matsunaga} N., {Tanab{\'e}}
  T., {Ita} Y., 2009, \mnras, 394, 795

\bibitem[{{Wood}(2015)}]{Wood2015}
{Wood} P.~R., 2015, \mnras, 448, 3829

\bibitem[{{Wood} {et~al}\mbox{.}(1999){Wood}, {Alcock}, {Allsman}, {Alves},
  {Axelrod}, {Becker}, {Bennett}, {Cook}, {Drake}, {Freeman}, {Griest}, {King},
  {Lehner}, {Marshall}, {Minniti}, {Peterson}, {Pratt}, {Quinn}, {Stubbs},
  {Sutherland}, {Tomaney}, {Vandehei}, \& {Welch}}]{Wood1999}
{Wood} P.~R. {et~al.}, 1999, in Asymptotic Giant Branch Stars, IAU Symposium
  191, Edited by T. Le Bertre, A. L\`ebre, and C. Waelkens. ISBN: 1-886733-90-2
  LOC: 99-62044., Vol. 191, p. 151

\bibitem[{Yuan {et~al}\mbox{.}(2018)Yuan, Macri, Javadi, Lin, \&
  Huang}]{Yuan2018}
Yuan W., Macri L.~M., Javadi A., Lin Z., Huang J.~Z., 2018, \aj, 156, 112

\bibitem[{{Zhukovska}, {Gail} \& {Trieloff}(2008){Zhukovska}, {Gail}, \&
  {Trieloff}}]{Zhukovska2008}
{Zhukovska} S., {Gail} H.~P., {Trieloff} M., 2008, \aap, 479, 453

\end{thebibliography}

%appendicies
\section*{Appendix A: Infrared lightcurves}

This appendix shows the phased lightcurves of the reliable-fit LPVs. IRAC 3.6\,$\mu$m (blue circles) lightcurves were fit with simple single-term sinusoidal functions using the Lomb-Scargle algorithm. Also shown are the corresponding 4.5\,$\mu$m magnitudes (red squares). Simulated 3.6\,$\mu$m photometry (see \S\ref{sect: identifying LPVs}) are shown as open circles. Figure \ref{fig:histogram} shows the number of epochs for our reliable-fit and low-confidence LPVs. \vspace{-0.4cm}

\counterwithin{figure}{section}
\renewcommand{\thefigure}{A\arabic{figure}}
\setcounter{figure}{0}

\begin{figure*}
  \begin{center}
    %\vspace{-0.15cm}
	\includegraphics[width=6cm]{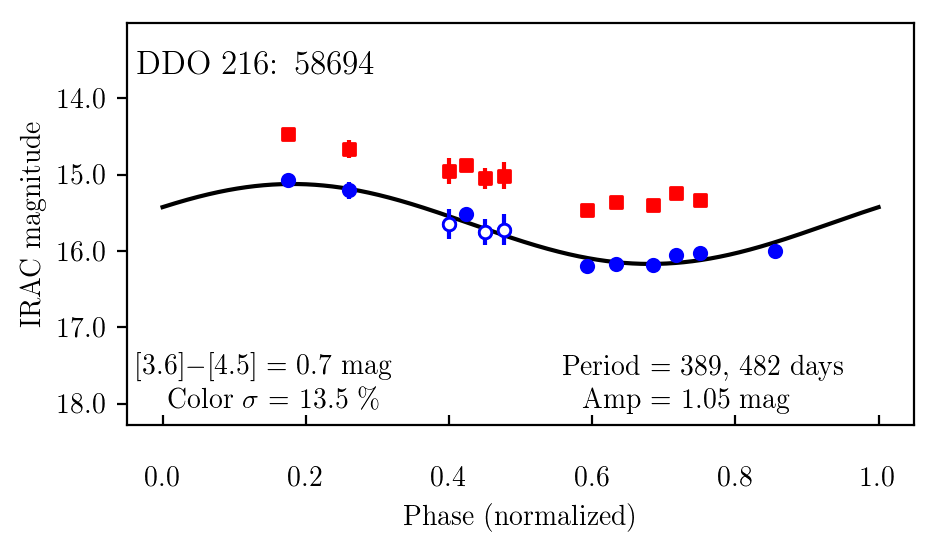} \hspace{-0.32cm}
	\includegraphics[width=6cm]{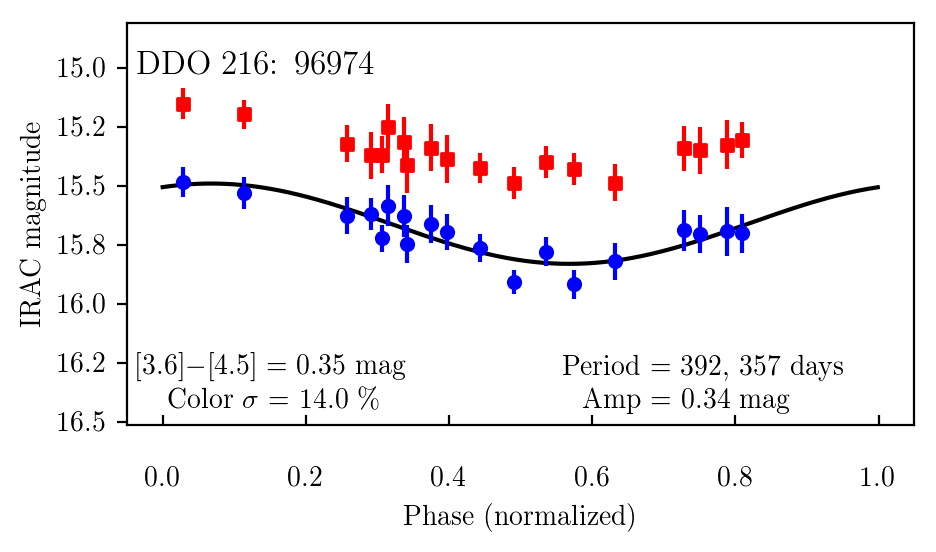} \hspace{-0.32cm}
	\includegraphics[width=6cm]{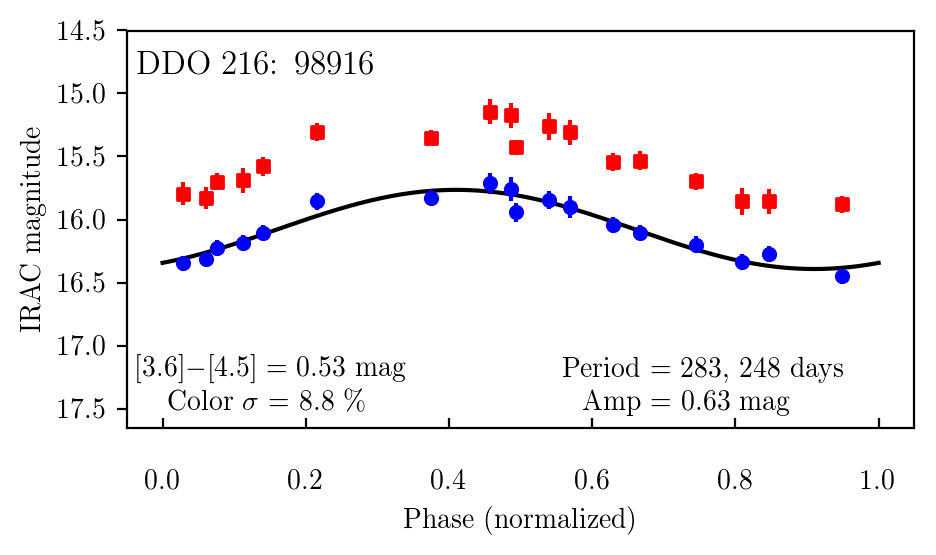} \vspace{-0.75cm} \\ 

	\includegraphics[width=6cm]{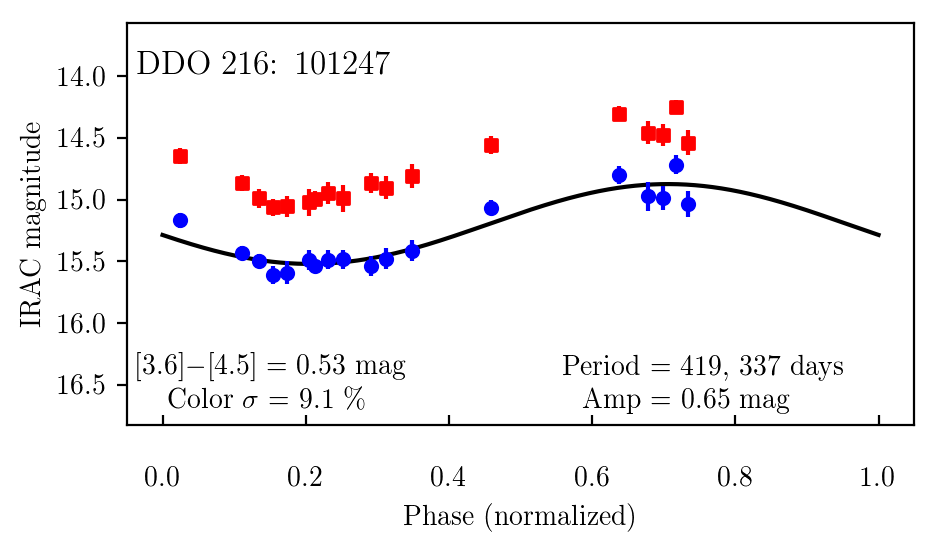} \hspace{-0.32cm}
	\includegraphics[width=6cm]{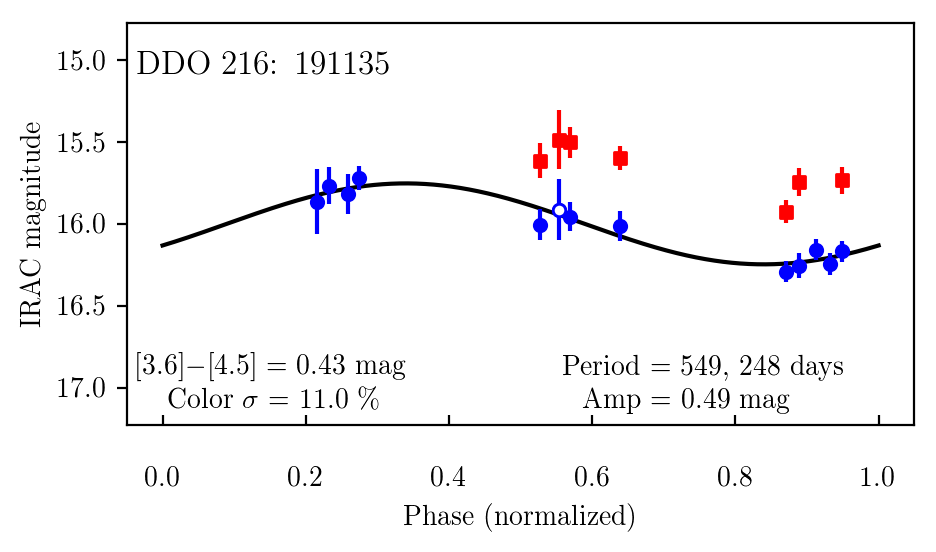} \hspace{-0.32cm}
	\includegraphics[width=6cm]{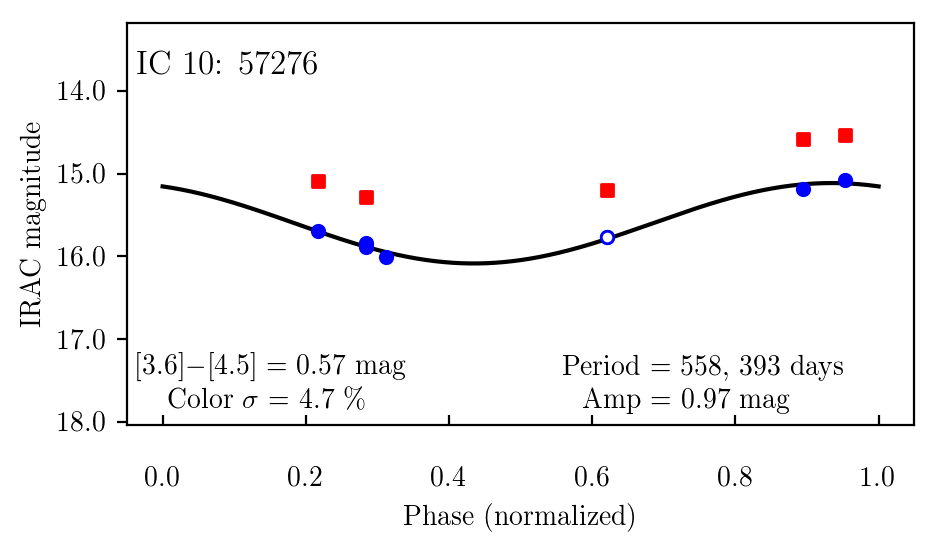} \vspace{-0.75cm}\\

	\includegraphics[width=6cm]{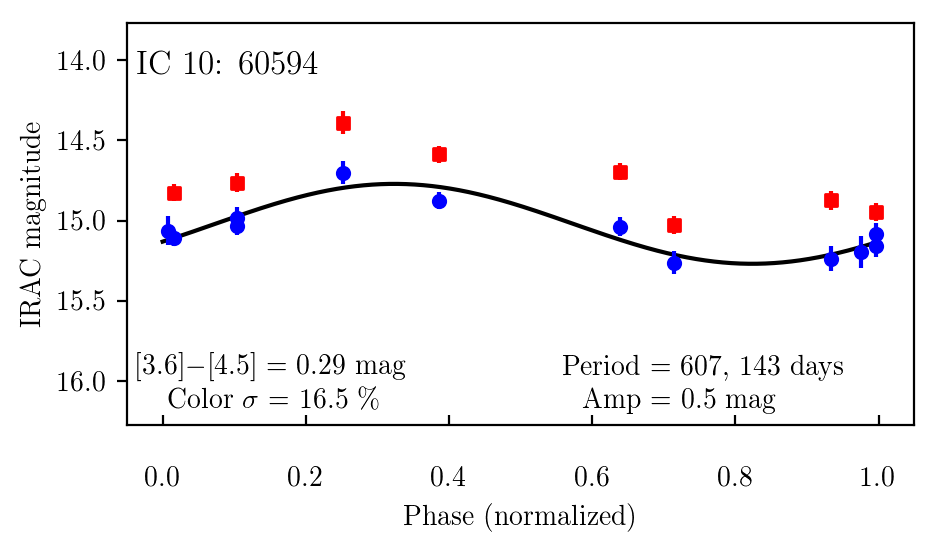} \hspace{-0.32cm}
	\includegraphics[width=6cm]{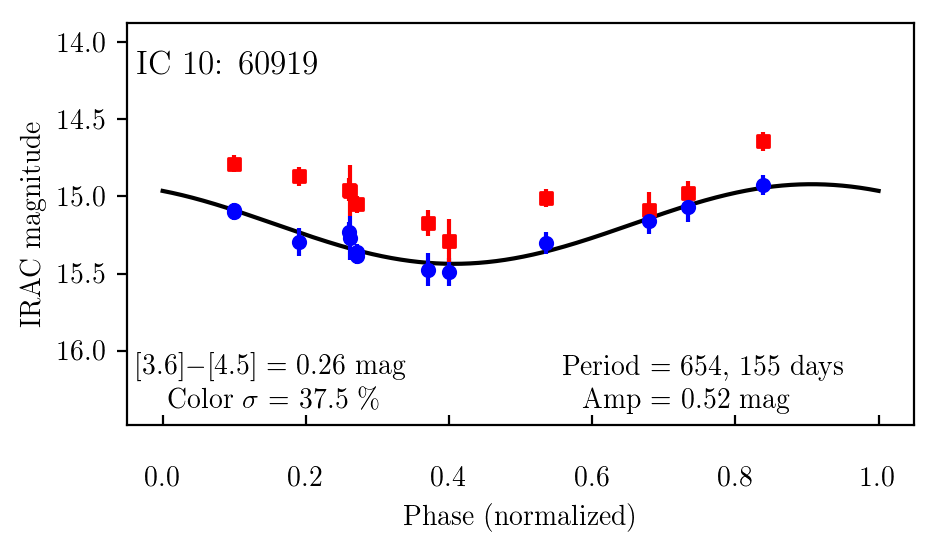} \hspace{-0.32cm}
	\includegraphics[width=6cm]{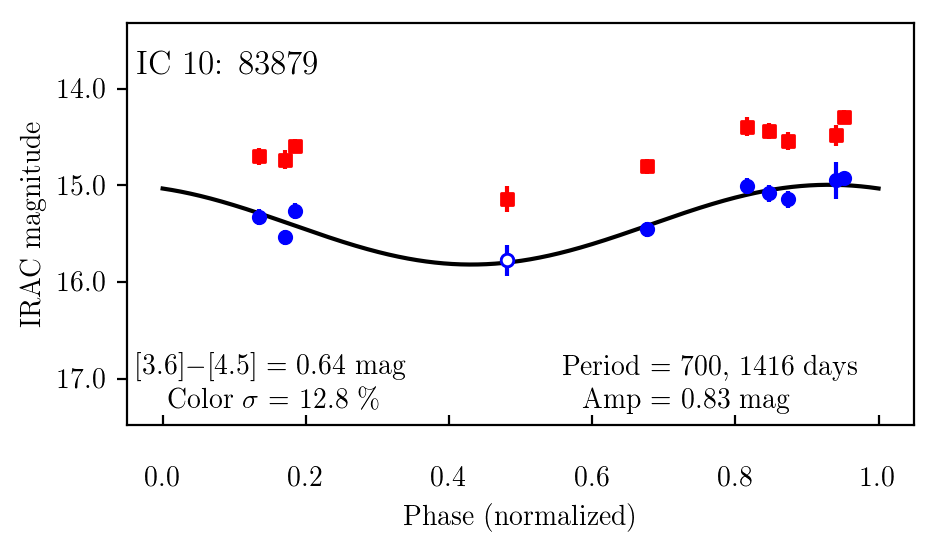} \vspace{-0.75cm}\\

	\includegraphics[width=6cm]{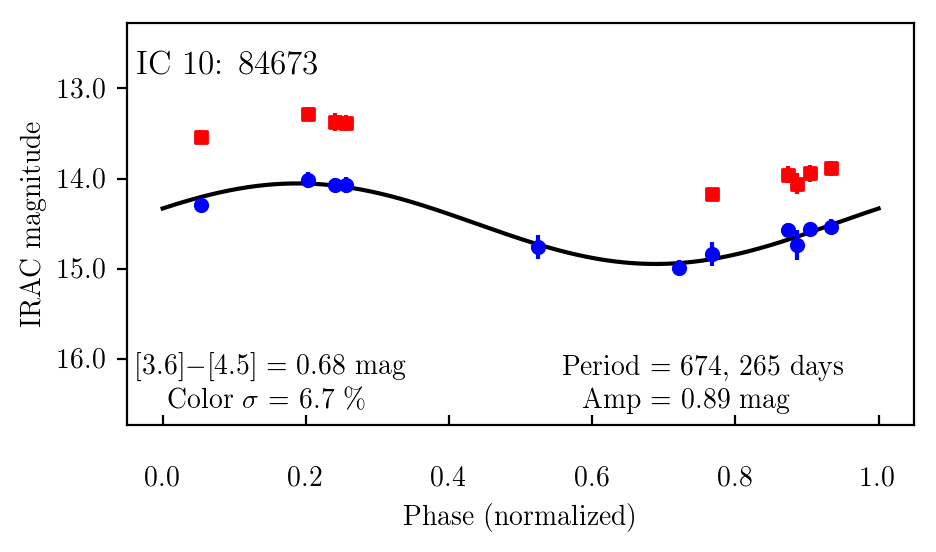} \hspace{-0.32cm}
	\includegraphics[width=6cm]{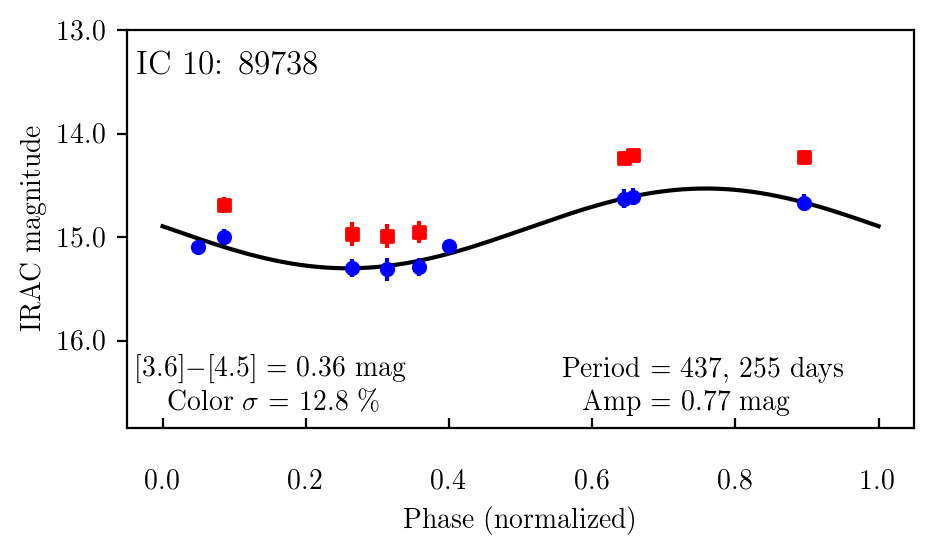} \hspace{-0.32cm}
	\includegraphics[width=6cm]{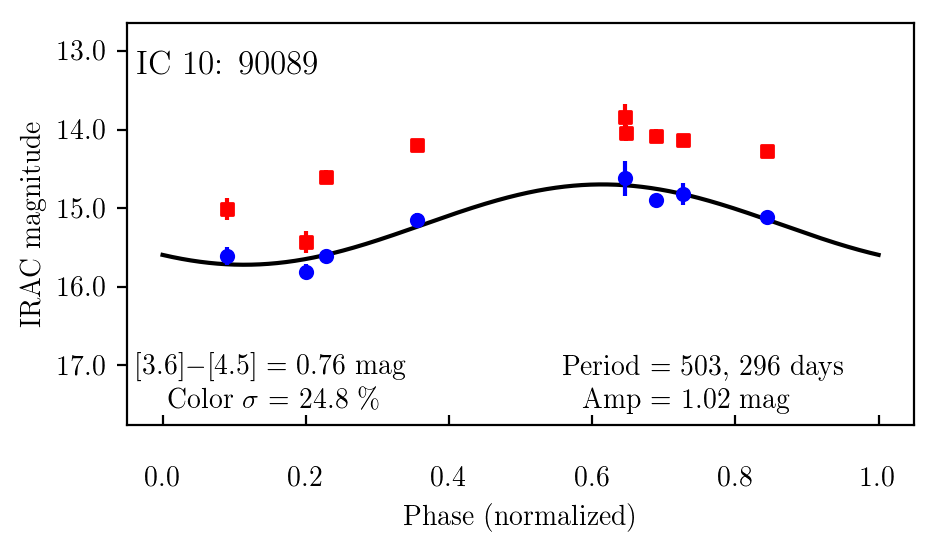} \vspace{-0.75cm}\\

	\includegraphics[width=6cm]{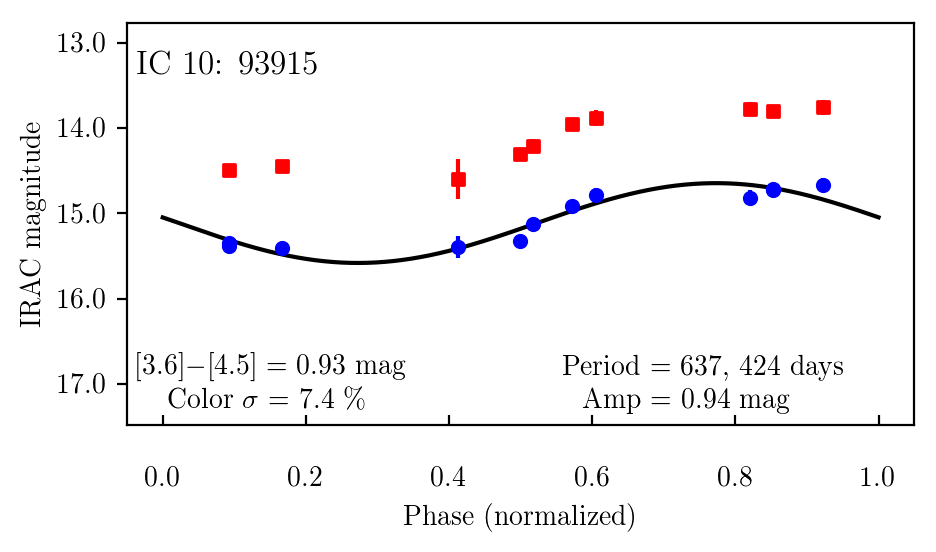} \hspace{-0.32cm}
	\includegraphics[width=6cm]{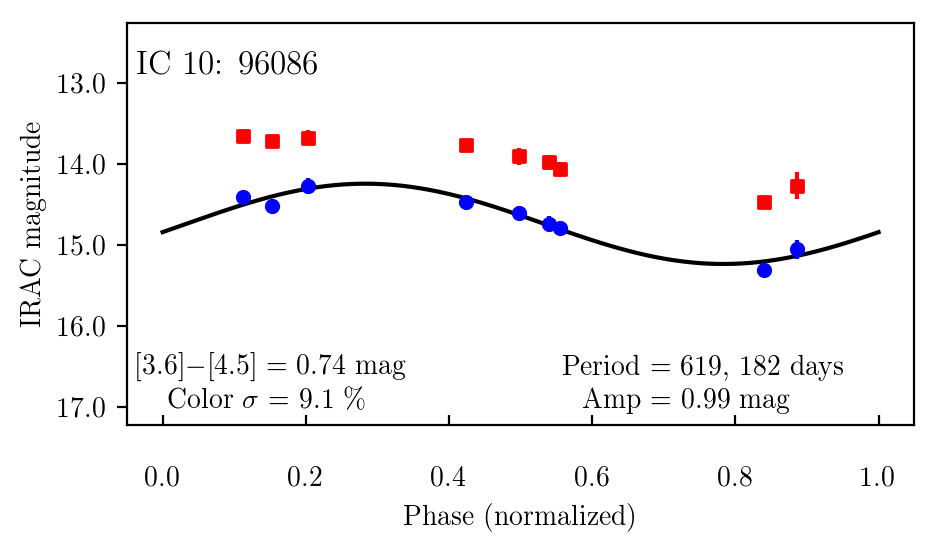} \hspace{-0.32cm}
	\includegraphics[width=6cm]{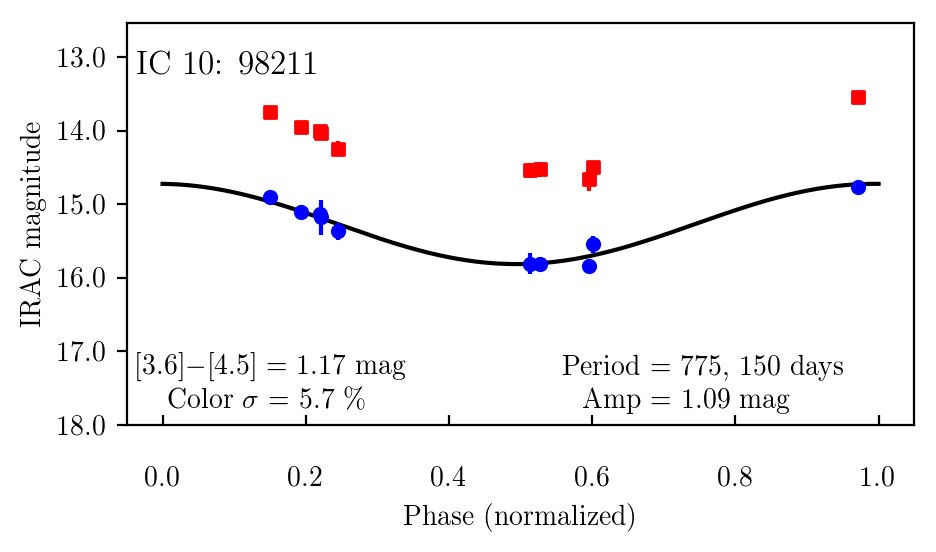} \vspace{-0.75cm}\\

	\includegraphics[width=6cm]{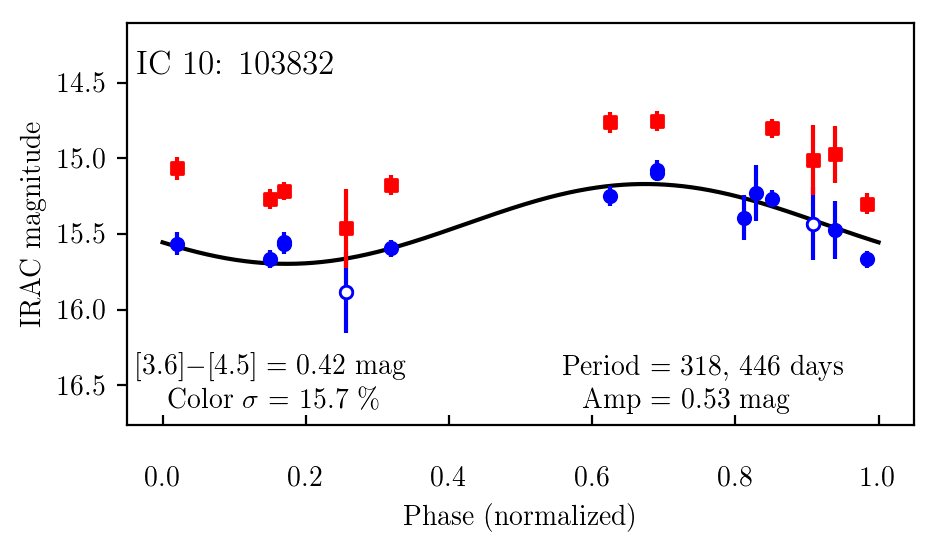} \hspace{-0.32cm}
	\includegraphics[width=6cm]{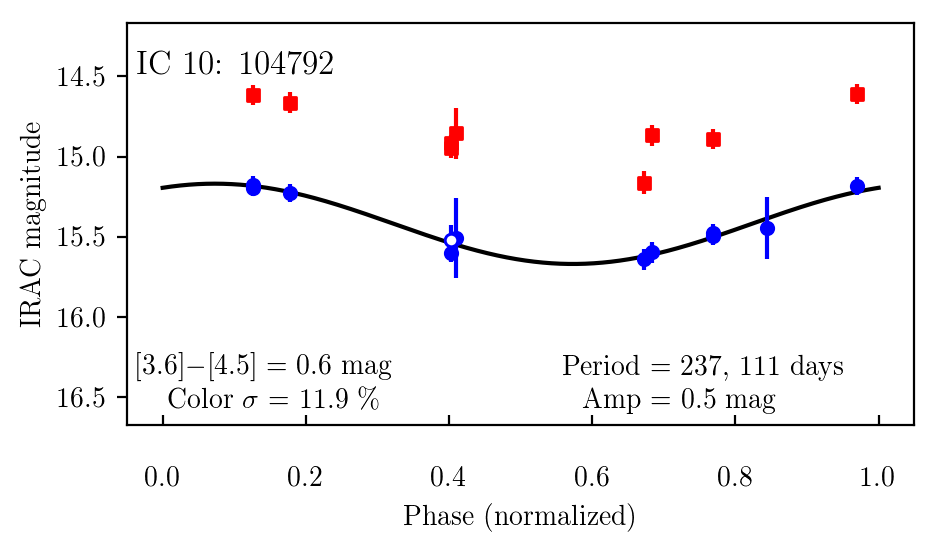} \hspace{-0.32cm}
	\includegraphics[width=6cm]{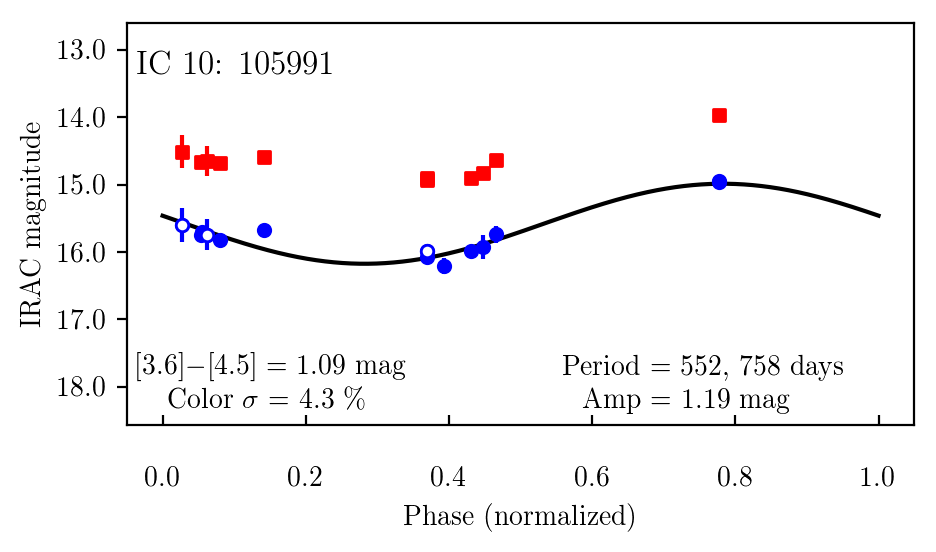} \vspace{-0.75cm}\\

	\includegraphics[width=6cm]{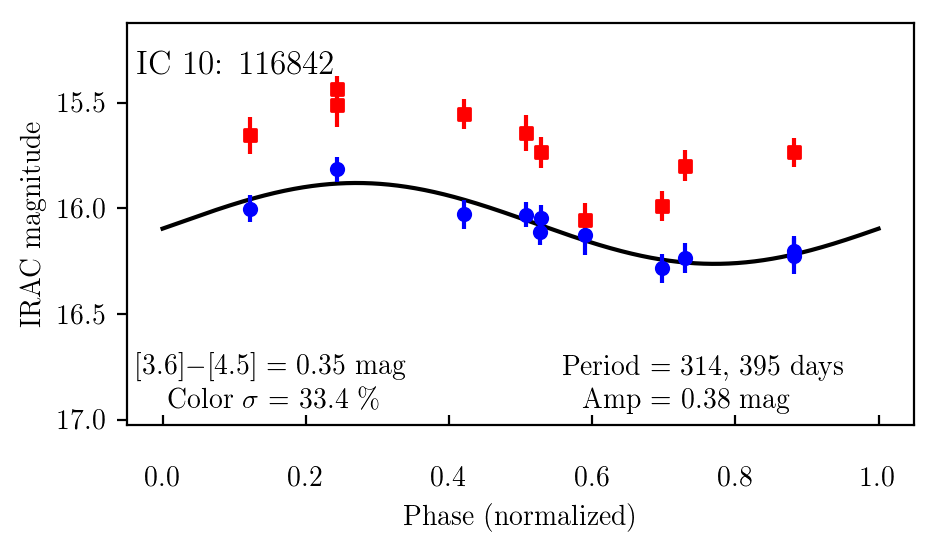} \hspace{-0.32cm}
	\includegraphics[width=6cm]{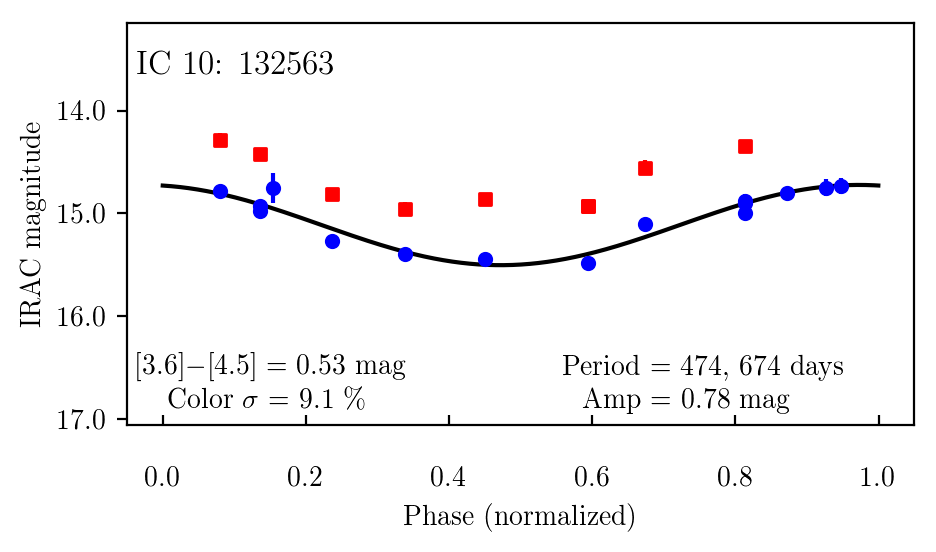} \hspace{-0.32cm}
	\includegraphics[width=6cm]{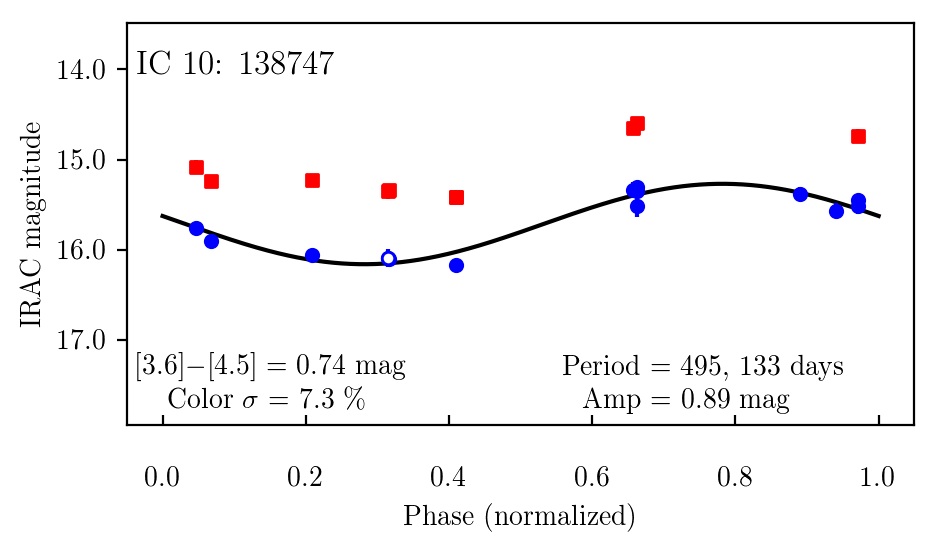} \\
  \end{center}
  % \vspace{-0.25cm}
  \caption{Our sample of high-confidence fit lightcurves. Shown are the IRAC 3.6\,$\mu$m (blue) and IRAC 4.5\,$\mu$m (red) photometry, and the best-fitting periodogram, fitted to the [3.6] data. Unless otherwise indicated, the data error bars are smaller than the plotting symbols. Also shown is our ``simulated'' photometry denoted using open circles (see Section \ref{sect: identifying LPVs}). The two numbers listed as period are the best- and second-best-fit values. Also shown is the best-fit 3.6\,$\mu$m amplitude. Lightcurves are available in the electronic version\label{fig:high_confidence_lightcurves}.}
\end{figure*}

\addtocounter{figure}{-1}
\begin{figure*}
  \begin{center}
	\includegraphics[width=6cm]{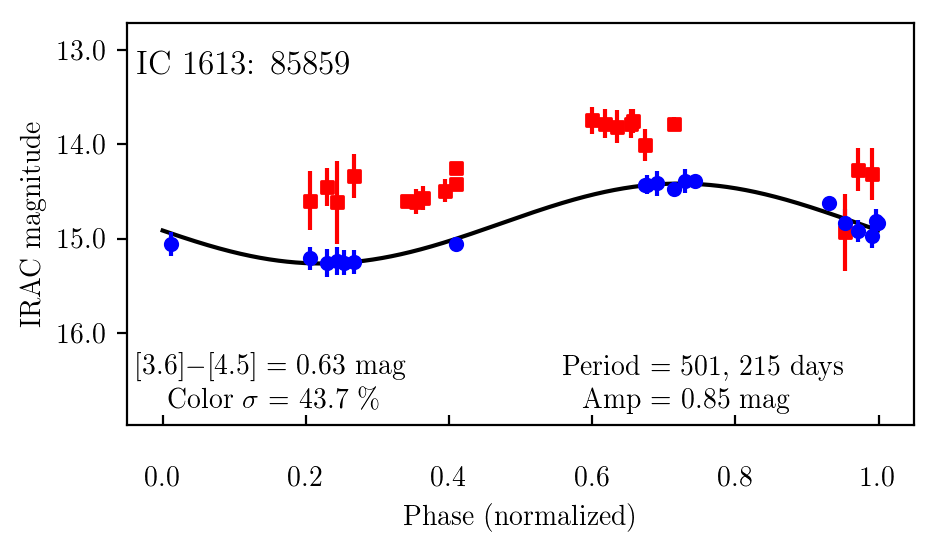} \hspace{-0.32cm}
	\includegraphics[width=6cm]{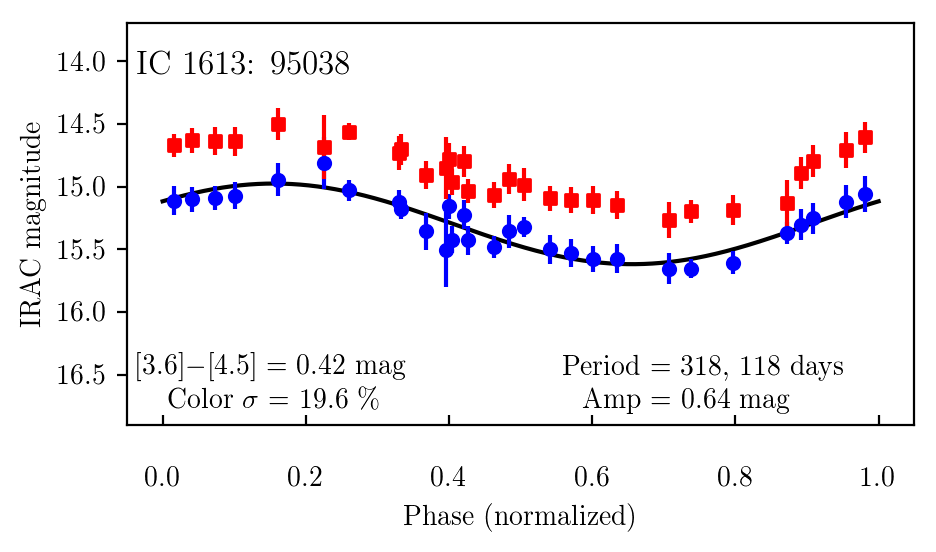} \hspace{-0.32cm}
	\includegraphics[width=6cm]{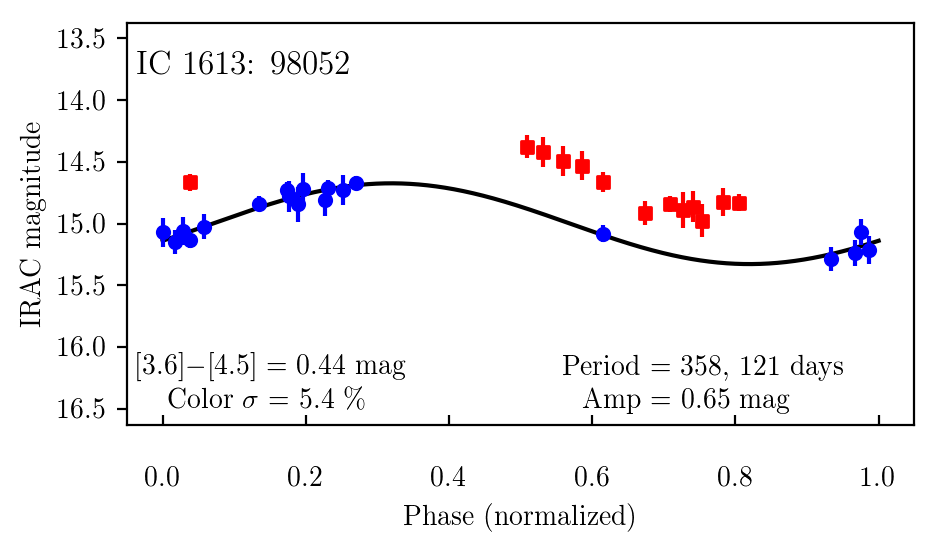} \vspace{-0.75cm}\\ 

	\includegraphics[width=6cm]{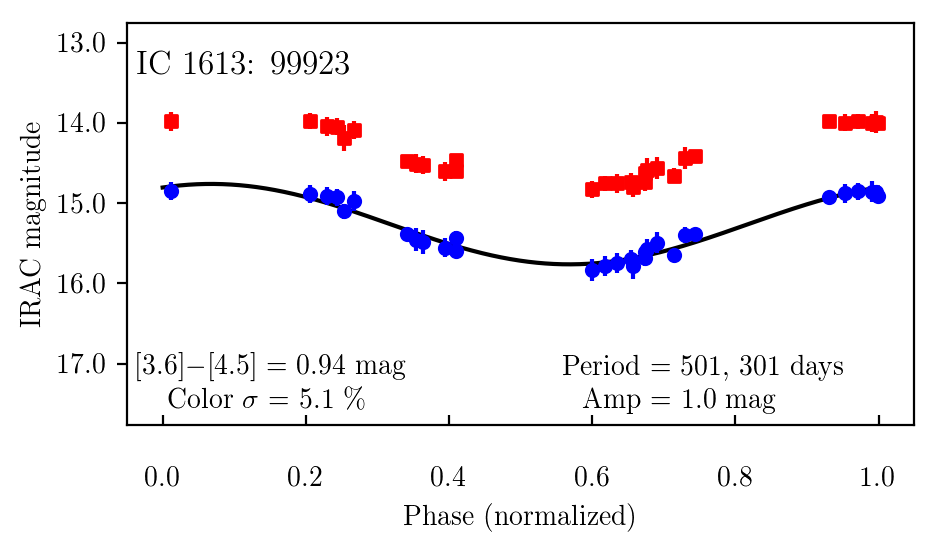} \hspace{-0.32cm}
	\includegraphics[width=6cm]{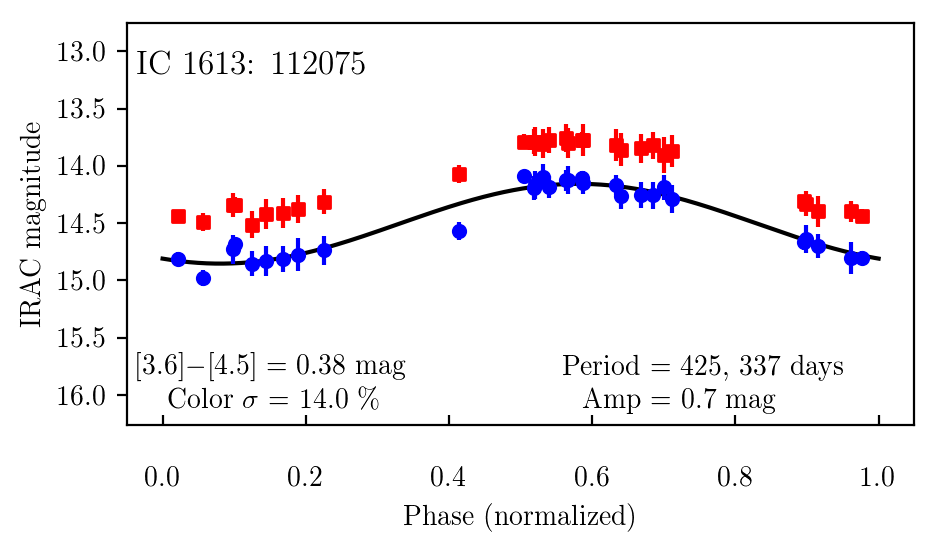} \hspace{-0.32cm}
	\includegraphics[width=6cm]{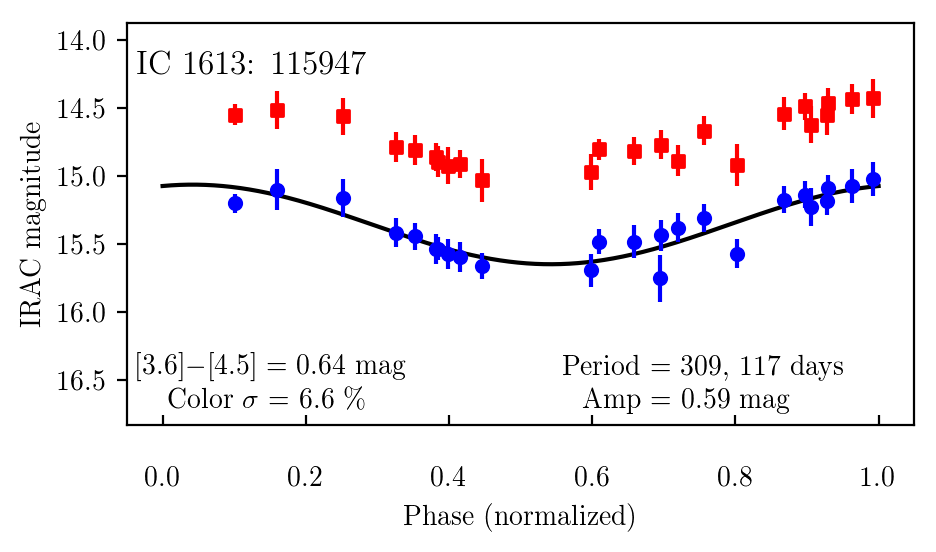} \vspace{-0.75cm}\\

	\includegraphics[width=6cm]{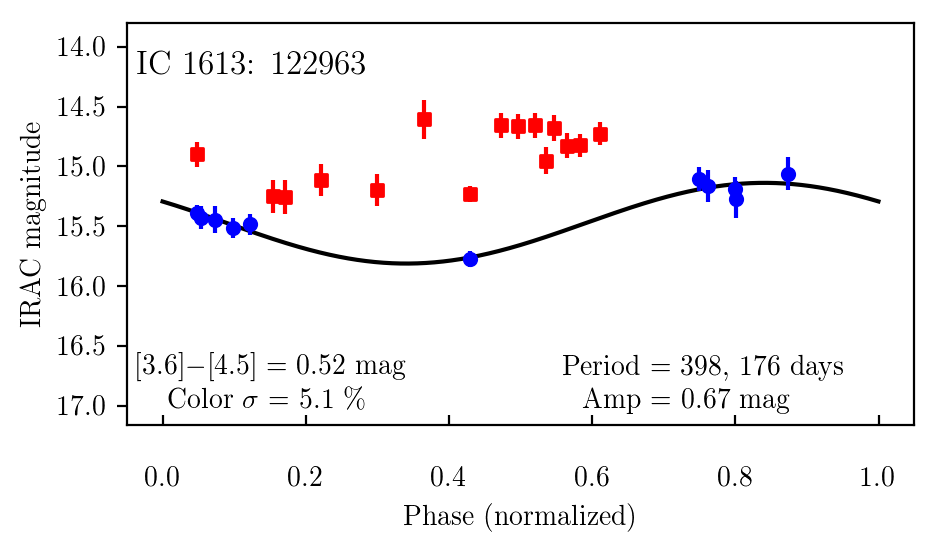} \hspace{-0.32cm}
	\includegraphics[width=6cm]{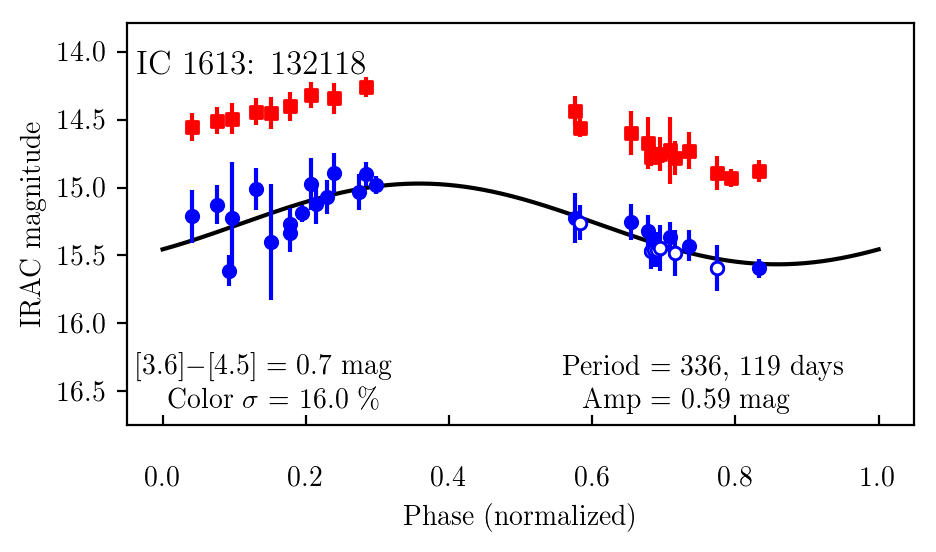} \hspace{-0.32cm}
	\includegraphics[width=6cm]{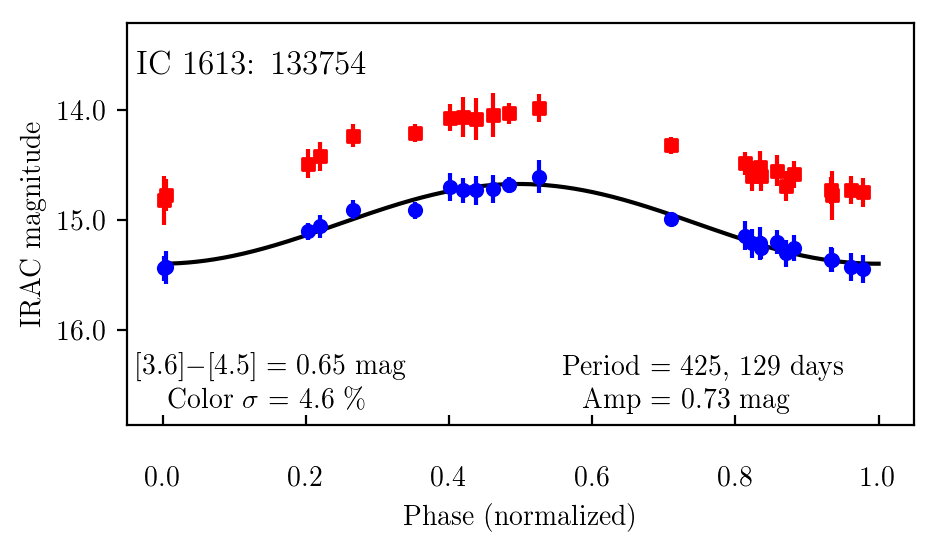} \vspace{-0.75cm}\\

	\includegraphics[width=6cm]{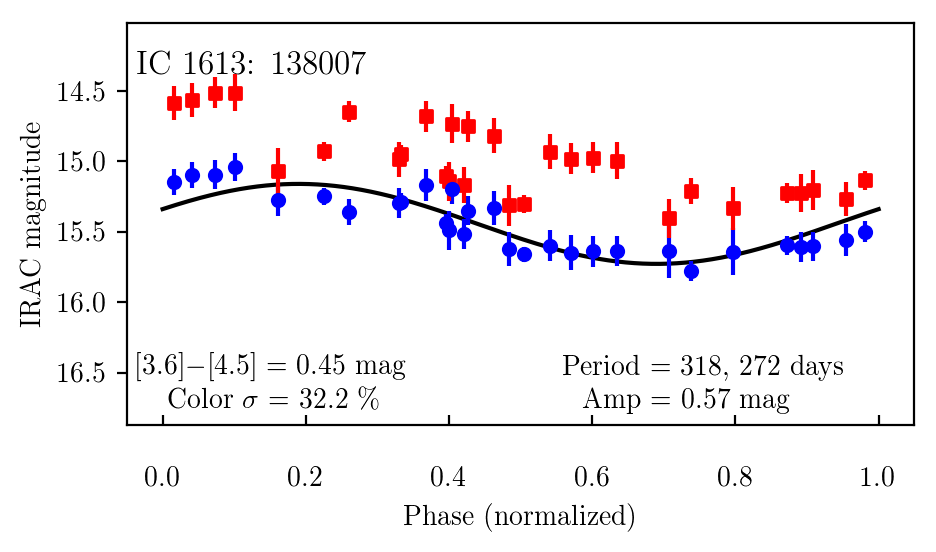} \hspace{-0.32cm}
	\includegraphics[width=6cm]{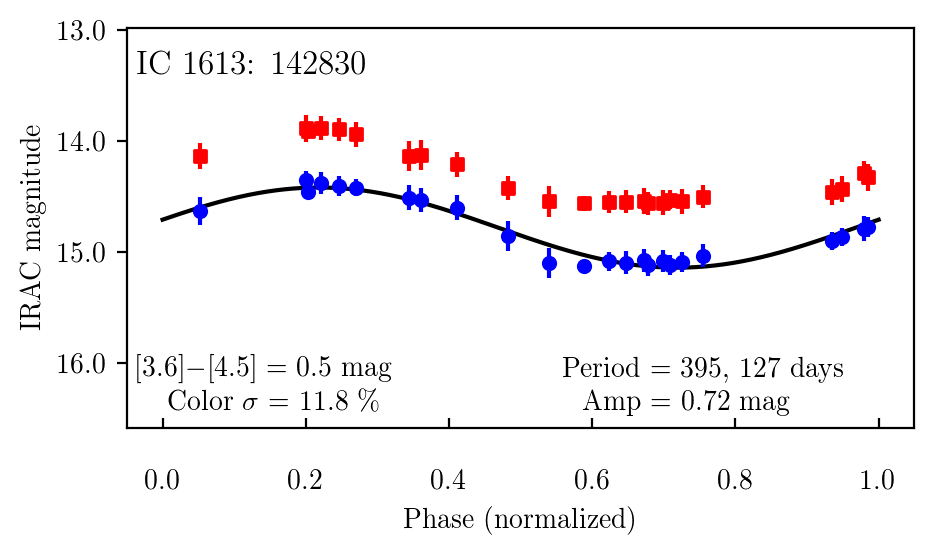} \hspace{-0.32cm}
	\includegraphics[width=6cm]{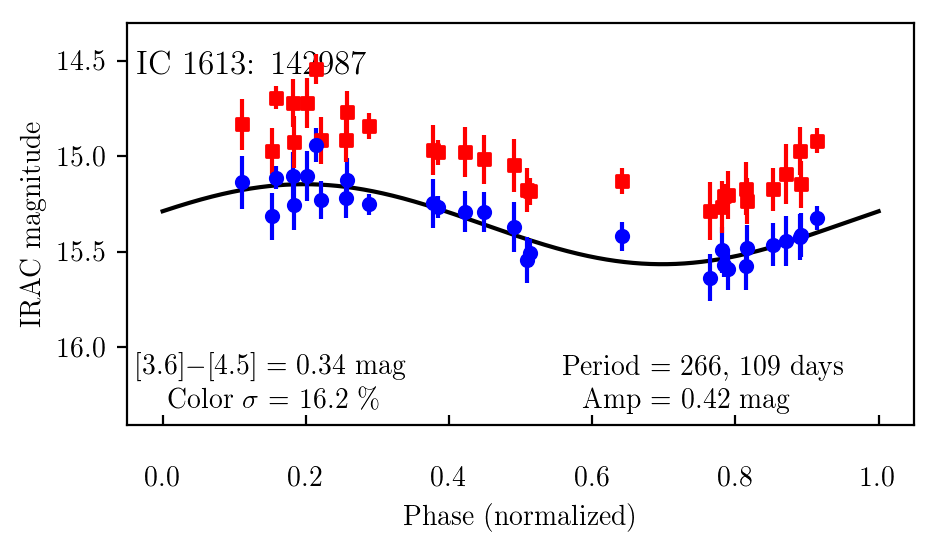} \vspace{-0.75cm}\\

	\includegraphics[width=6cm]{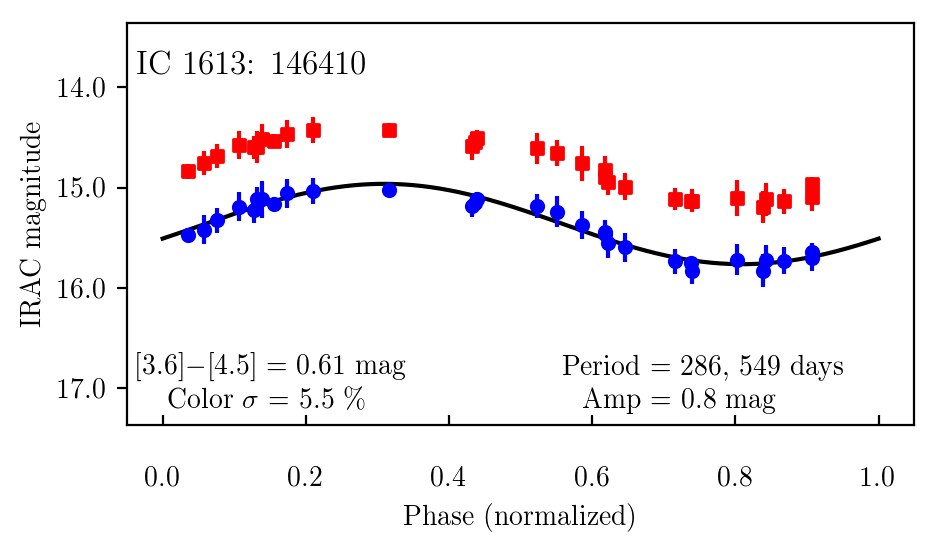} \hspace{-0.32cm}
	\includegraphics[width=6cm]{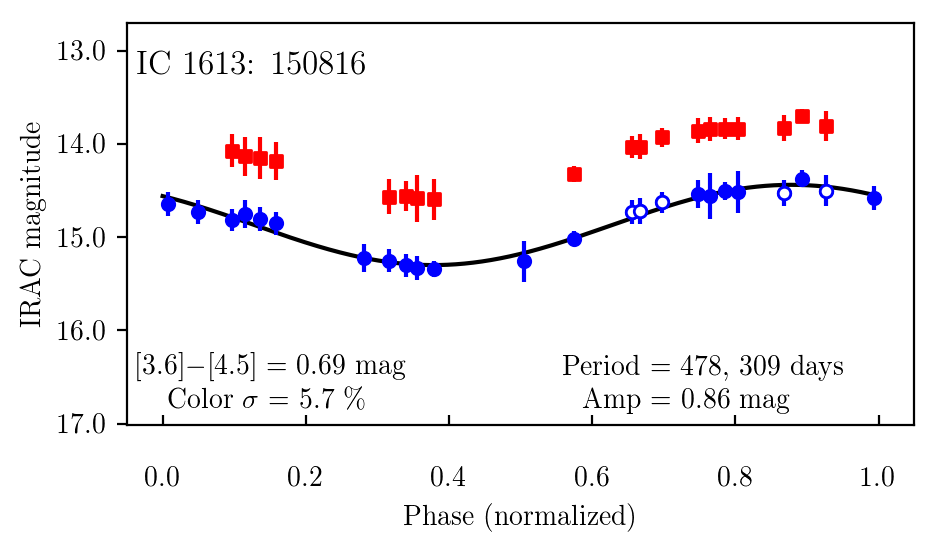} \hspace{-0.32cm}
	\includegraphics[width=6cm]{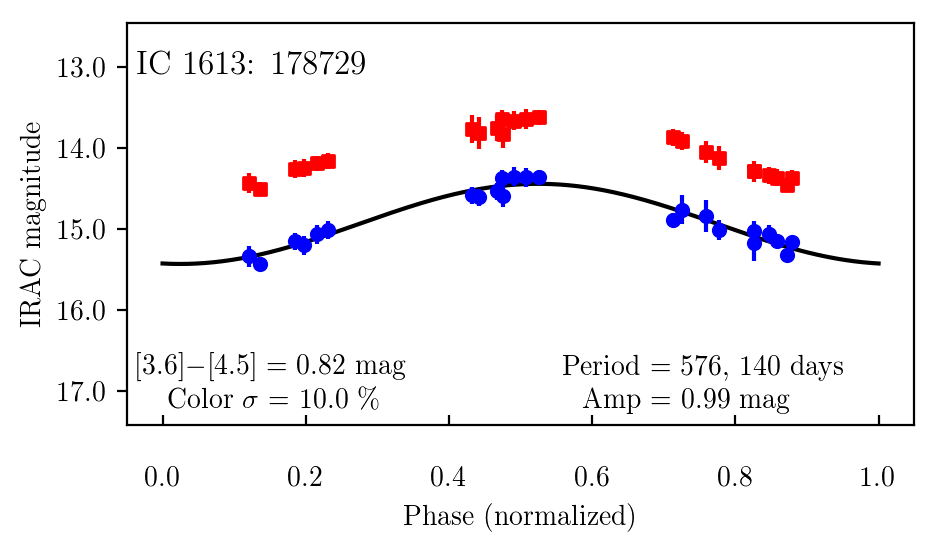} \vspace{-0.75cm}\\

	\includegraphics[width=6cm]{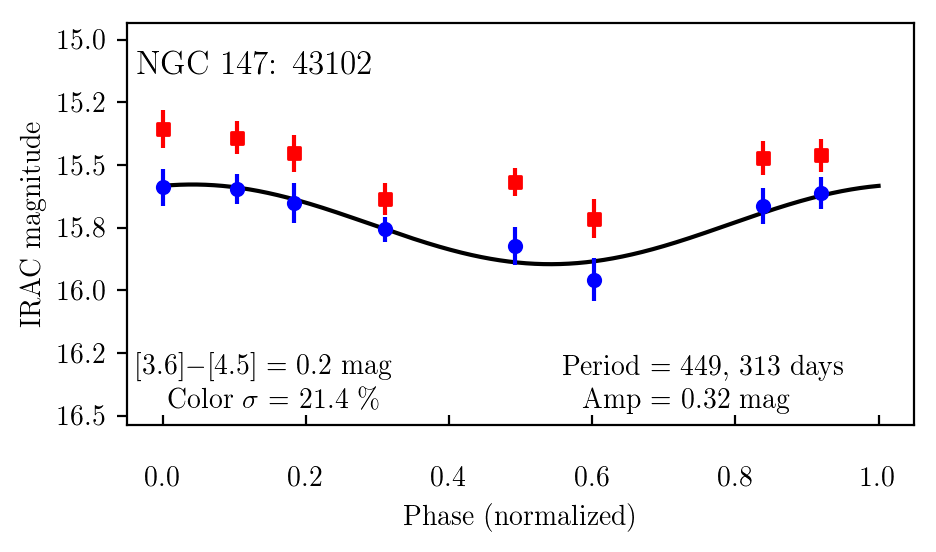} \hspace{-0.32cm}
	\includegraphics[width=6cm]{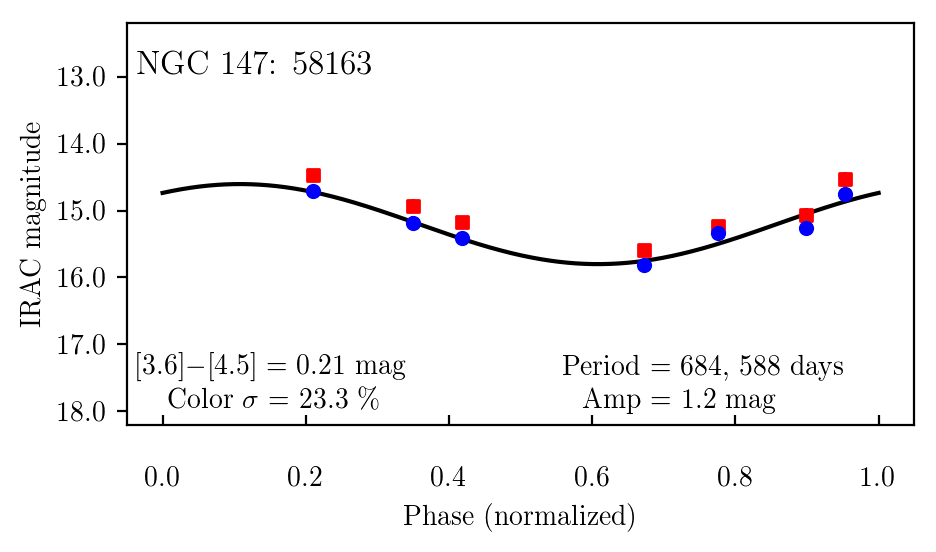} \hspace{-0.32cm}
	\includegraphics[width=6cm]{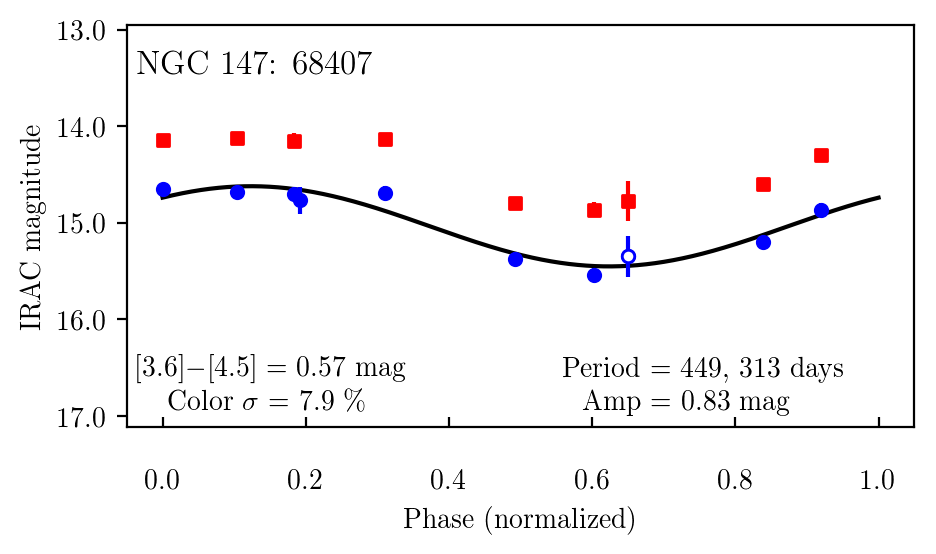} \vspace{-0.75cm}\\

	\includegraphics[width=6cm]{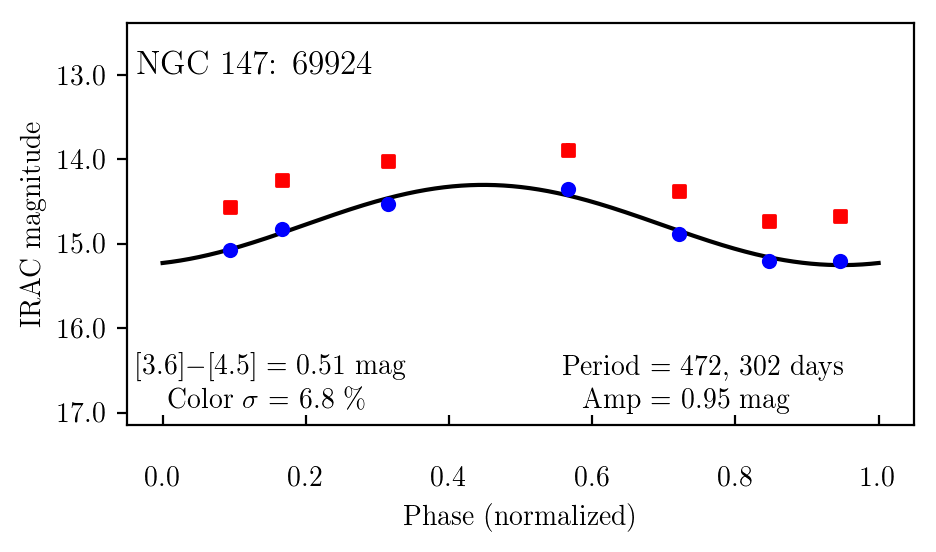} \hspace{-0.32cm}
	\includegraphics[width=6cm]{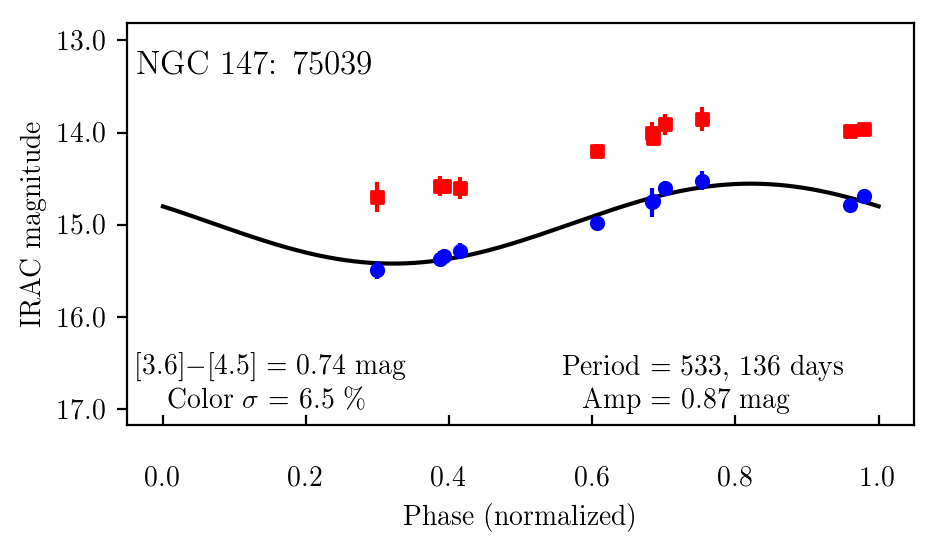} \hspace{-0.32cm}
	\includegraphics[width=6cm]{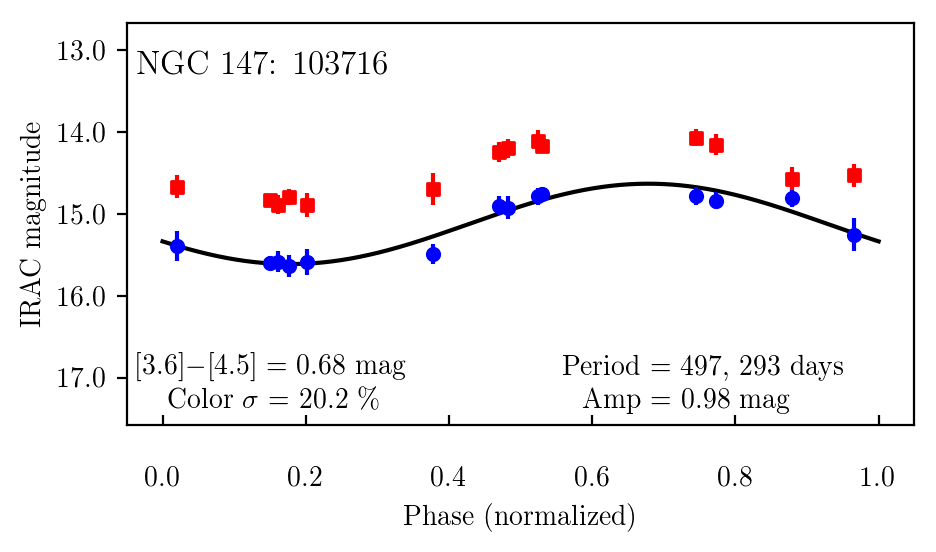} \\
   \end{center}
   % \vspace{-0.45cm}
   \caption{continued}
\end{figure*}

\clearpage	

\addtocounter{figure}{-1}
\begin{figure*}
  \begin{center}
	\includegraphics[width=6cm]{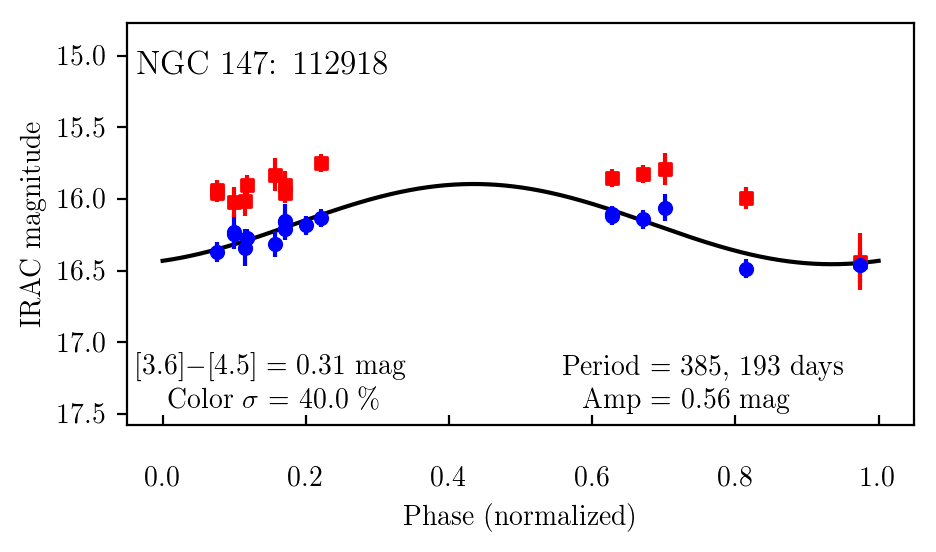} \hspace{-0.32cm}
	\includegraphics[width=6cm]{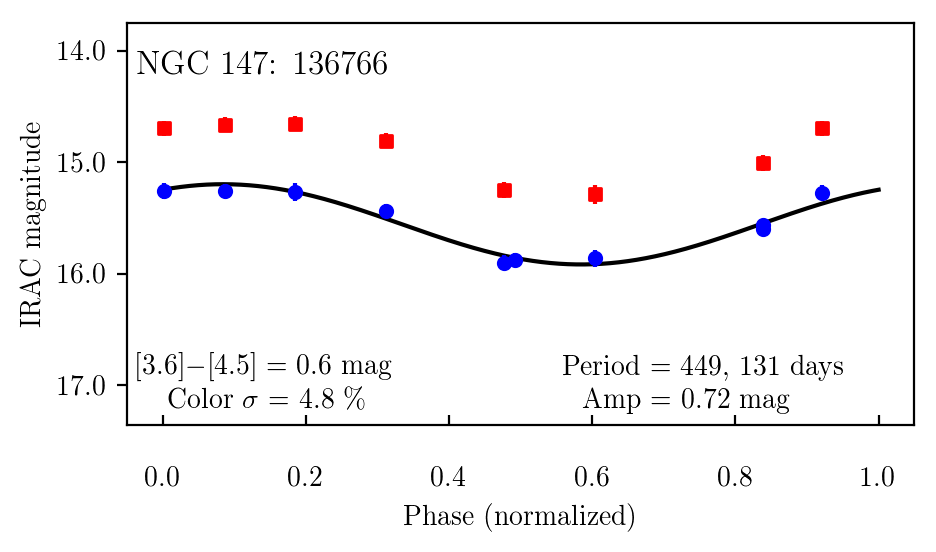} \hspace{-0.32cm}
	\includegraphics[width=6cm]{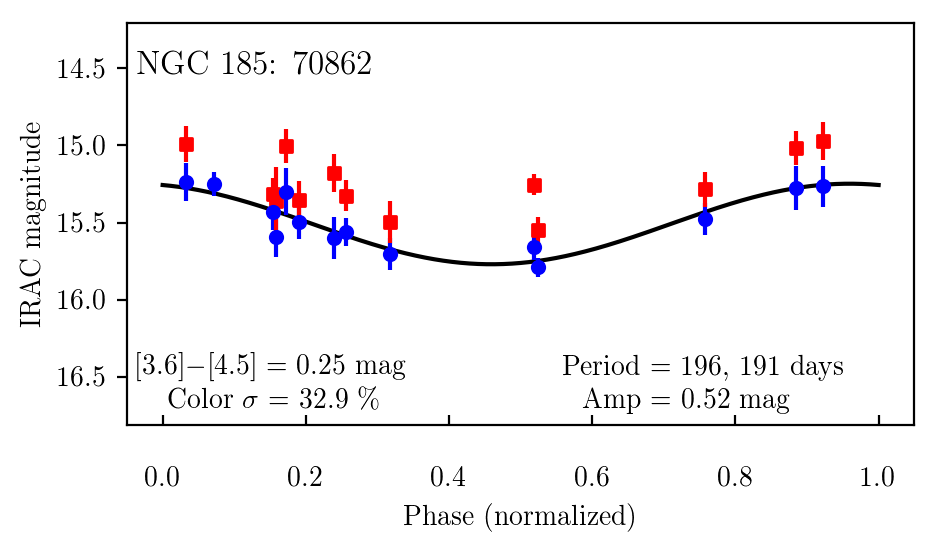} \vspace{-0.75cm}\\ 

	\includegraphics[width=6cm]{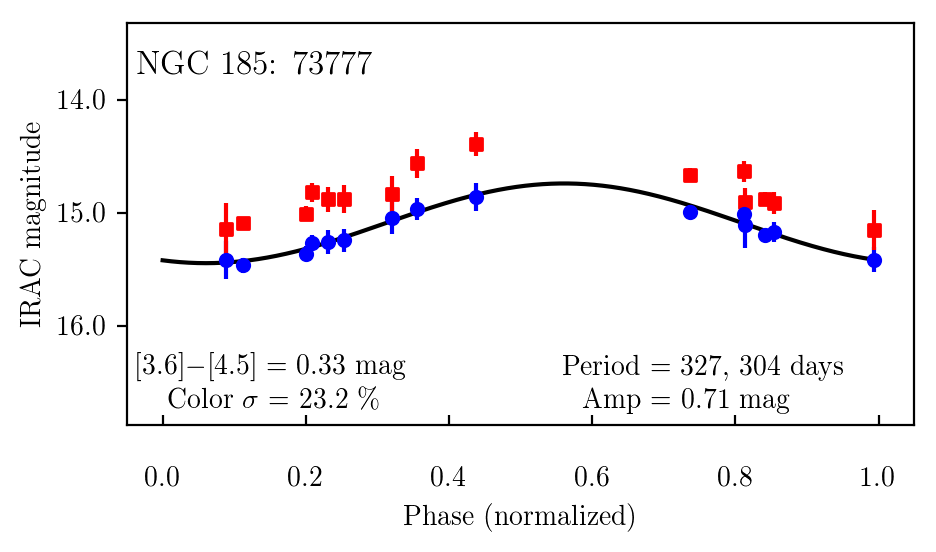} \hspace{-0.32cm}
	\includegraphics[width=6cm]{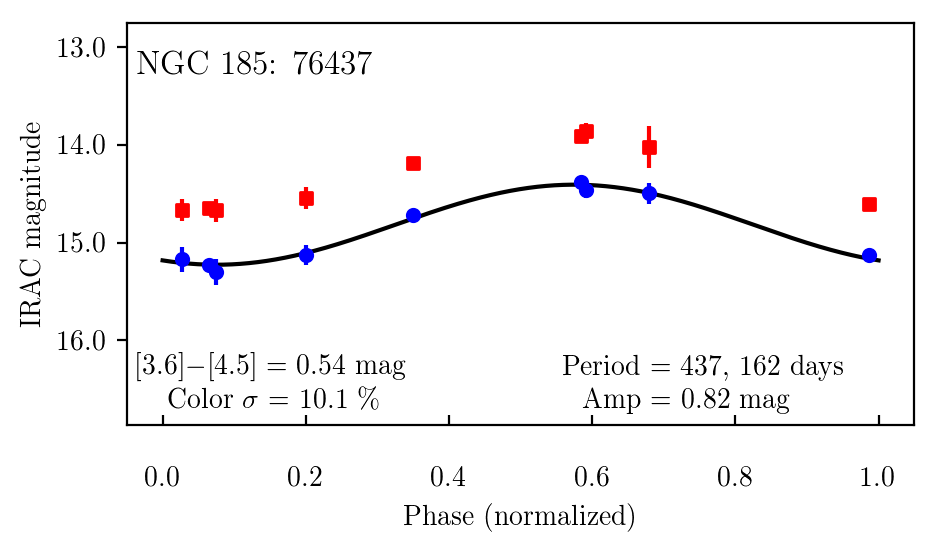} \hspace{-0.32cm}
	\includegraphics[width=6cm]{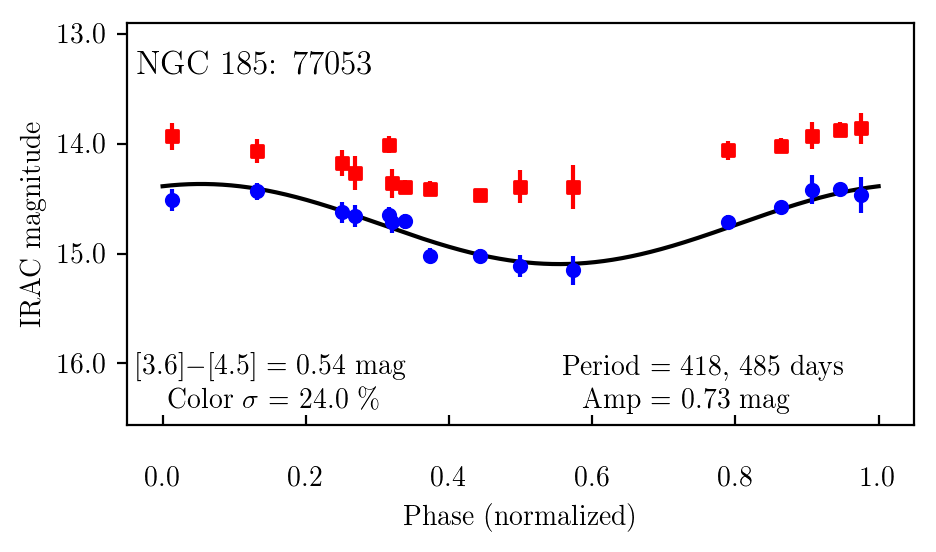} \vspace{-0.75cm}\\

	\includegraphics[width=6cm]{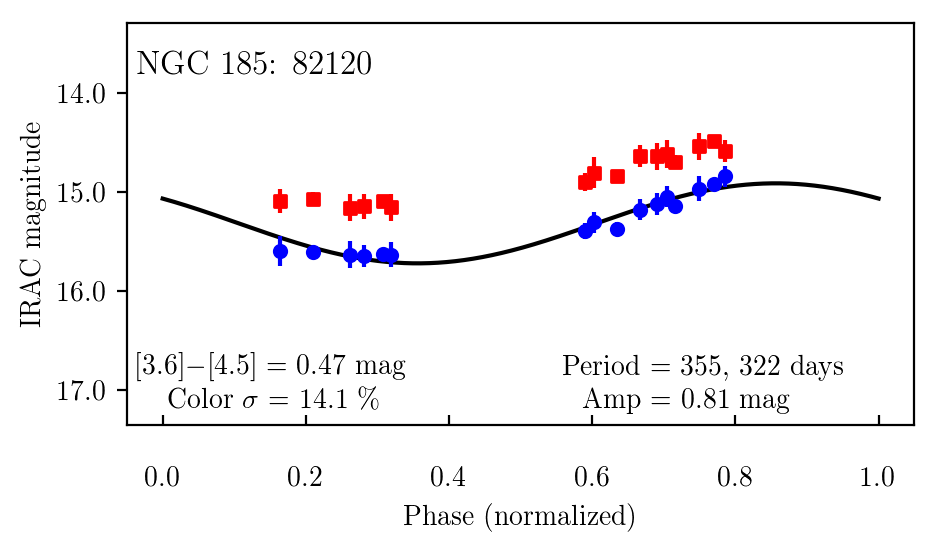} \hspace{-0.32cm}
	\includegraphics[width=6cm]{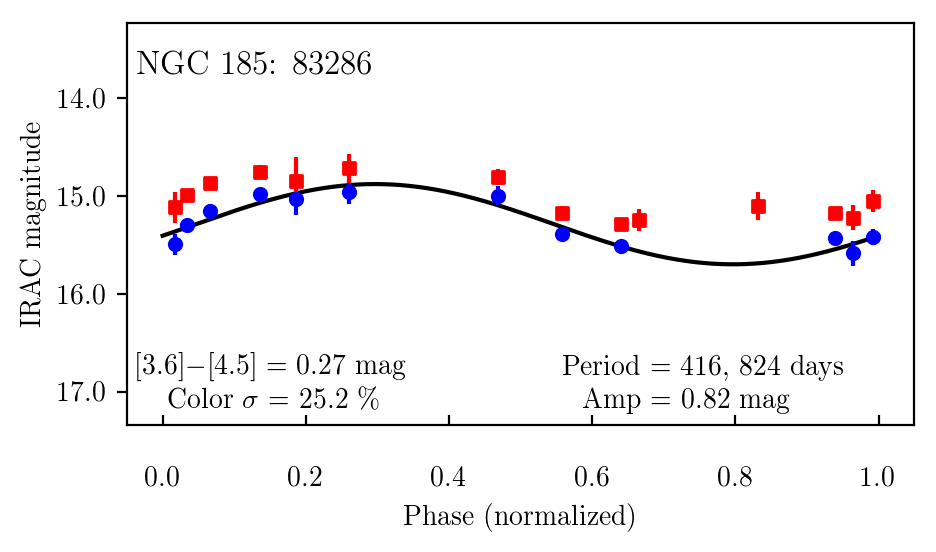} \hspace{-0.32cm}
	\includegraphics[width=6cm]{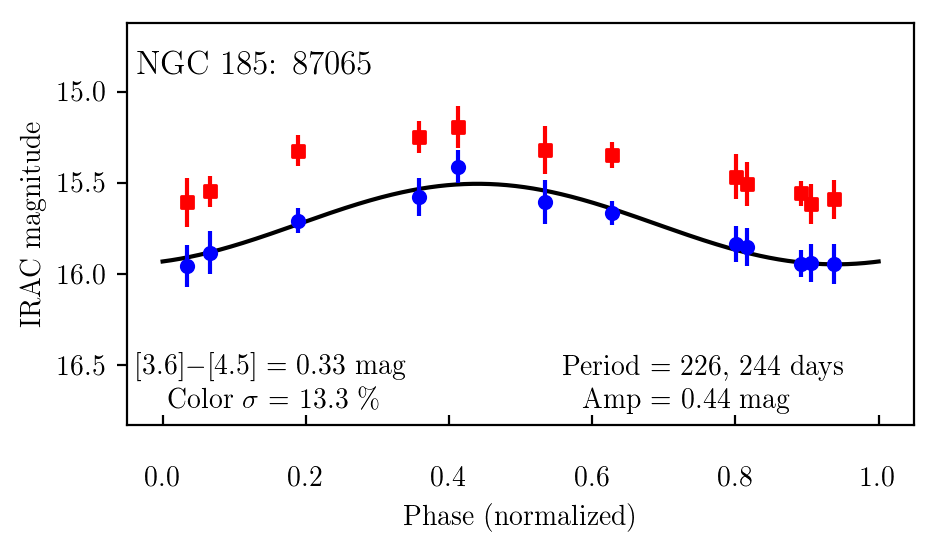} \vspace{-0.75cm}\\

	\includegraphics[width=6cm]{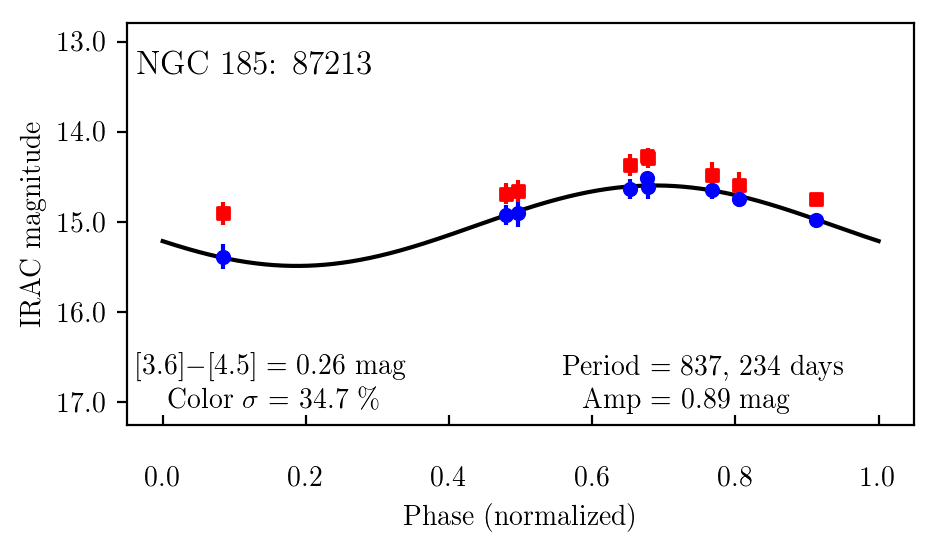} \hspace{-0.32cm}
	\includegraphics[width=6cm]{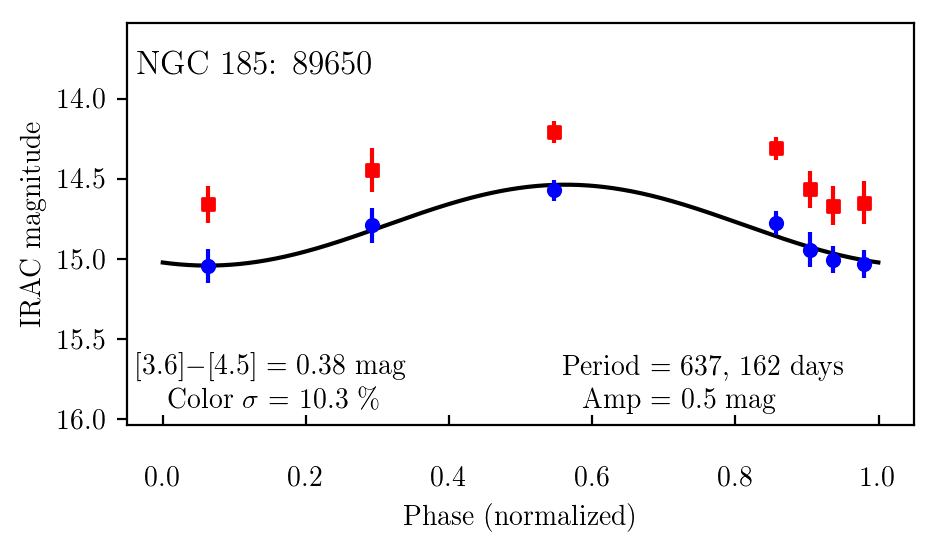} \hspace{-0.32cm}
	\includegraphics[width=6cm]{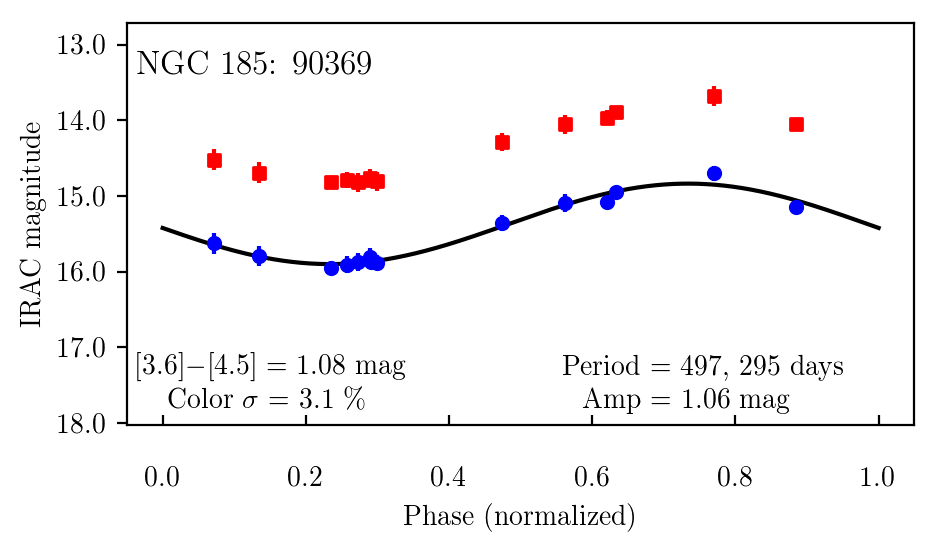} \vspace{-0.75cm}\\

	\includegraphics[width=6cm]{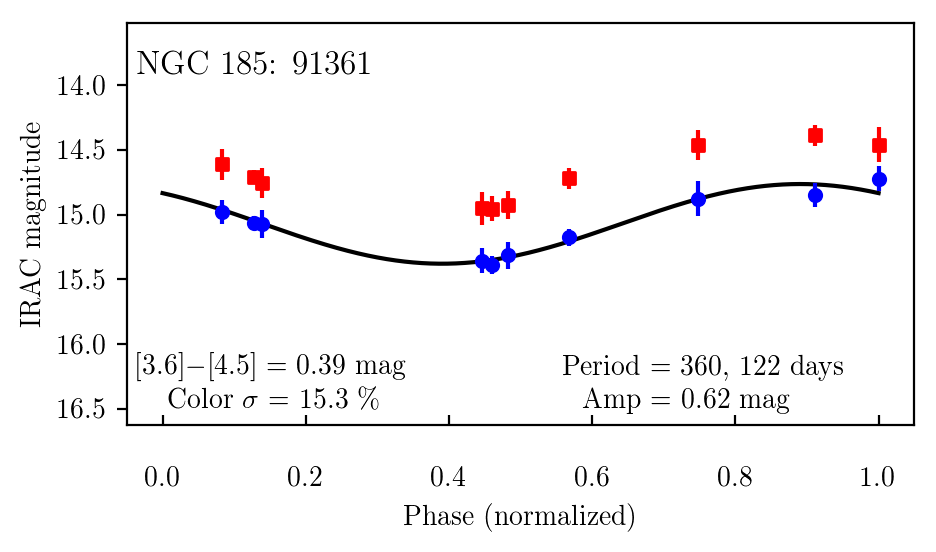} \hspace{-0.32cm}
	\includegraphics[width=6cm]{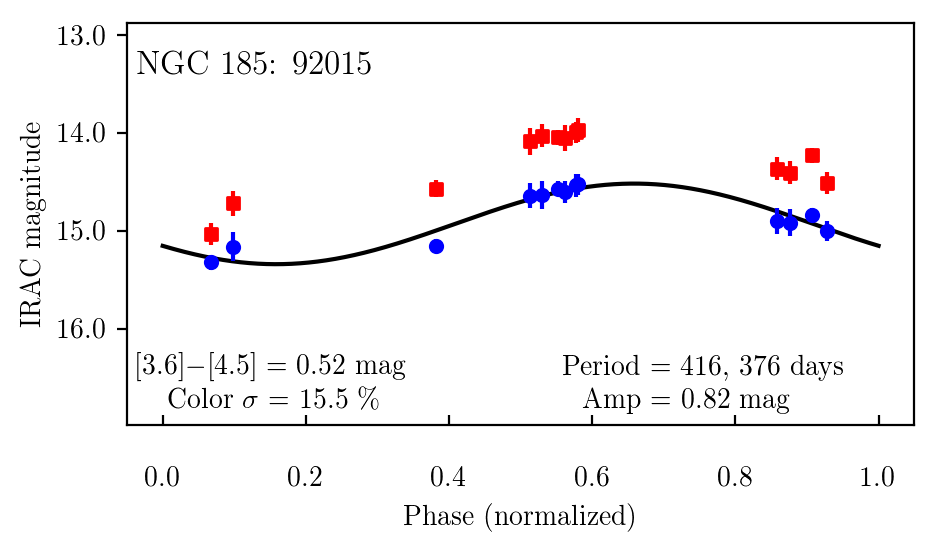} \hspace{-0.32cm}
	\includegraphics[width=6cm]{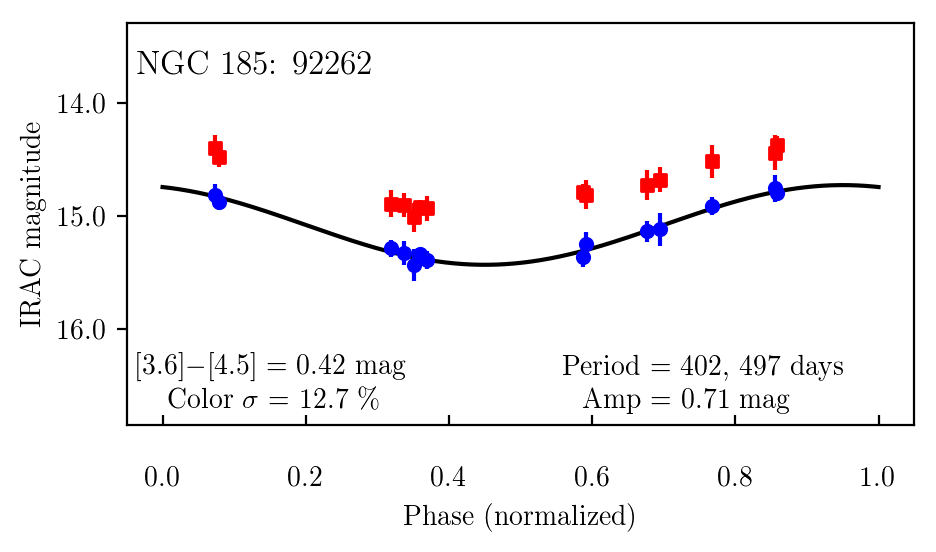} \vspace{-0.75cm}\\

	\includegraphics[width=6cm]{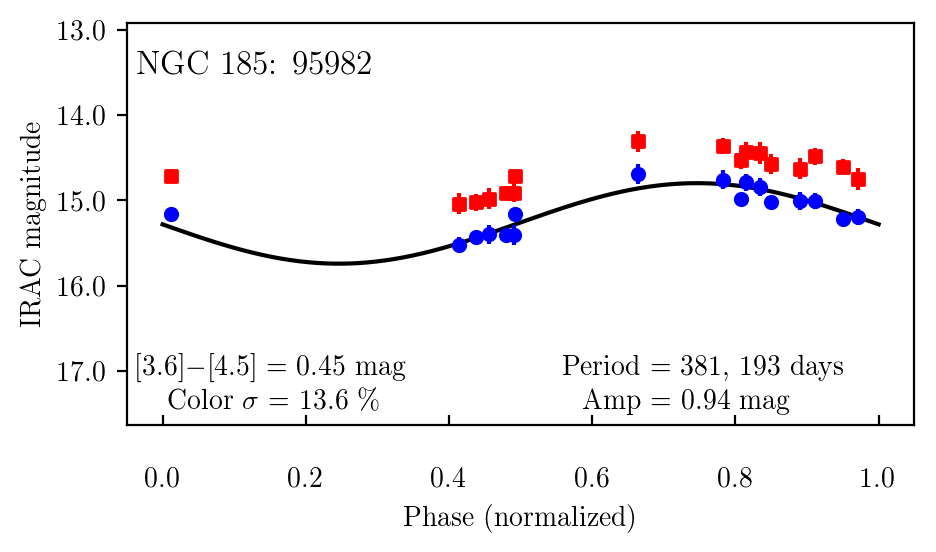} \hspace{-0.32cm}
	\includegraphics[width=6cm]{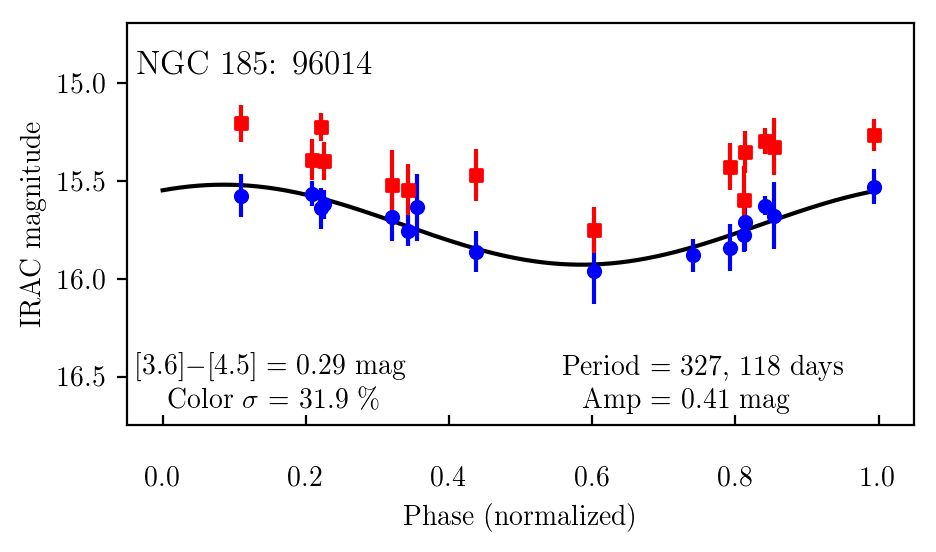} \hspace{-0.32cm}
	\includegraphics[width=6cm]{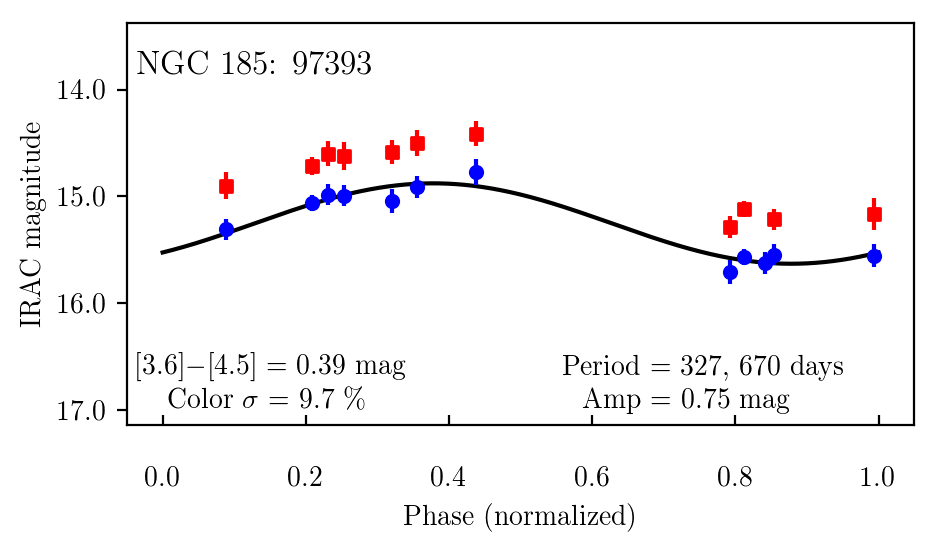} \vspace{-0.75cm}\\

	\includegraphics[width=6cm]{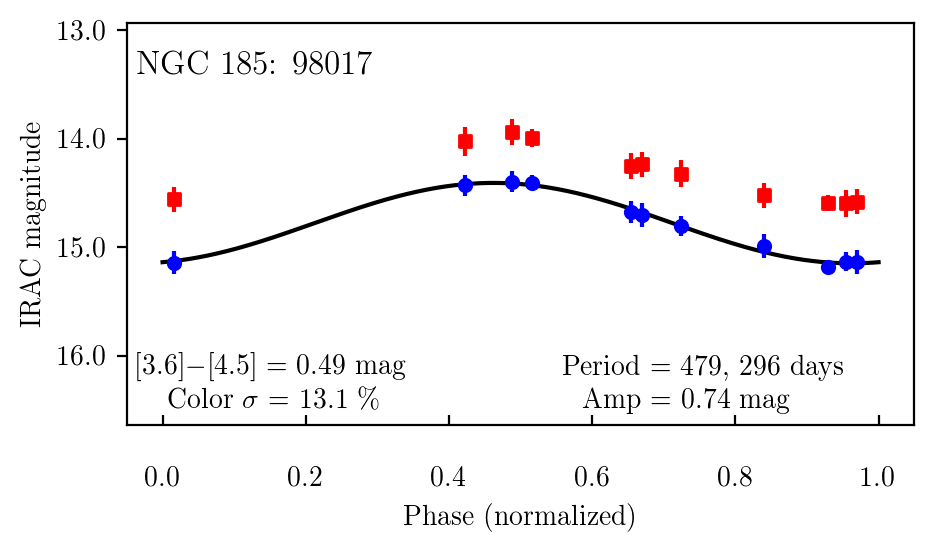} \hspace{-0.32cm}
	\includegraphics[width=6cm]{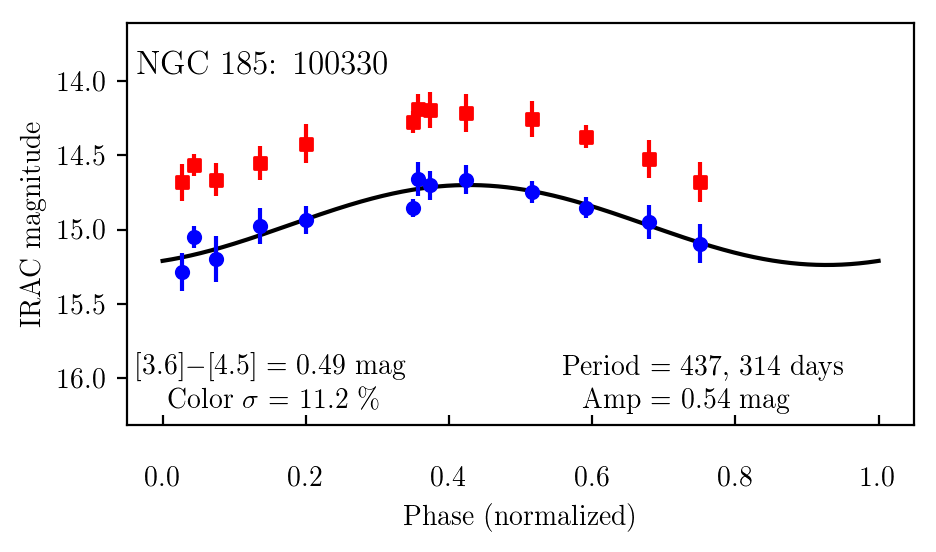} \hspace{-0.32cm}
	\includegraphics[width=6cm]{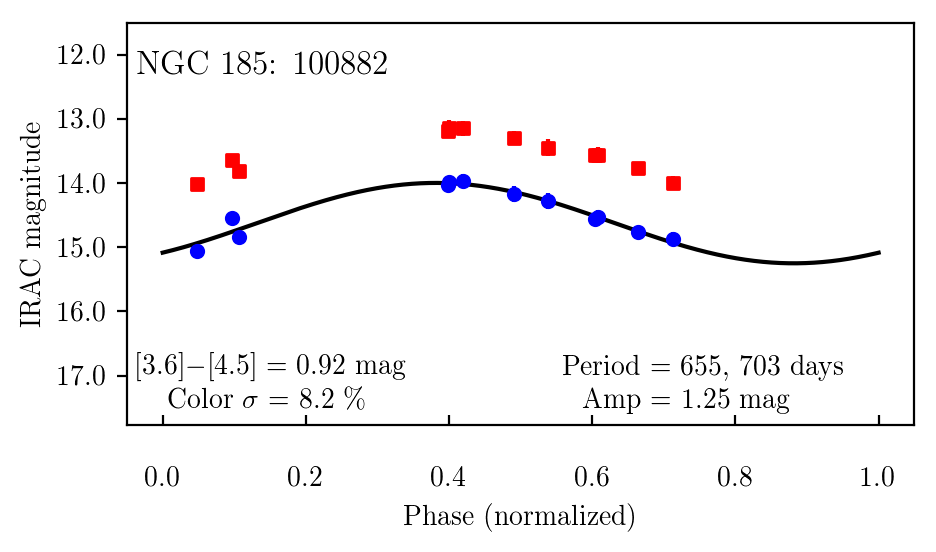} \\
   \end{center}
   % \vspace{-0.45cm}
   \caption{continued}
\end{figure*}

\clearpage

\addtocounter{figure}{-1}
\begin{figure*}
  \begin{center}
	\includegraphics[width=6cm]{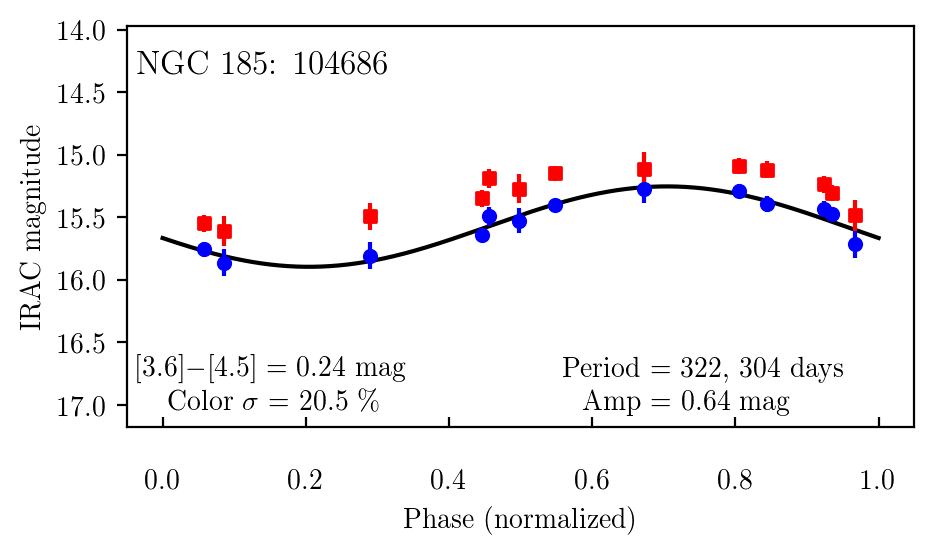} \hspace{-0.32cm}
	\includegraphics[width=6cm]{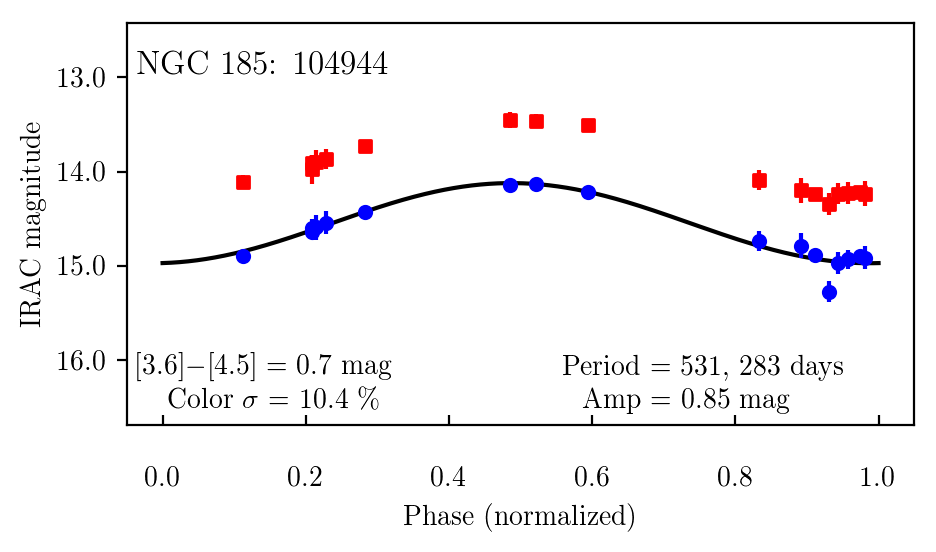} \hspace{-0.32cm}
	\includegraphics[width=6cm]{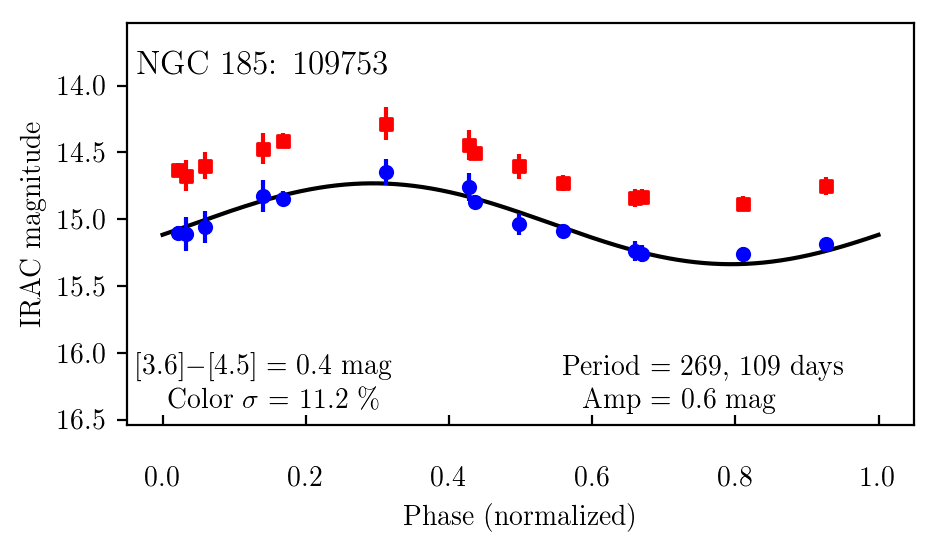} \vspace{-0.75cm}\\ 

	\includegraphics[width=6cm]{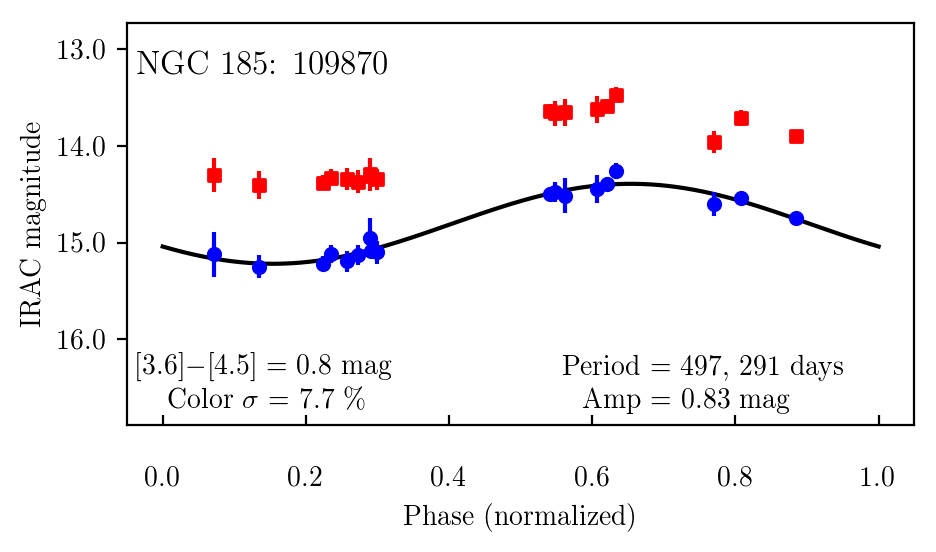} \hspace{-0.32cm}
	\includegraphics[width=6cm]{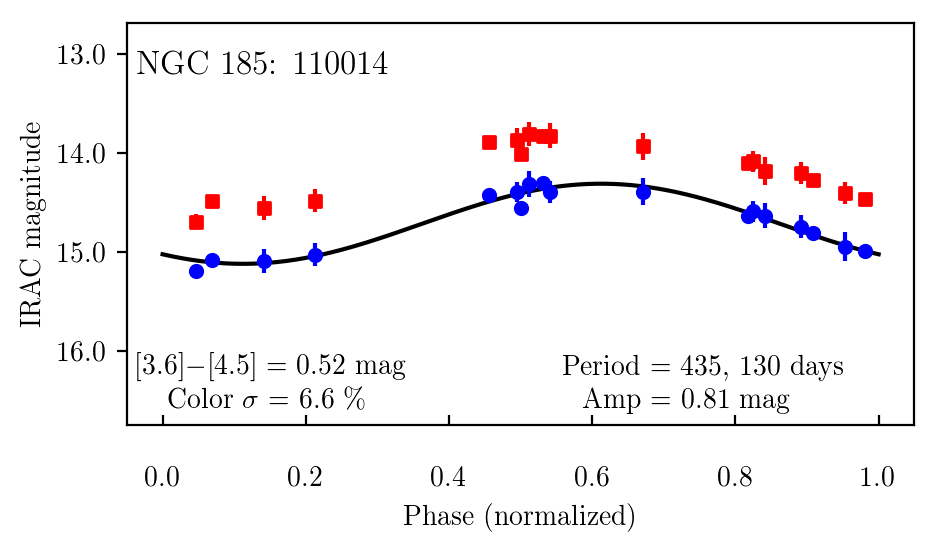} \hspace{-0.32cm}
	\includegraphics[width=6cm]{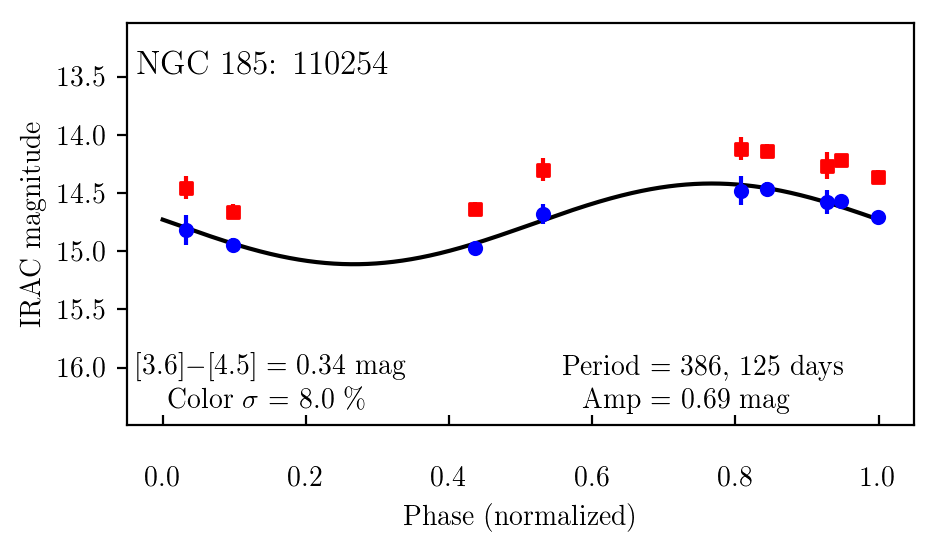} \vspace{-0.75cm}\\

	\includegraphics[width=6cm]{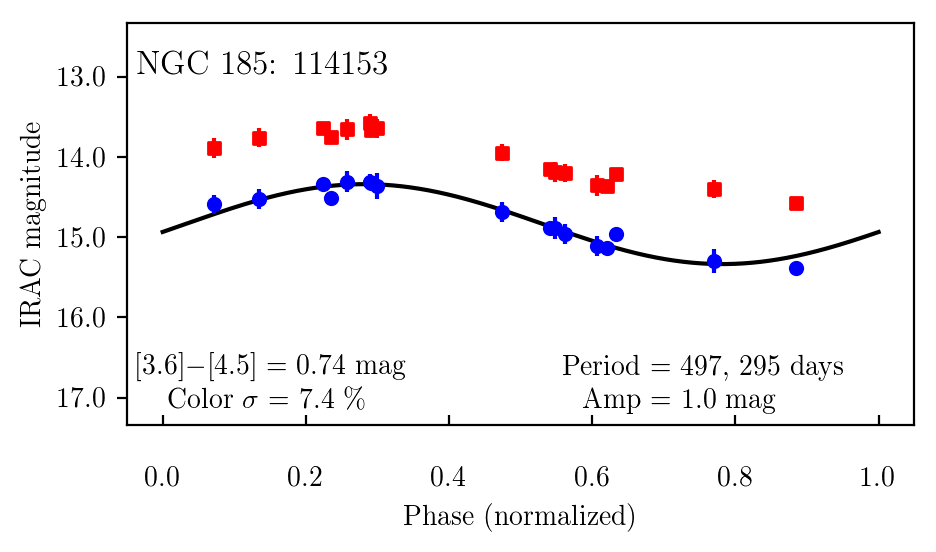} \hspace{-0.32cm}
	\includegraphics[width=6cm]{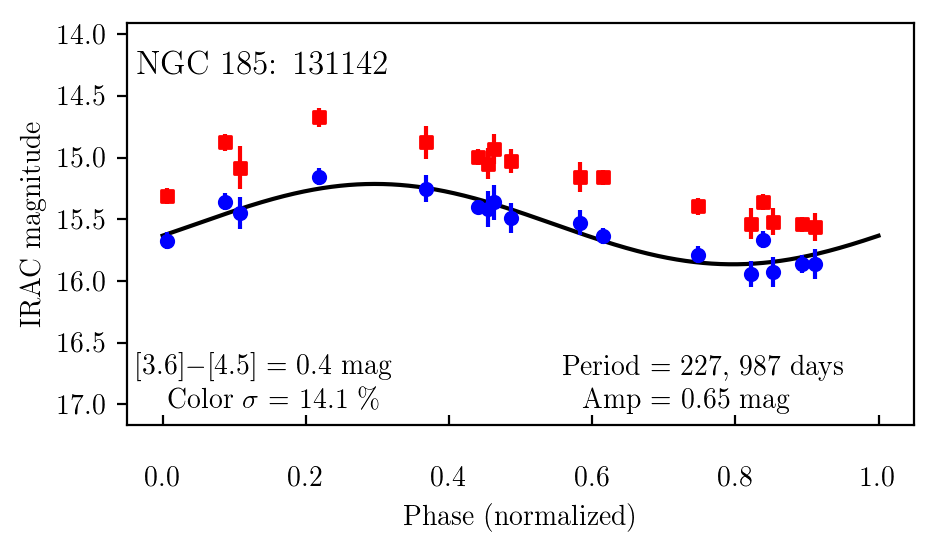} \hspace{-0.32cm}
	\includegraphics[width=6cm]{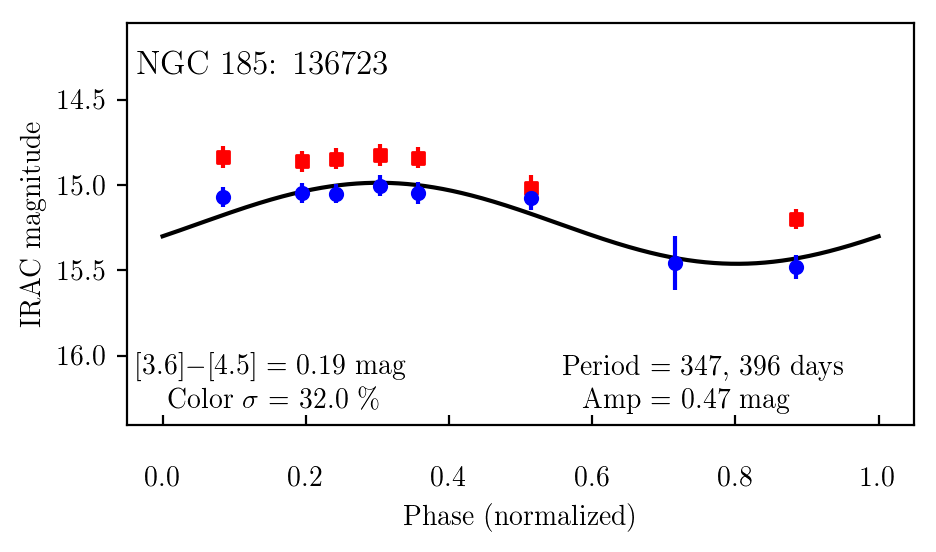} \vspace{-0.75cm}\\

	\includegraphics[width=6cm]{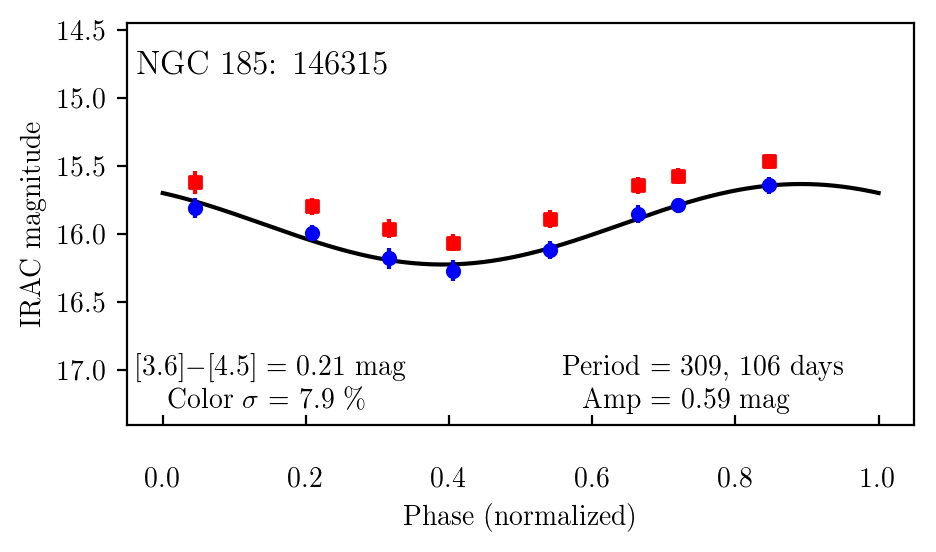} \hspace{-0.32cm}
	\includegraphics[width=6cm]{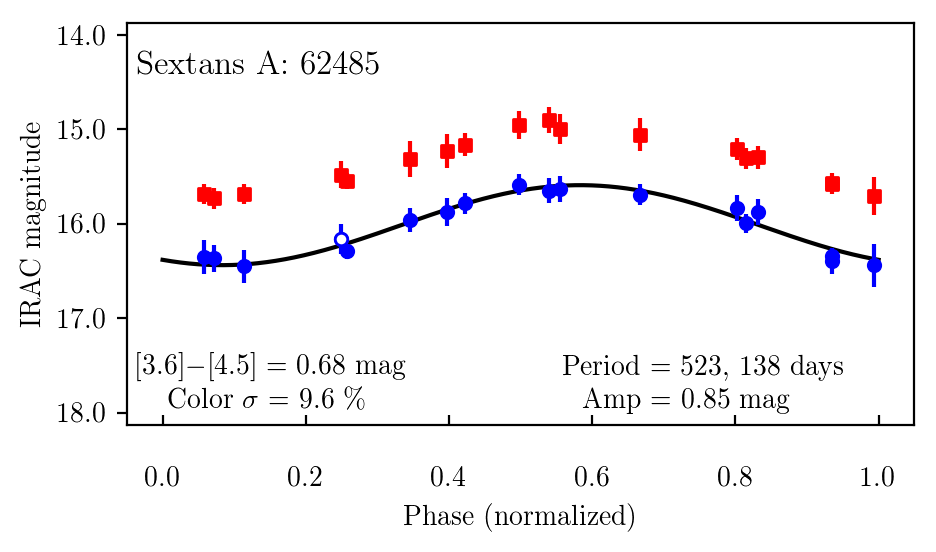} \hspace{-0.32cm}
	\includegraphics[width=6cm]{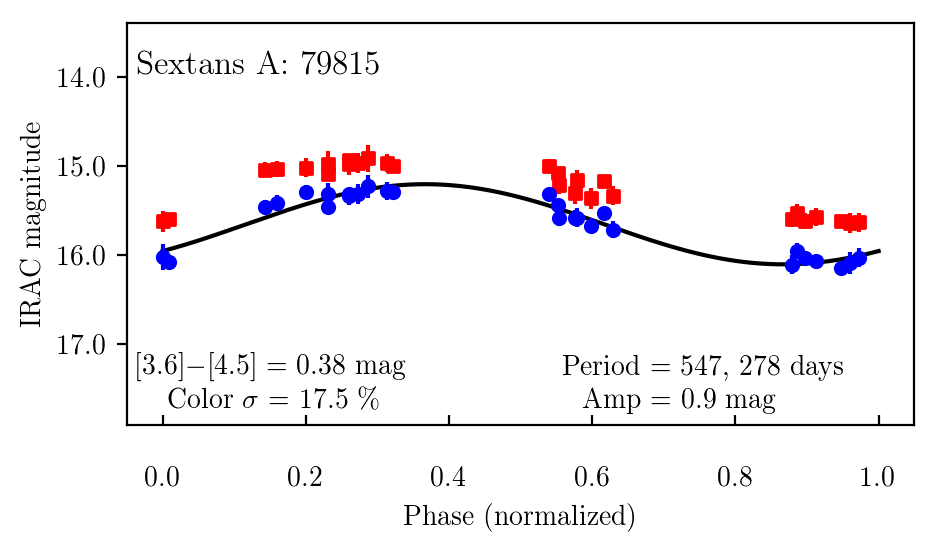} \vspace{-0.75cm}\\

	\includegraphics[width=6cm]{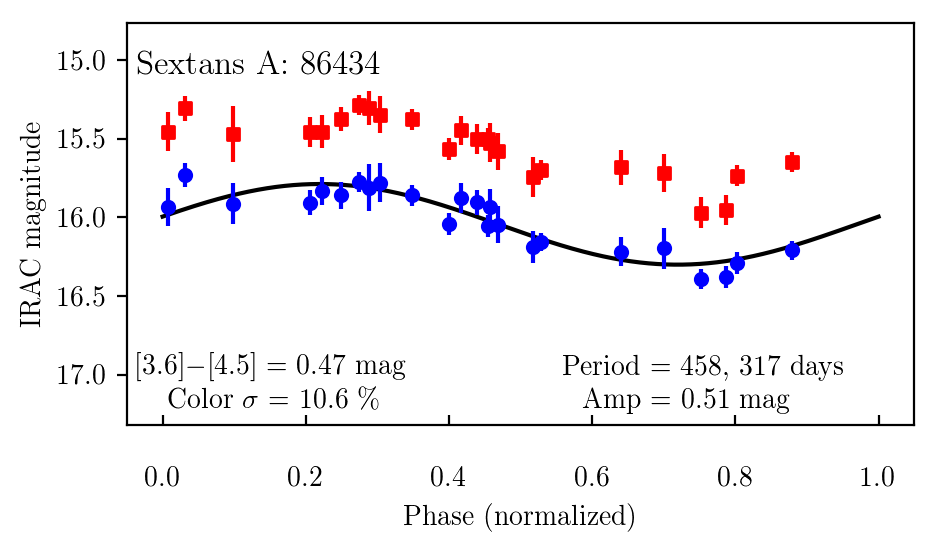} \hspace{-0.32cm}
	\includegraphics[width=6cm]{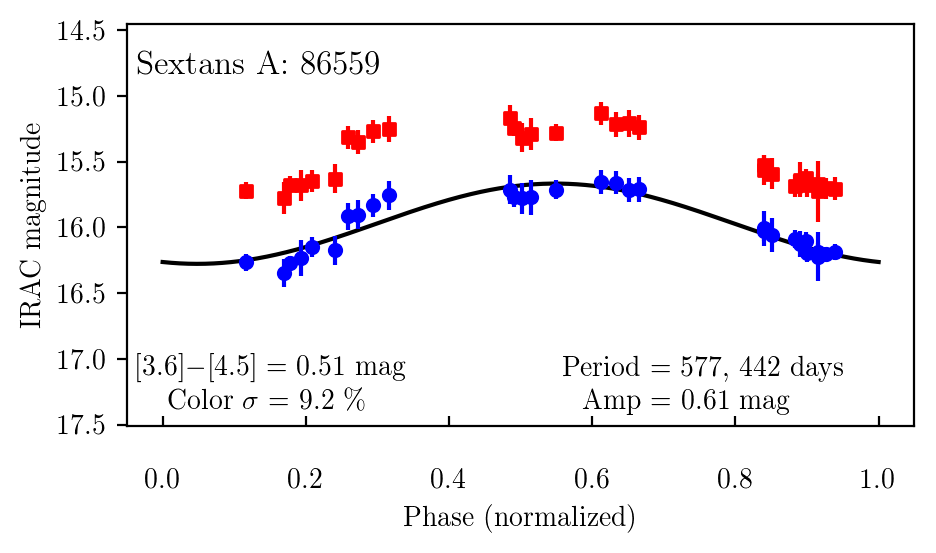} \hspace{-0.32cm}
	\includegraphics[width=6cm]{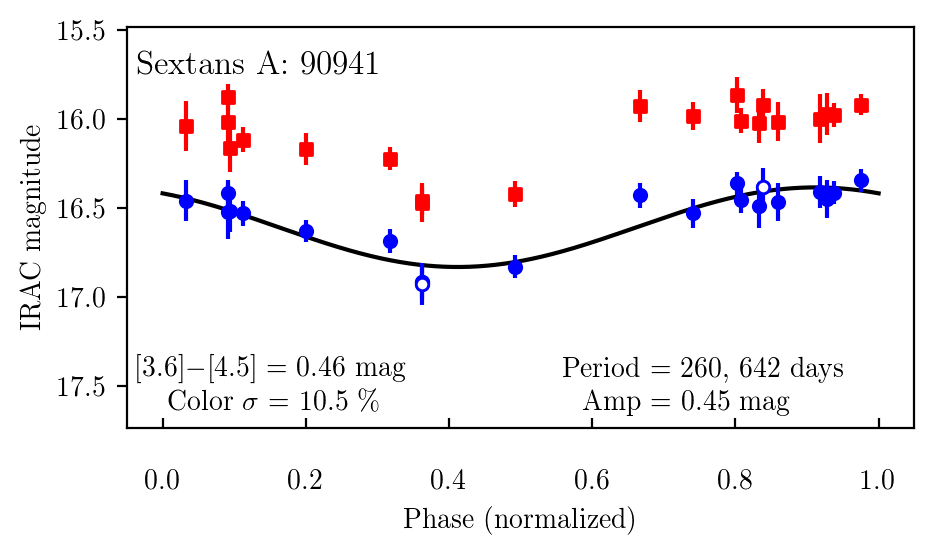} \vspace{-0.75cm}\\

	\includegraphics[width=6cm]{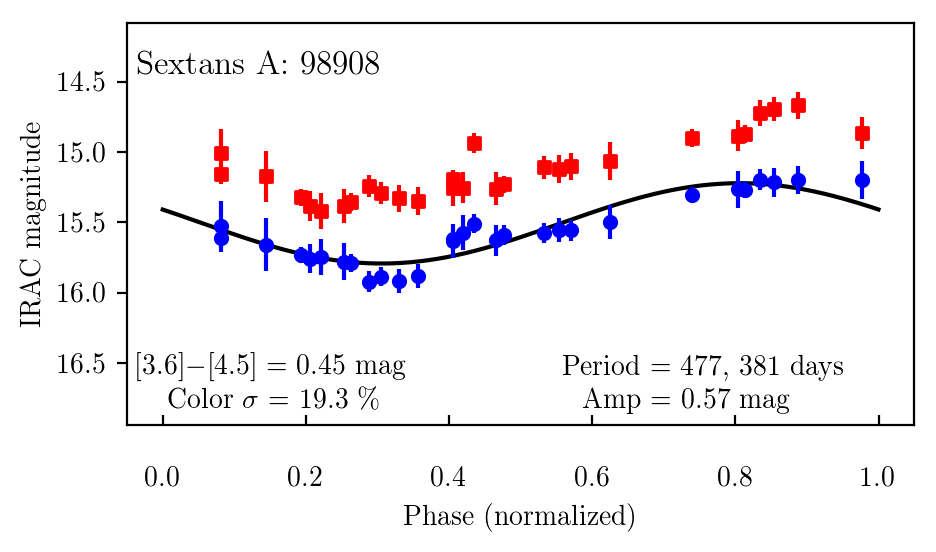} \hspace{-0.32cm}
	\includegraphics[width=6cm]{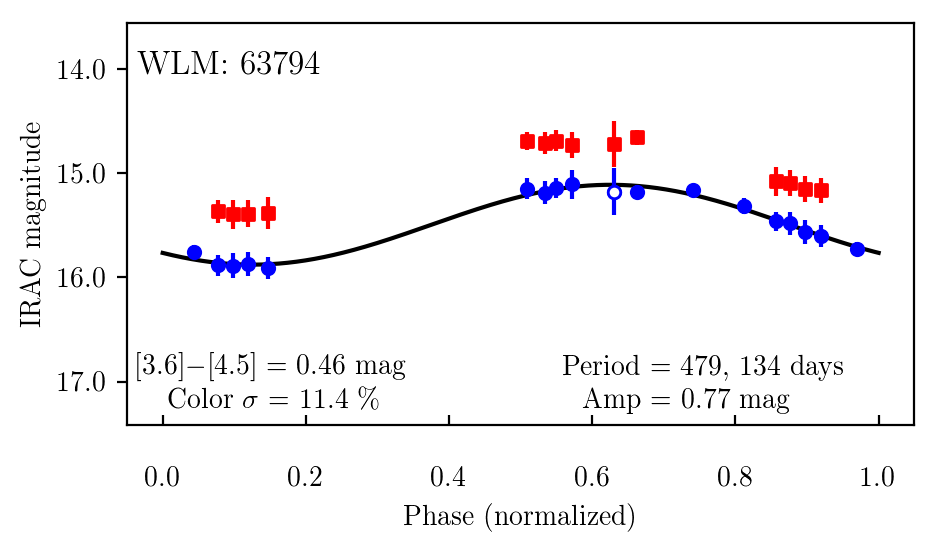} \hspace{-0.32cm}
	\includegraphics[width=6cm]{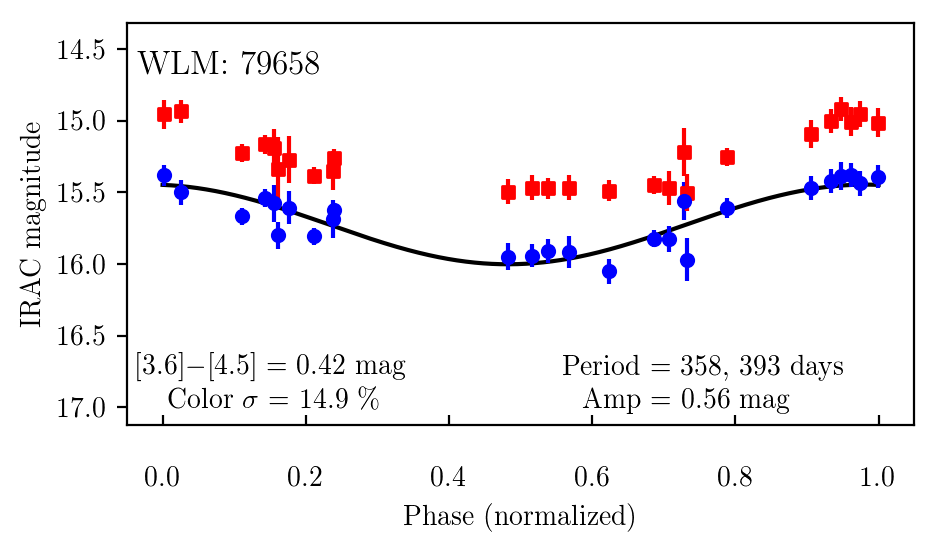} \vspace{-0.75cm}\\

	\includegraphics[width=6cm]{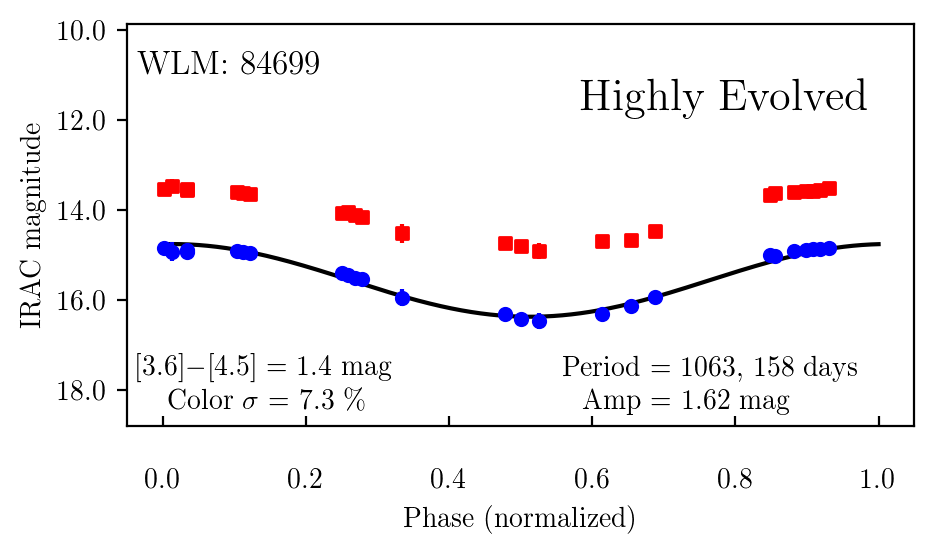} \hspace{-0.32cm}
	\includegraphics[width=6cm]{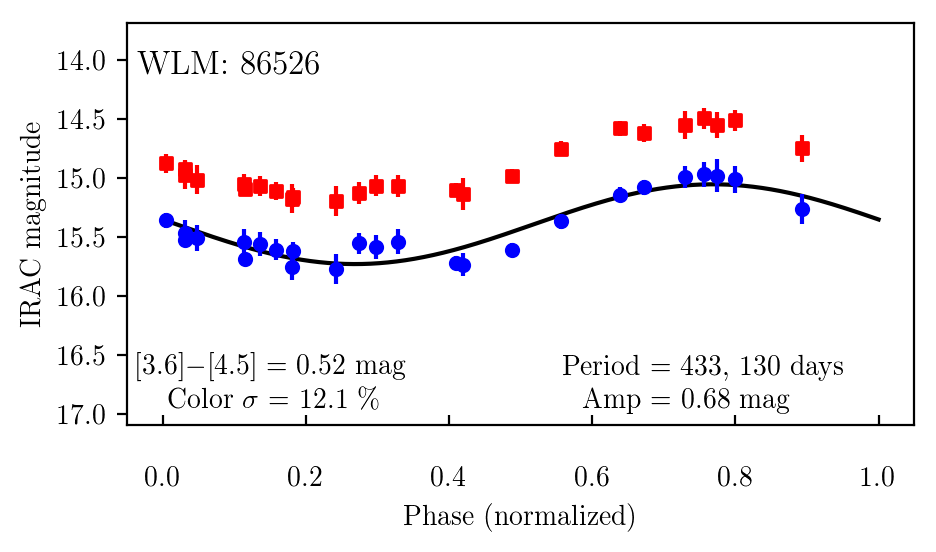} \hspace{-0.32cm}
	\includegraphics[width=6cm]{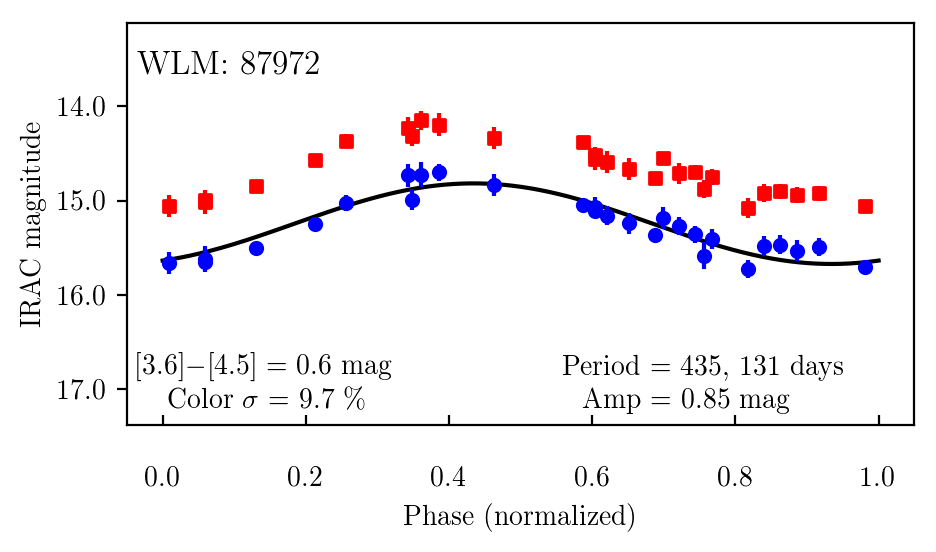} \\
   \end{center}
   % \vspace{-0.45cm}
   \caption{continued}
\end{figure*}

\clearpage

\addtocounter{figure}{-1}
\begin{figure*}
  \begin{center}
	\includegraphics[width=6cm]{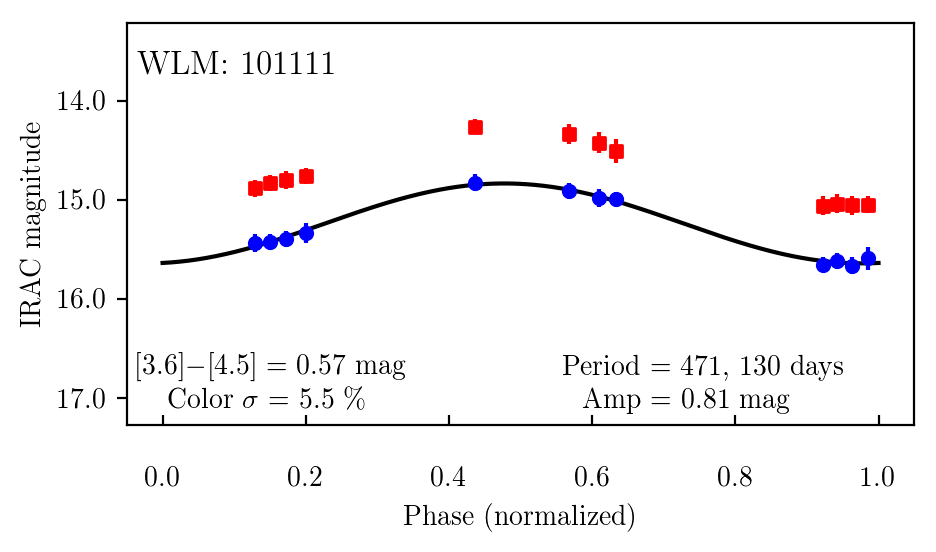} \hspace{-0.32cm}
	\includegraphics[width=6cm]{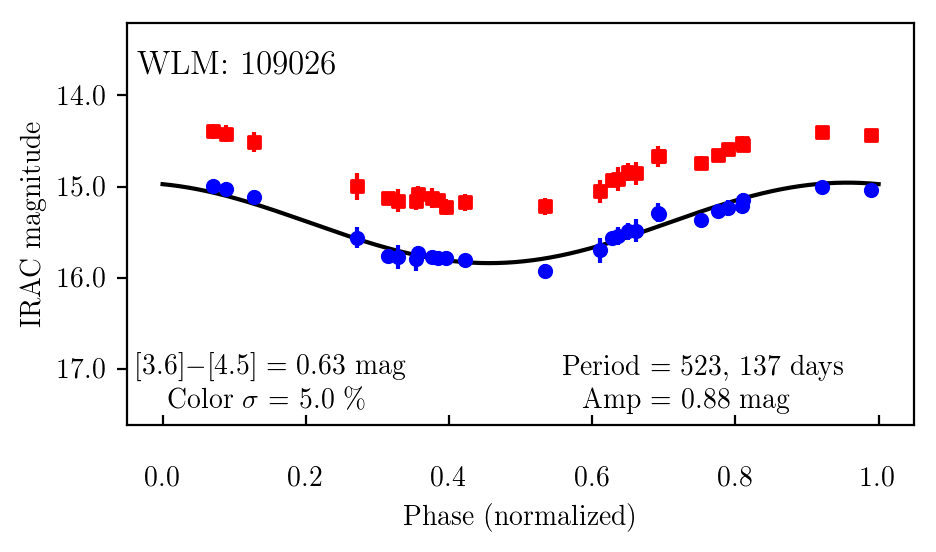} \hspace{-0.32cm}
	\includegraphics[width=6cm]{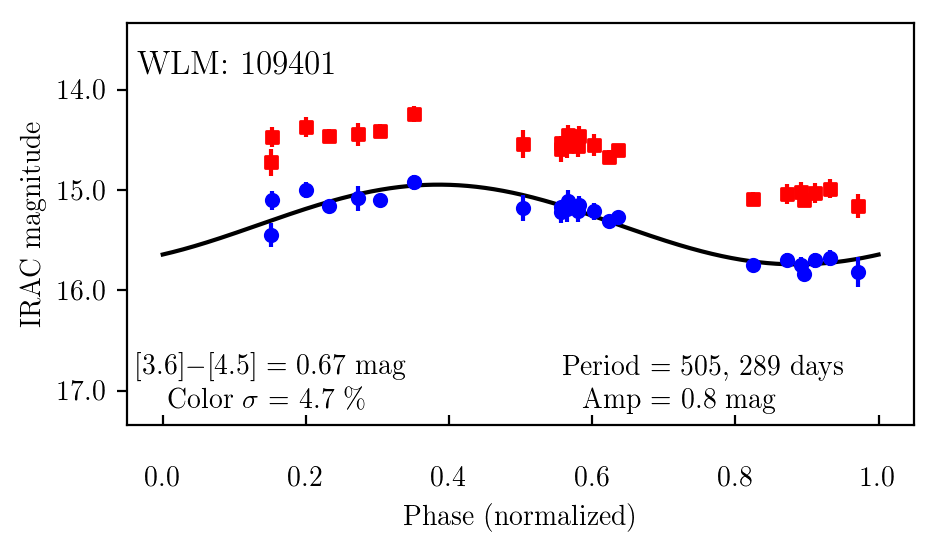} \vspace{-0.75cm}\\ 

	\includegraphics[width=6cm]{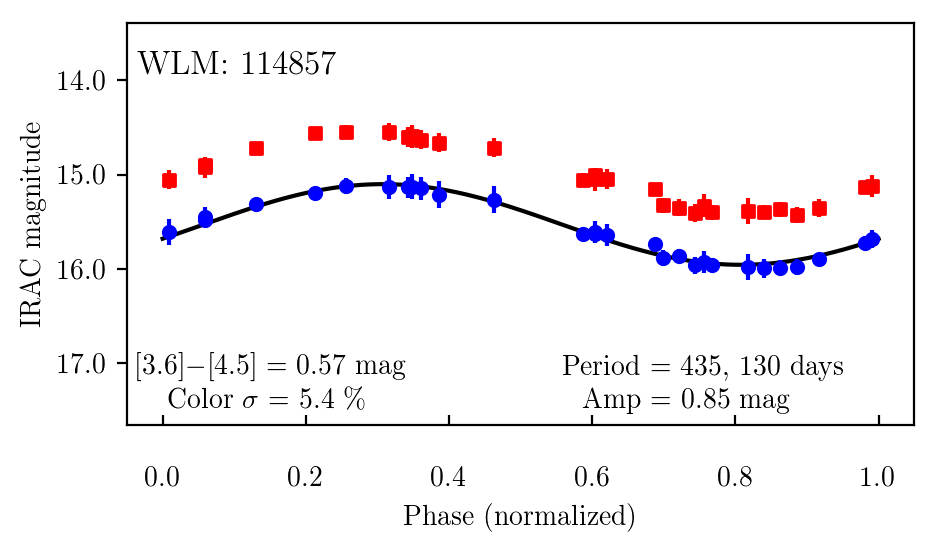} \hspace{-0.32cm}
	\transparent{0}\includegraphics[width=6cm]{blank.pdf} \hspace{-0.32cm}
	\transparent{0}\includegraphics[width=6cm]{blank.pdf} \\
	\end{center}
   \caption{continued}
\end{figure*}

\begin{figure}[h]
\begin{centering} \hspace{-1.2cm}
\includegraphics[width=0.7\columnwidth]{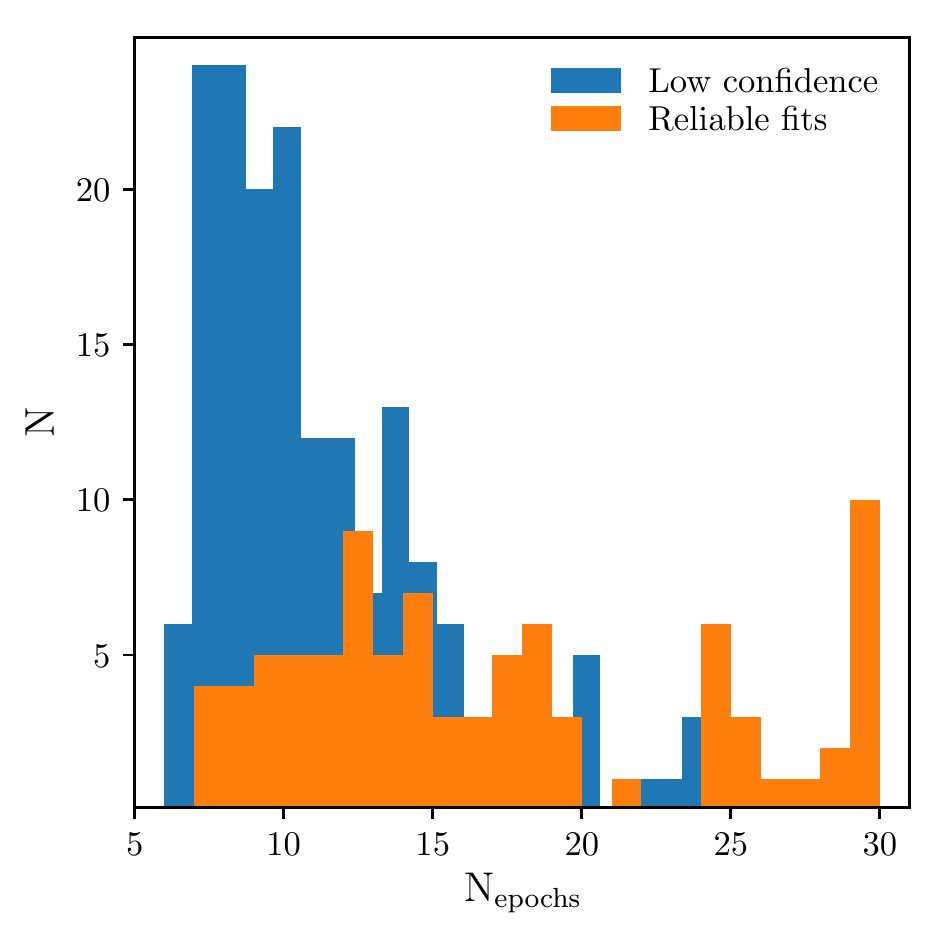}
\caption{A histogram of the number of epochs for the reliable-fit and low-confidence LPVs. Lightcurves with less than 10 epochs are typically categorized as low-confidence LPVs. \\}
\label{fig:histogram}
\end{centering}
\end{figure}

	% \transparent{0}\includegraphics[width=6cm]{blank.pdf} \\

\counterwithin{figure}{section} 
\renewcommand{\thefigure}{B\arabic{figure}}
\setcounter{figure}{0}
\section*{Appendix B: Low-confidence DUSTiNGS variables}

\begin{figure*}
\begin{flushright}
\includegraphics[width=0.97\linewidth]{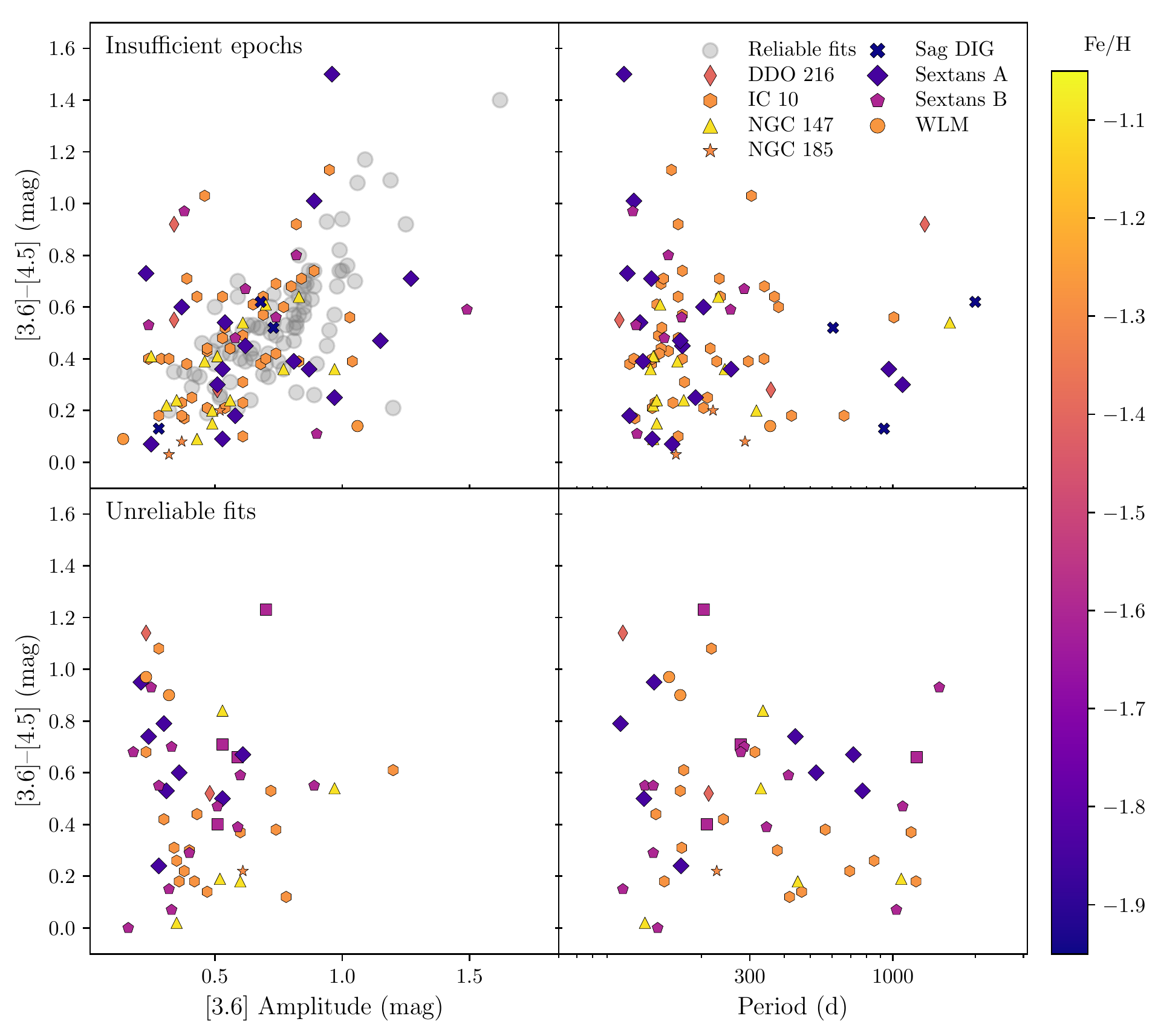}
\end{flushright}
\caption{The same as Figure \ref{fig:color_subplot} but showing the sources designated as insufficient epochs (IE; Top) and unreliable fit (UF; Bottom). While accurate periods could not be measured, the amplitudes are expected to be more accurate. }
\label{fig:color_subplot_low_confidence}
\end{figure*}

\begin{figure*}
\begin{center}
\vspace{-1.5cm}
\includegraphics[width=8.1cm]{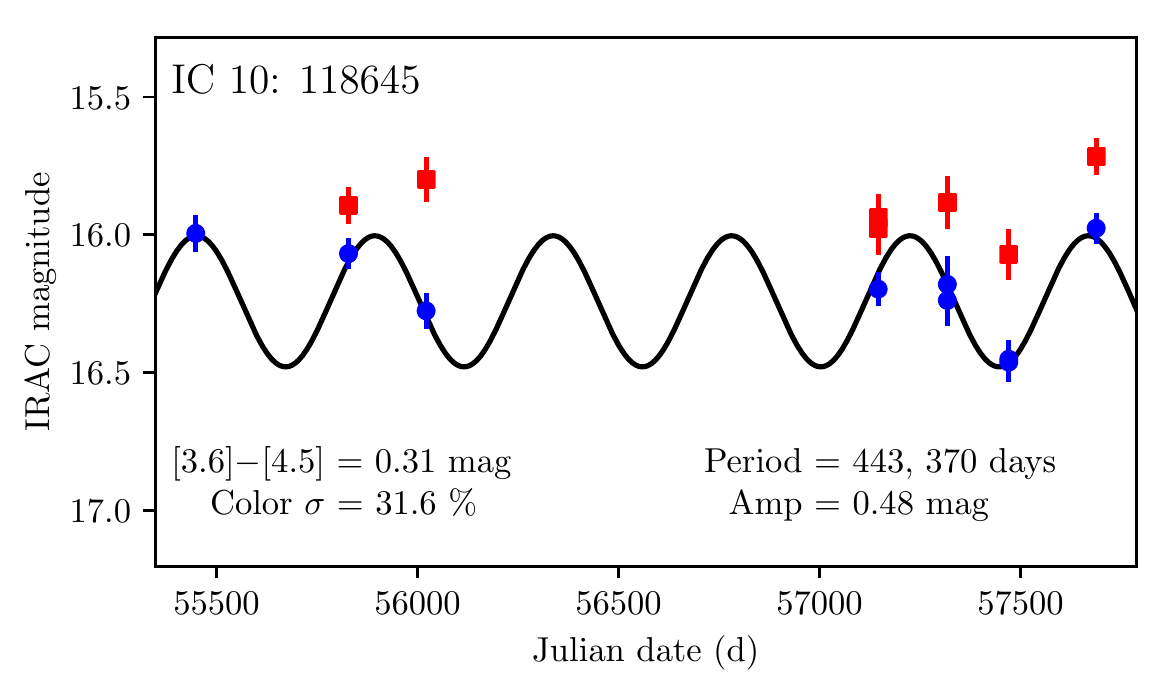}
\includegraphics[width=8.1cm]{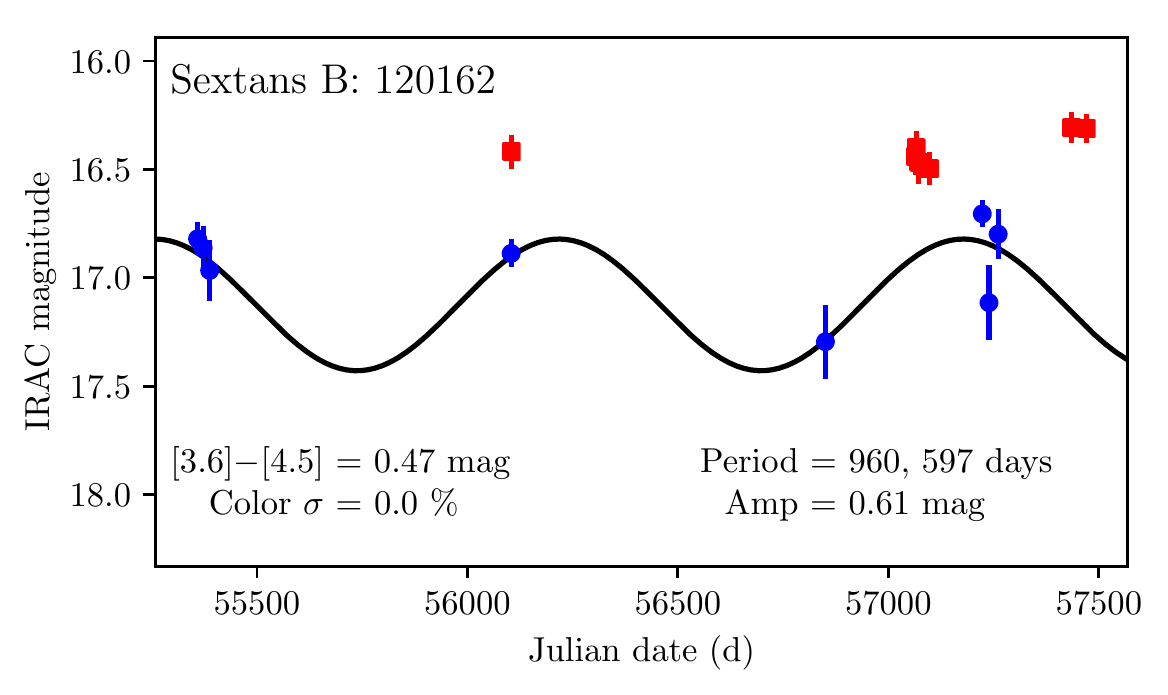}
\caption{Examples of lightcurves with insufficient epochs (Left) and an unreliable fit (Right). The classification of insufficient epochs is only made if a reliable-fit (RF) source could plausibly be fit with a shorter or longer period.  \label{fig:low_confidence_lightcurves}}
\end{center}
\end{figure*}

\begin{figure*}
 \centering
 \includegraphics[width=\linewidth]{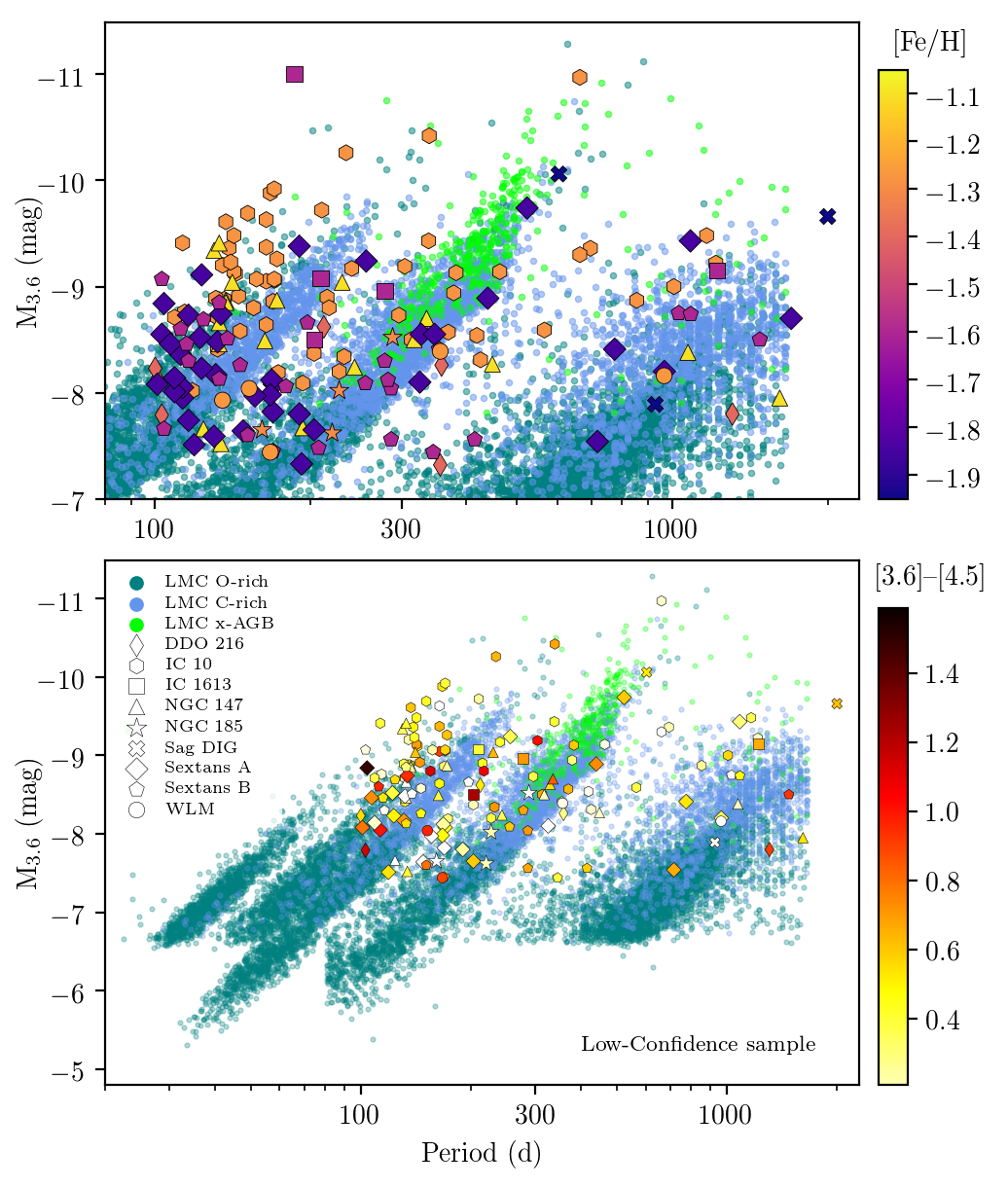}
 \caption{The same as Figure \ref{fig:PL_relation} but showing the low-confidence variables.}
 \label{fig:PL_relation_low_confidence} 
\end{figure*}

\begin{figure*}
 \centering
 \includegraphics[width=\linewidth]{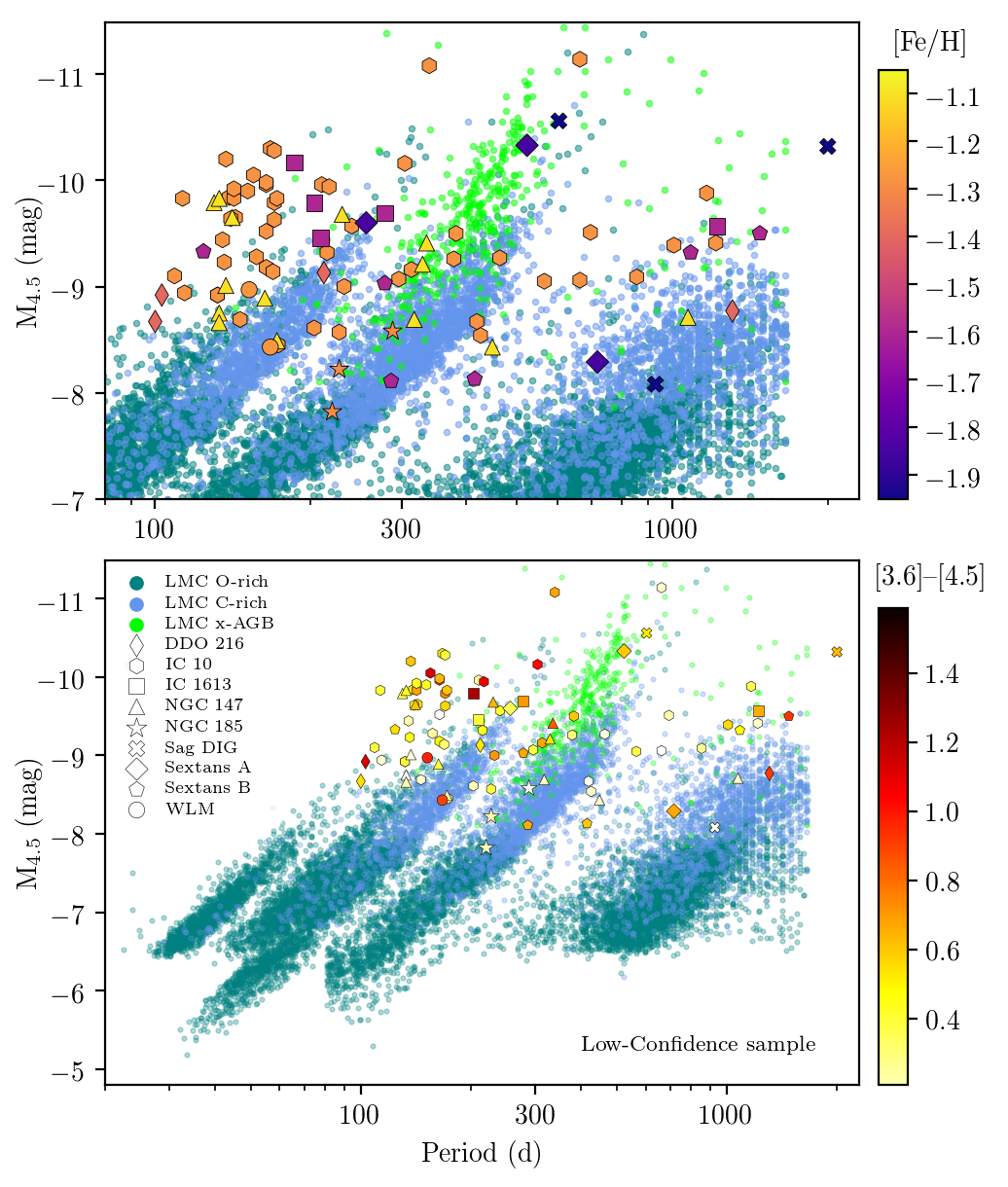}
 \caption{The same as Figure \ref{fig:PL_relation_low_confidence} but showing the 4.5\,$\mu$m magnitudes.}
 \label{fig:PL_relation_low_45} 
\end{figure*}

From our visual examination of our sources we have categorized them into two groups: high- and low-confidence LPVs (described in \ref{section:lightcurve_classification}). The low-confidence variables are those with insufficient epochs to constrain the lightcurve or a poor fit of the model to the data. For the low-confidence variables we show the $P$--$L$ relation and how the pulsation behavior is affected by the [3.6]$-$[4.5] color in Figures \ref{fig:color_subplot_low_confidence}, \ref{fig:PL_relation_low_confidence}, and \ref{fig:PL_relation_low_45}. Examples of low-confidence lightcurves are shown in Figure \ref{fig:low_confidence_lightcurves}; high-confidence variables are shown in Figure \ref{fig:PL_relation}.

\clearpage

\clearpage

\counterwithin{figure}{section} 
\renewcommand{\thefigure}{C\arabic{figure}}
\setcounter{figure}{0}

\section*{Appendix C: DUSTiNGS spatial distribution}

We have mapped the spatial distribution of the high- and low-confidence DUSTiNGS LPVs on the \textit{Spitzer} mosaics from Paper I (Figure \ref{fig:spitzer_mosaics}). Also shown are the footprints of the \textit{HST} observations used in Paper IV to disentangle the oxygen- and carbon-rich evolved AGB stars, and the intersecting region that covers all six of the Cycle 11 observations. We have compared the LPVs detected in these intersection regions to the results of Paper II (Table \ref{table:D2_comparison}) to understand how additional epochs have identified high-confidence variables. Paper II identified 2- and 3-sigma variables using two epochs of data. This survey confirmed a subset of those variables (Table \ref{table:D2_comparison}). Varying spatial coverage between epochs prevented all of the Paper II variables from being confirmed.  However, the increase in the number of epochs in some regions resulted in the discovery of new variables (from 2 to 46 per galaxy) that were not identified in Paper II.  Given our spatially and temporally uneven coverage, the variable stars we confirm here are likely a small subset of the total variable population in each galaxy.

LPVs detected outside of these regions have sporadic temporal coverage and poorly constrained lightcurves, with the exception of a few sources with additional archival data. There is also a high number of low-confidence variables in Sextans A and Sextans B above the intersection region. This is due to a higher number of epochs covering these regions as opposed to in the South.

\setcounter{table}{0}
\renewcommand{\thetable}{C\arabic{table}}

\begin{deluxetable*}{lccccccc}
\tablewidth{\linewidth}
\tabletypesize{\normalsize}
\tablecolumns{8}
\tablecaption{DUSTiNGS results that were covered in every epoch of the Cycle 11 DUSTiNGS Supplementary data  \label{table:D2_comparison}}

\tablehead{
\multirow{2}{*}{Galaxy} & 
\multirow{2}{*}{Reliable fit} &
\colhead{Insufficient} &
\multirow{2}{*}{Unreliable fit} &
\multirow{2}{*}{LPV\,$5000\Plus$} &
\colhead{Paper II } &
\colhead{Paper II } &
\colhead{3$\sigma$ x-AGBS detected} \\
 & 
 &
\colhead{epochs} &
 &
 &
\colhead{2$\sigma$ x-AGBs} &
\colhead{3$\sigma$ x-AGBs} &
\colhead{in this work} 
}

\startdata
And IX & 0 & 0 & 0 & 0 & 0 & 2 & $\ldots$ \\
DDO 216 & 3 & 1 & 2 & 0 & 0 & 5 & \llap{6}0\,\%  \\
IC 10 & 10 & 31 & 16 & 2 & 11 & 122 & 8\,\%  \\
IC 1613 & 5 & 0 & 0 & 0 & 1 & 10 & \llap{5}0\,\%  \\
NGC 147 & 5 & 10 & 1 & 0 & 2 & 36 & \llap{1}3\,\%  \\
NGC 185 & 16 & 2 & 0 & 0 & 2 & 28 & \llap{5}7\,\%  \\
Sag DIG & 0 & 3 & 0 & 0 & 0 & 5 &  $\ldots$ \\
Sextans A & 5 & 1 & 1 & 0 & 0 & 21 & \llap{2}3\,\%  \\
Sextans B & 0 & 3 & 4 & 0 & 2 & 19 & $\ldots$ \\
WLM & 7 & 0 & 1 & 1 & 1 & 18 & \llap{3}8\,\%  \\
\textit{Total} & 51 & $\ldots$&$\ldots$ &$\ldots$ &$\ldots$ & 266 & \llap{1}9\,\%
\enddata

\tablenotetext{}{\small{\textbf{Note.} Here we list a subset of the results of the lightcurve fitting, that lie in the regions covered by all epochs in the Cycle 11 DUSTiNGS Supplementary data (shown in blue in Figure \ref{fig:spitzer_mosaics}). Also shown are the Paper II x-AGB stars that were also found in those regions, and the percentage of those that were confirmed in this work. \\}}
\end{deluxetable*}

\begin{figure*}
 \centering
 \includegraphics[height=7.7cm]{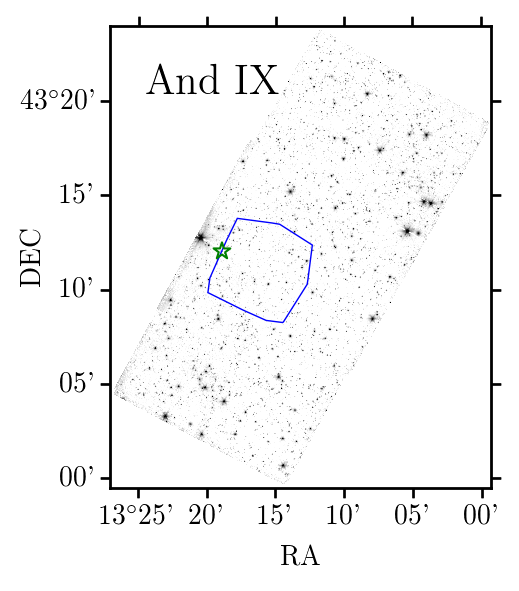} \hspace{-0.2cm}
 \includegraphics[height=7.7cm]{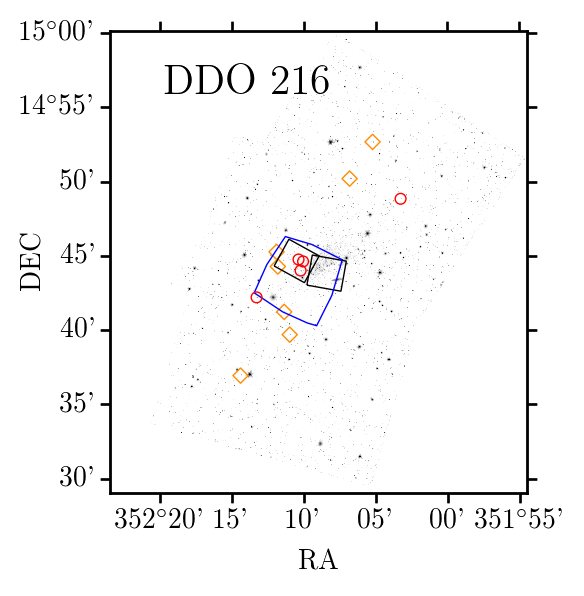} \\ \vspace{-0.25cm} 
 \includegraphics[height=7.7cm]{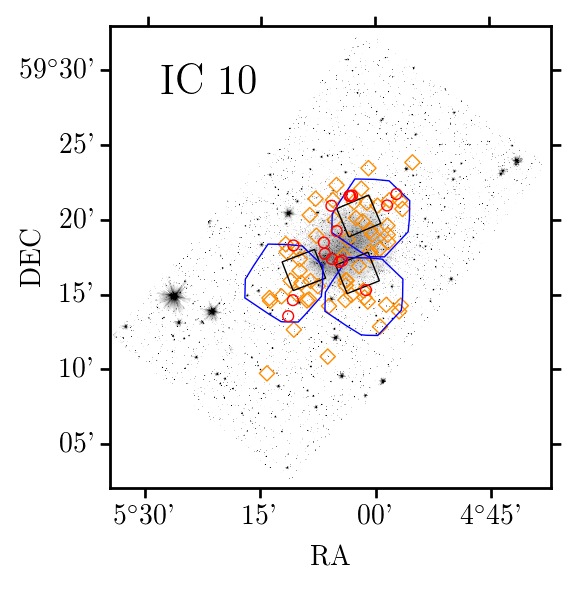} \hspace{-0.2cm} 
 \includegraphics[height=7.7cm]{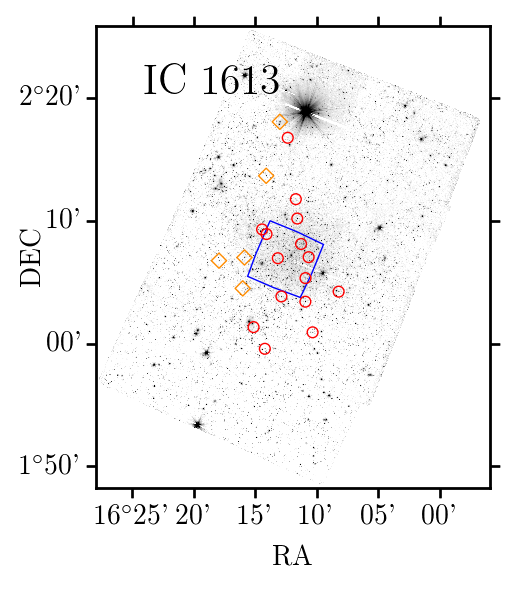} \\ \vspace{-0.25cm} 
 \includegraphics[height=7.7cm]{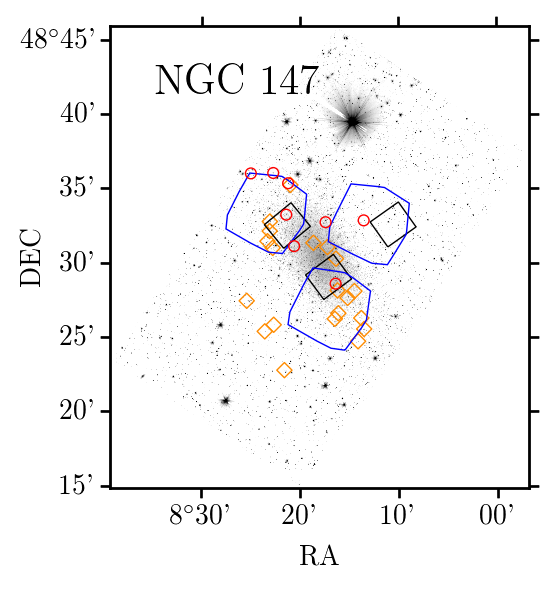} \hspace{-0.2cm} 
  \includegraphics[height=7.7cm]{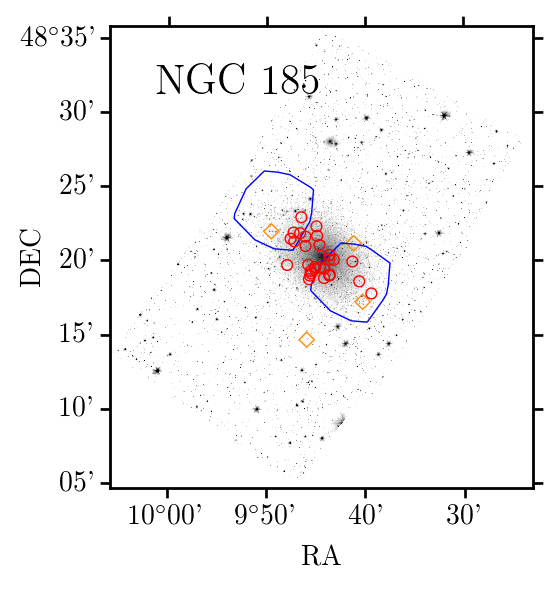} \\ 

 \caption{Spitzer images of the DUSTiNGS galaxies showing the spatial distribution of the high-confidence variables (red), low-confidence variables (orange), the footprint of the \textit{Hubble} observations (black), and intersection region that is covered by all six of the Cycle 11 DUSTiNGS supplementary observations (blue). The source And\,IX 46835 is shown with a green star.}
 \label{fig:spitzer_mosaics} 
\end{figure*}

\addtocounter{figure}{-1} 
\begin{figure*} 
 \centering
 \includegraphics[height=7.7cm]{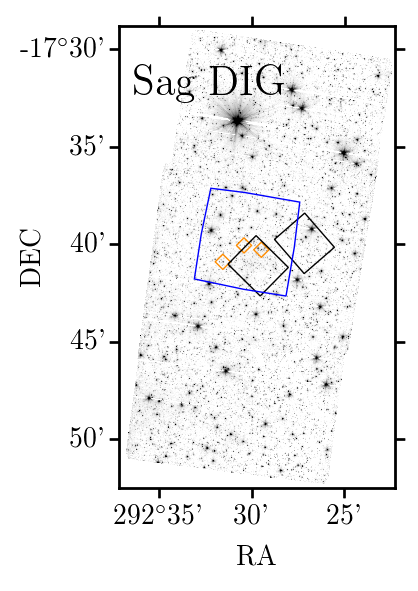} \hspace{-0.2cm} 
 \includegraphics[height=7.7cm]{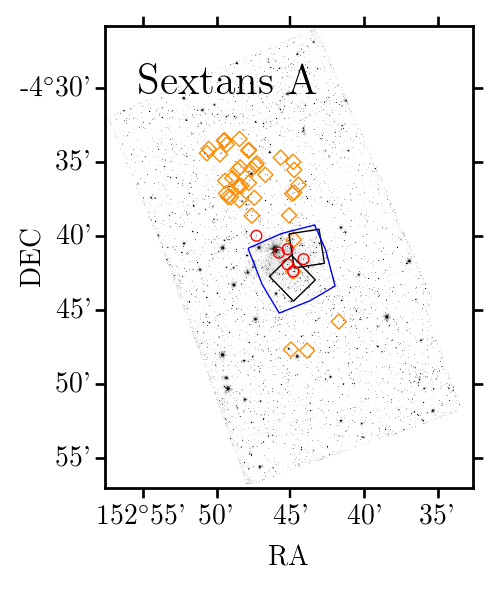} \\ \vspace{-0.25cm} 
 \includegraphics[height=7.7cm]{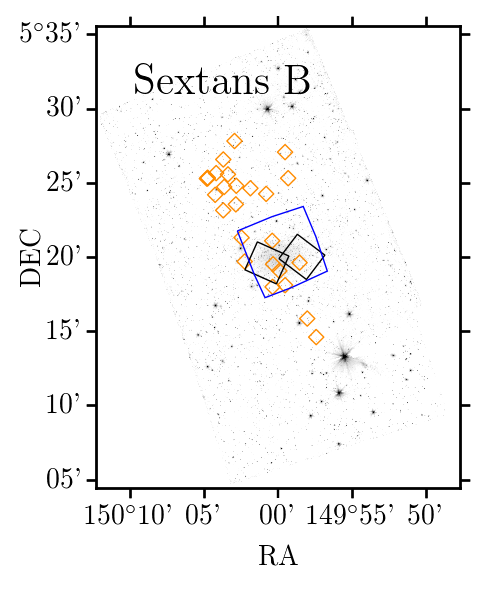} \hspace{-0.2cm} 
  \includegraphics[height=7.7cm]{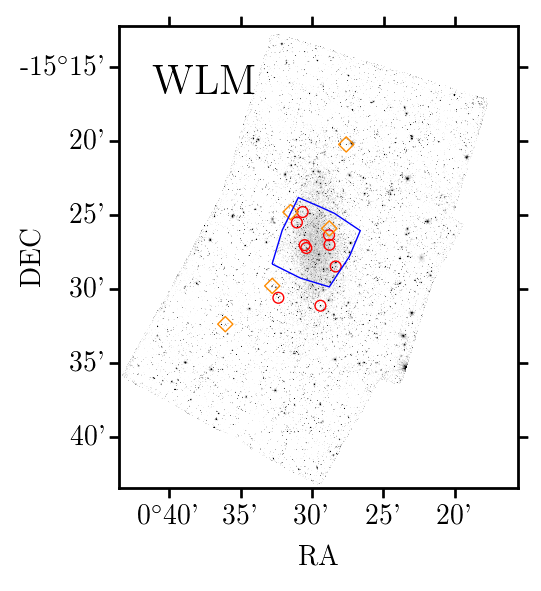} \\ \vspace{-0.25cm} 
 \caption{continued}
\end{figure*}

\end{document}